\documentclass[11pt,a4paper]{article}
\pdfoutput=1
\usepackage{jheppub}
\usepackage{amssymb}
\usepackage[figuresright]{rotating}
\usepackage{color}
\usepackage{amsmath} 
\usepackage{amssymb} 
\usepackage{graphics}
\usepackage{xspace}
\usepackage[gen]{eurosym}
\usepackage{latexsym}
\usepackage[mathscr]{eucal} 
\usepackage{latexsym}
\usepackage[left]{lineno}
\usepackage{rotating}
\usepackage{multirow}

\newcommand{\nubare}{\ensuremath{\overline{\nu}_{e}}\xspace}

\newcommand{\nubarenubare}{\ensuremath{\nubare \rightarrow \nubare\xspace}}

\newcommand{\deltatre}{\ensuremath{\Delta m^2_{31}}\xspace}
\newcommand{\deltadue}{\ensuremath{\Delta m^2_{32}}\xspace}

\newcommand{\boss}[2]{\ensuremath{\rlap{\kern-2.5pt\ensuremath{\overset{\scriptscriptstyle(-)}{\phantom{#1}}}}{\ensuremath{{#1}_{#2}}}}}

\newcommand{\rOMAI}     {INFN, Sezione di Roma Tre, 00146 Roma, Italy}
\newcommand{\pADOVAI}          {INFN, Sezione di Padova, 35131 Padova, Italy}
\newcommand{\rOMAU}     {Dipartimento di Fisica dell'Universit\`a  di Roma Tre, 00146 Roma, Italy}
\newcommand{\pADOVAU}          {Dipartimento di Fisica e Astronomia dell'Universit\`a  di Padova, 35131 Padova, Italy}
\newcommand{\msu}  {Department of Physics, Moscow State University, 119991 Moscow, Russia}
\newcommand{\inr}  {Institute for Nuclear Research RAS, 117312 Moscow, Russia}

\begin{document}


\title{A new way to determine the neutrino mass hierarchy at reactors}
\vskip 10pt
\author[a]{L.~Stanco}
\author[b,c]{G.~Salamanna}
\author[d,e]{A.~Lokhov}
\author[a,f]{F.~Sawy}
\author[a,f]{C.~Sirignano}

\affiliation[a]{\pADOVAI}
\affiliation[b]{\rOMAU}
\affiliation[c]{\rOMAI}
\affiliation[d]{\msu}
\affiliation[e]{\inr}
\affiliation[f]{\pADOVAU}

\vskip10pt

\date{\today}

\vskip 20pt
\abstract{
The determination of the neutrino mass ordering is currently pursued by several experiments and proposals. A very challenging
one is its evaluation from reactor experiments based on the tiny interference effect between the \deltatre and  \deltadue
oscillations. 
Current analyses require
several years of data taking and an extreme energy resolution
to achieve anyhow less than 5 $\sigma$.
Referring to the JUNO experimental conditions we developed a completely new technique that would provide
a robust 5 $\sigma$ measurement in less than six years of running.
The two orderings could be discriminated at the price of allowing for two different values of \deltatre. This degeneracy on \deltatre (around $12\times 
10^{-5}$ eV$^2$) can however be measured at an unprecedented accuracy of much less than 1\%, i.e. $10^{-5}$ eV$^2$,
within the same analysis. 
Analogies with the usual $\chi^2$ analysis, where the \deltatre degeneracy is much more important,  are discussed. 
Evaluation and inclusion of systematic errors and backgrounds have been performed, 
the most relevant among them 
being the addition of the two remote reactor plants 250 km away. Baselines of each contributing
reactor core and its spatial resolution have been taken into account. 
Possible results after two years of running and the foreseen initially-reduced available reactor power have been studied, too.
These results confirm the very positive perspectives for JUNO to determine the mass ordering in a vacuum-oscillation 
dominated region.
}

\maketitle


\section{Introduction}\label{sec:intro}

The almost coherent picture established until now for neutrinos
describes the mixing of three neutrino flavour--states, $\nu_e$, $\nu_{\mu}$ and $\nu_{\tau}$, with three  $\nu_1$, $\nu_2$ and $\nu_3$ mass eigenstates~\cite{pdg}. The full coherence is currently missing due to the still uncertain presence of sterile states at the experimental level.
Furthermore, some fundamental parameters have not been determined for the 3-$\nu$ paradigm, yet. 
Namely, the absolute masses of neutrinos
and the amount of the possible leptonic CP violation. On top of that the neutrino mass ordering (MO) still remains undetermined. 
Many basic neutrino observables that are planned to be measured in
currently running and/or upcoming neutrino experiments depend critically on MO, like its Dirac or Majorana nature.
The pursue of MO is therefore a major
effort that many groups are undertaking, from long baseline neutrino oscillation experiments~\cite{nova-nue} to astroparticle studies~\cite{astro}.

Following the initial idea of~\cite{petcov1} the neutrino mass ordering can be evaluated in medium baseline experiments with reactor
anti-neutrinos. In such experiments a spectral distortion of the oscillation pattern driven by the so called atmospheric mass, $\Delta m_{31}^2=m_3^2 - m_1^2$
for the normal hierarchy (NH) or $\Delta m_{23}^2=m_2^2 - m_3^2$ for the inverted hierarchy (IH), can be observed
(see e.g.~\cite{lucas}).
Contrary to the accelerator-based experiments a side benefit of the neutrino-reactor experiments is the independence of 
the oscillation probabilities from $\delta_{CP}$, the CP violating phase as well as from strong matter effects in the contemplated energy region. 
However, as it will be shown later, the current uncertainty on the value of atmospheric mass introduces important degeneracies
in the neutrino oscillations at reactor experiments.
The 20 kton JUNO underground liquid-scintillator (LS) detector~\cite{juno} has been proposed and approved for realization in the south of China,
being the mass ordering evaluation one of its main goals.

Initial studies were dedicated to the MO extraction at reactor neutrino experiments in~\cite{petcov1,petcov2} and subsequently
in~\cite{pet21,pet22,pet23,pet24}, followed by a long list of technical 
evaluations~\cite{pet25,pet26,pet27,pet28,pet29,pet30,pet31,pet32,pet33,petcov3} once the $\theta_{13}$ mixing angle was found
in 2012 to be not very small. 
The JUNO collaboration underwent a rigorous and detailed study in its reference paper~\cite{juno}, too.

MO sensitivity is usually quoted in terms of the chi-square difference:

\begin{equation}
\Delta\chi^2= \chi^2_{min}({\rm IH})-\chi^2_{min}({\rm NH}),
\end{equation}
\noindent where the two minima are evaluated spanning the uncertainties on the three-neutrino oscillation parameters,
namely the solar mass $\delta m_{21}^2=m_2^2 - m_1^2$, the atmospheric mass $\Delta m_{32(13)}^2$,
the CP phase $\delta_{CP}$ and 
the mixing angles $\theta_{12}$, $\theta_{23}$,
 $\theta_{13}$ in the standard parameterization.
For reactor anti-neutrinos there is no dependence on $\theta_{23}$ and, more relevant, on $\delta_{CP}$ because 
the
probability to remain in the same flavour
is measured. Dependences on $\theta_{12}$, $\theta_{13}$ and $\delta m_{21}^2$ uncertainties are not large
but they can be intriguing.
In~\cite{juno} conclusions about MO  bring to a $\Delta\chi^2$ around 15,
for the benchmark of six running years
at the full reactor power and an unprecedented 3\%$/\sqrt{E(MeV)}$ energy resolution.
Up to 3-4 $\Delta\chi^2$ units can be lost when detector systematics, 
different baselines from the reactor cores, flux shape and background uncertainties are included.
JUNO claims that the loss can be recovered using an atmospheric mass precision at 1\% level, to be inferred by other experiments~\cite{juno}.

The $\Delta\chi^2$ evaluation is based on two distinct hypotheses, NH and IH. For each MO  
the best solution is found: the  
 $\chi^2_{min}$ comes from two different best-fit values for NH and IH, and the $\Delta\chi^2$ is the result of the internal
adjustments of the two separate fits. No real understanding of the weight from single contributions is possible.
For example, the analysis in~\cite{pet29} shows that the major contribution to $\Delta\chi^2$ comes from the atmospheric-mass 
parameter, there defined in terms of the effective mass $\Delta m_{ee}^2=\cos^2\theta_{12}\Delta m_{31}^2+\sin^2\theta_{12}\Delta m_{32}^2$ (see e.g. fig.~2
of~\cite{pet29}). From that analysis the sensitivity on NH/IH  could be extremely high
if the same $\Delta m_{atm}^2$ value
is chosen for both hypotheses. Thus, it is valuable to try to identify an estimator that couples NH/IH and decouples the 
$\Delta m_{atm}^2$ dependance. 

This issue was extremely relevant in the early days of the studies about the MH extraction from neutrino reactors
with the introduction of the effective masses~\cite{parke}. Nevertheless, the solution found at that time can be overcome
with a complementary approach as outlined in this paper. In particular, the exact dependence on $\Delta m_{atm}^2$,
which brings to degenerated solutions, has to be clearly identified. For the $\chi^2$ analysis the degeneracy is
hidden in the effective mass definition. Then, the solution was to factorize out that dependence and consider only the
evolution of the so called effective {\em phase}. That, however, corresponded to discard part of the information.
More details are given in appendix~\ref{app:C}. A new way to present results for the $\Delta\chi^2$ is also given there. 
In our view that should motivate the conclusion that more studies are needed to deeper understand the $\Delta\chi^2$ behaviour 
for the MH extraction at reactors.

Our new strategy identifies a new test statistic, following the suggestion in~\cite{lucas}, and its successful application
to the NOvA results in~\cite{lucas1}. 
Coupling of the NH/IH hypotheses is performed by constructing an estimator that includes information from both
of them, as already exploited in other studies of particle physics, e.g. in the determination of the Higgs boson mass~\cite{cls}.
Therefore, it is no longer needed at the final stage to compare the two NH/IH hypotheses, 
which would incur into errors of type I and II~\cite{pdg}. The sensitivity is instead evaluated as rejection of the wrong 
hypothesis.\footnote{
Broader discussions on the $\Delta\chi^2$ test statistic and the way to approach analyses on the mass hierarchy can be found 
in section 3 of~\cite{lucas} and references therein.}
The new identified estimator is bi-dimensional. A procedure to quote the expected sensitivity
has been developed, too.

In the next sections the reactor environment for anti-neutrino oscillations is recalled. The JUNO configuration is then considered,
the new estimator is introduced and its capability to decouple the single parameter dependence is described. 
A side result is the evaluation
of $\Delta m_{31(23)}^2$ at an unprecedented accuracy with reactor neutrinos only.
In the following section results are reported, including detailed studies of the effects due to different systematic sources 
and backgrounds. Conclusions are finally drawn. The appendix~\ref{app:A} reports numerical details of the sensitivities obtained
for different experiment configurations. In appendix~\ref{app:B} a technical reasoning of the procedure used for the $F$ 
estimator, which give insight about the strong correlation between $\Delta m_{31(23)}^2$ and the mass hierarchy
discrimination at neutrino reactor experiments, is reported.  In appendix~\ref{app:C} comparisons with the $\chi^2$ method
are shown, together with a critical discussion and a new way to present the $\chi^2$ results.

\section{The anti-neutrino reactor oscillations}\label{sec:reactor}

In a nuclear-reactor neutrinos are mainly produced from the fission products of  four radio-active isotopes,
$^{235}$U, $^{238}$U, $^{239}$Pu and $^{241}$Pu. Their flux $\phi$ depends on the kind of fission as function of the energy E$_{\nu}$.
For a total thermal power $T$ the rate of  events produced in a proton detector, 
$\sigma_{\bar{\nu}_ep}$ being the anti-neutrino cross section, is
\begin{equation}\label{eq:2}
dN/dE_{\nu}=T\times\phi(E_{\nu})\times \sigma_{\bar{\nu}_ep} \times P_{\nubarenubare},
\end{equation}
modulated by  the oscillation survival probability $P_{\nubarenubare}$.
In scintillator and water Cherenkov detectors anti-neutrino are detected via the inverse beta-decay
$\bar{\nu}_e + p \rightarrow e^+ + n$, with a threshold of 1.8 MeV  for a proton at rest.
The \nubare survival probability
is expressed as
\begin{eqnarray}\label{eq:3}
P_{\nubarenubare}&=&\left|\sum_{i=1}^3 U_{ei}\exp \left(-i\frac{m_i^2}{2E_i}\right)U^{\ast}_{ei}\right|^2\\
&=&1-\cos^4\theta_{13}\sin^2 2\theta_{12} \sin^2 (\Delta_{21}) \nonumber \\
& &\quad -\cos^2\theta_{12}\sin^2 2\theta_{13} \sin^2 (\Delta_{31}) \nonumber \\
& &\quad -\sin^2\theta_{12}\sin^2 2\theta_{13} \sin^2 (\Delta_{32}) \nonumber 
\end{eqnarray}
where $U_{ei}$ is the neutrino mixing-matrix element relating the electronic neutrino to the mass eigenstate $\nu_i$. 
The quantities
$m_i$ and $E_i$ are the mass and the energy of the corresponding mass eigenstate, while $\theta_{ij}$ are the neutrino 
mixing-angles. The oscillation phases $\Delta_{ij}$ are defined as
\[
\Delta_{ij}\equiv\frac{\Delta m^2_{ij} L}{4 E_{\nu}}, \quad (\Delta m^2_{ij}\equiv m^2_i-m^2_j),
\]
with  the baseline length $L$. The matter effect can be neglected to a good approximation for the concerned energy interval and 
baselines~\cite{matter}.
In obtaining the second expression in~(\ref{eq:3}) the tiny energy differences between the three mass eigenstates have been ignored,
$E_{\nu}\sim E_1 \sim E_2\sim E_3$.

In a widely used convention (PDG notation~\cite{pdg}) the three neutrino mass eigenstates are numbered as
\begin{eqnarray}\label{eq:two}
(NH)&\quad m_1 < m_2 < m_3, &\quad \Delta m^2_{atm} \equiv\Delta m^2_{31}\\
(IH)&\quad m_3 < m_1 < m_2, &\quad \Delta m^2_{atm} \equiv - \Delta m^2_{32}
\end{eqnarray}
for the normal and the inverted normal ordering of the spectrum, respectively. We will also use the notation
$\delta m^2_{sol}\equiv \Delta m^2_{12}$. 

Another convention, also widely used, is based on a ri-parameterization
of the atmospheric and solar masses~\cite{mee}, bringing to the definitions of the effective mass-squared differences, 
$\Delta m_{ee}^2$ and $\Delta m_{\mu\mu}^2$, and a $\phi$ phase. 
This parametrization can be quite useful when a fitting procedure 
is applied, due to the monotonic (in neutrino energy) behavior of  $\phi$. 
However, in our procedure no fit will be performed. The parameters will be all marginalized out except for the 
atmospheric one, $\Delta m_{atm}^2$. It can be easily demonstrated that the two parameterizations are analytically
equivalent when applied to the F estimator.

Following e.g. the development in~\cite{petcov2} the survival probability can be re-written as

\begin{eqnarray}\label{eq:6}
P_{NH}&(\nubarenubare)=&1-\frac{1}{2}\sin^2 2\theta_{13}\left(1-\cos\frac{L\Delta m^2_{atm}}{2E_{\nu}}\right)  \\
&& -\frac{1}{2}\cos^4 \theta_{13}\sin^2 2\theta_{12}\left(1-\cos\frac{L\delta m^2_{sol}}{2E_{\nu}}\right) \nonumber \\
+\frac{1}{2}&\sin^2 2\theta_{13}\sin^2 \theta_{12}&\left(\cos\frac{L}{2E_{\nu}}\left(\Delta m^2_{atm}-\delta m^2_{sol}\right)
-\cos\frac{L\delta m^2_{atm}}{2E_{\nu}}\right), \nonumber 
\end{eqnarray}
\begin{eqnarray}\label{eq:7}
P_{IH}&(\nubarenubare)=&1-\frac{1}{2}\sin^2 2\theta_{13}\left(1-\cos\frac{L\Delta m^2_{atm}}{2E_{\nu}}\right)  \\
&& -\frac{1}{2}\cos^4 \theta_{13}\sin^2 2\theta_{12}\left(1-\cos\frac{L\delta m^2_{sol}}{2E_{\nu}}\right) \nonumber \\
+\frac{1}{2}&\sin^2 2\theta_{13}\cos^2 \theta_{12}&\left(\cos\frac{L}{2E_{\nu}}\left(\Delta m^2_{atm}-\delta m^2_{sol}\right)
-\cos\frac{L\delta m^2_{atm}}{2E_{\nu}}\right). \nonumber 
\end{eqnarray}
The only difference between $P_{NH}$ and $P_{IH}$ is in a coefficient of the last terms, 
being either $\sin^2\theta_{12}$ or
$\cos^2\theta_{12}$ for NH and IH, respectively.

For the sake of the following discussion it is relevant to report  the difference between the two expected
event rates for NH and IH (from equation~(\ref{eq:2})):

\begin{eqnarray}\label{eq:8}
\Delta N(E_{\nu})\equiv (dN/dE_{\nu})_{IH}-(dN/dE_{\nu})_{NH}=&\\
T\times\phi(E_{\nu})\times\sigma_{\bar{\nu}_ep}
\times\frac{1}{2}\sin^2 2\theta_{13}\cos 2\theta_{12} \nonumber& \\
\times\left(\cos\frac{L}{2E_{\nu}}\left(\Delta m^2_{atm}-\delta m^2_{sol}\right)
-\cos\frac{L\Delta m^2_{atm}}{2E_{\nu}}\right)&. \nonumber
\end{eqnarray}

The last expression can be easily rewritten as product of different factors:

\begin{eqnarray}\label{eq:9}
\Delta N(E_{\nu})_{IH-NH}=T\times\phi(E_{\nu})\times\sigma_{\bar{\nu}_ep}\times& \\
\sin^2 2\theta_{13}\cos 2\theta_{12}\times\sin\frac{L\delta m^2_{sol}/2}{2E_{\nu}} \nonumber& \\
\times\sin\left[\frac{L}{2E_{\nu}}\left(\Delta m^2_{atm}-\delta m^2_{sol}/2\right)\right]. & \nonumber
\end{eqnarray}

A new estimator that couples NH/IH will get a fast modulation in  $E_{\nu}$ due to the last factor 
of~(\ref{eq:9}), all the other factors showing a slow variation in $E_{\nu}$. 
For example, the modulation due to $\delta m^2_{sol}$ is very smooth because the first minimum
for a generic baseline of 50 km and the current value of $\delta m^2_{sol}$ occurs around 1 MeV, its fast modulation 
being below the 1.8 MeV cutoff.
In principle, for this new estimator, named $F$ from now on, only three parameters will matter: the baseline $L$, 
the atmospheric and the solar masses. Any background contribution slowly varying with $E_{\nu}$ will be factorized out,
as well as the thermal power, the flux and the cross-section. This is a key feature of the new procedure here suggested.
No fit computation will be performed, all the latter quantities being integrated over their uncertainty distributions.

A toy simulation has been developed on a single event basis, having in mind the JUNO configuration as reference.
The oscillation parameters have been taken from the most recent global fits~\cite{lisi-last}.
They are reported in table~\ref{table:1} for convenience.

\begin{table}[h]
\begin{center}
\begin{tabular}{lcc}
\hline
parameter  & best fit & 1 $\sigma$ range \\
\hline
$\sin^2\theta_{12}$ & 0.297 & $\pm 0.017$ \\
$\sin^2\theta_{13}$ & 0.0215 & $\pm 0.0007$ \\
$\delta m^2_{sol}$ & 7.37$\times 10^{-5}$ & 0.16$\times 10^{-5}$ \\
$\Delta m^2_{31} (NH)$ & 256.2$\times 10^{-5}$ & (-3.0+4.3)$\times 10^{-5}$ \\
$\Delta m^2_{23} (IH)$ & 254.5$\times 10^{-5}$ & (-3.2+3.4)$\times 10^{-5}$ \\
\hline
\end{tabular}
\end{center}
\caption{\label{table:1} The most recent best-fit values for the oscillation parameters, as evaluated in~\cite{lisi-last}.
$\Delta m^2_{atm}$ values have been adjusted to fit the definitions here used.}
\end{table}

 It is noticed that the current uncertainty on $\Delta m^2_{atm}$ corresponds to about half 
 $\delta m^2_{sol}$.
 These quantities cause for the modulation of equation~(\ref{eq:9}) and the NH/IH discrimination.
 However, since $\Delta m^2_{atm}\gg\delta m^2_{sol}$ the modulation is driven
 by $\Delta m^2_{atm}$. 
 Therefore, the $F$-estimator should be very robust against $\Delta m^2_{atm}$.
As suggested above and as it will be demonstrated in the next section, the  solar mass $\delta m^2_{sol}$ 
and the baseline $L$ can be 
factorized out. Thus, only the atmospheric mass is left as key variable. We will demonstrate that the $F$-estimator
owns a fine-tuned controlled dependence on the atmospheric mass dependence, providing a unique sensitivity to the neutrino
mass ordering at reactors. Only a rather large degeneracy is left out, around $2\times \delta m^2_{sol}$.
In section~\ref{subsec:3a} we elaborate further on how the correlation between $F$ and $\Delta m^2_{atm}$ shows
in the ideal case; in section~\ref{bias} we hint at a possible strategy to resolve the mentioned degeneracy and determine
the value of $\Delta m^2_{atm}$ in a realistic case, once the mass ordering has been identified by means of $F$.

For the reactor flux we took the detailed simulation performed in appendix of~\cite{juno} and summarized in its fig.~10,
adapted with permission from~\cite{juno:108}. The cross-section has been computed in~\cite{vissani,juno:177}.
The total flux undergoes a 2-3\% uncertainty that may be lowered to 1\% with future studies. A subtlety is given by the
recent discovery of an excess around 5 MeV. We will show that this feature does not affect
the NH/IH sensitivity evaluated with the $F$-estimator. 
In summary, $F$ will be modulated only on the basis of the last factor in equation~(\ref{eq:9}), the other components acting as scaling
factors.

For the  baselines the JUNO configuration has been taken into account, reported in  table~\ref{table:2} for convenience.
Ten reactors in two different sites are considered, Yangjiang and Taishan, at about 52.5 km. In the table the
two remote reactor plants at Daya-Bay and Huizhou are also included. 
Due to their large power they give a sizable contribution (about 12\%) to the total flux. Their effect 
has been taken into account for the systematic studies. Anticipating part of the results, while the systematic uncertainties 
related to the reactor flux contribute to $F$ as simple scaling factors without changing the sensitivity on NH/IH,
the remote reactors constitute a coherent contribution leading to a slight decrease in  significance.

\begin{table}[h]
\begin{center}
\begin{tabular}{lcccccc}
\hline
Cores  & YJ-C1 & YJ-C2 & YJ-C3 & YJ-C4 & YJ-C5 & YJ-C6 \\
\hline
Power (GW) & 2.9 & 2.9 & 2.9 & 2.9 & 2.9 & 2.9 \\
Baseline(km) & 52.75 & 52.84 & 52.42 & 52.51 & 52.12 & 52.21 \\
\hline
Cores & TS-C1 & TS-C2 & TS-C3 & TS-C4 & DYB & HZ \\
Power (GW) & 4.6 & 4.6 & 4.6 & 4.6 & 17.4 & 17.4 \\
Baseline(km) & 52.76 & 52.63 & 52.32 & 52.20 & 215 & 265 \\
\hline
\end{tabular}
\end{center}
\caption{\label{table:2} The thermal power and baseline for the JUNO detector of the
Yangjiang (YJ) and Taishan (TS) reactor cores, as well as the remote reactors of Daya-Bay (DYB) and 
Huizhou (HZ). In our simulation the uncertainty of the source has been assumed to be
$[-5, +5]$ m, uniformly distributed, for the near reactors. A $[-0.5, +0.5]$ km uniform dispersion has been considered 
for the two remote reactors.
Taken from~\cite{juno}, table 2. Courtesy of JUNO collaboration.}
\end{table}

\begin{figure}[htbp]
\begin{center}
\includegraphics[width=9cm,height=9cm]{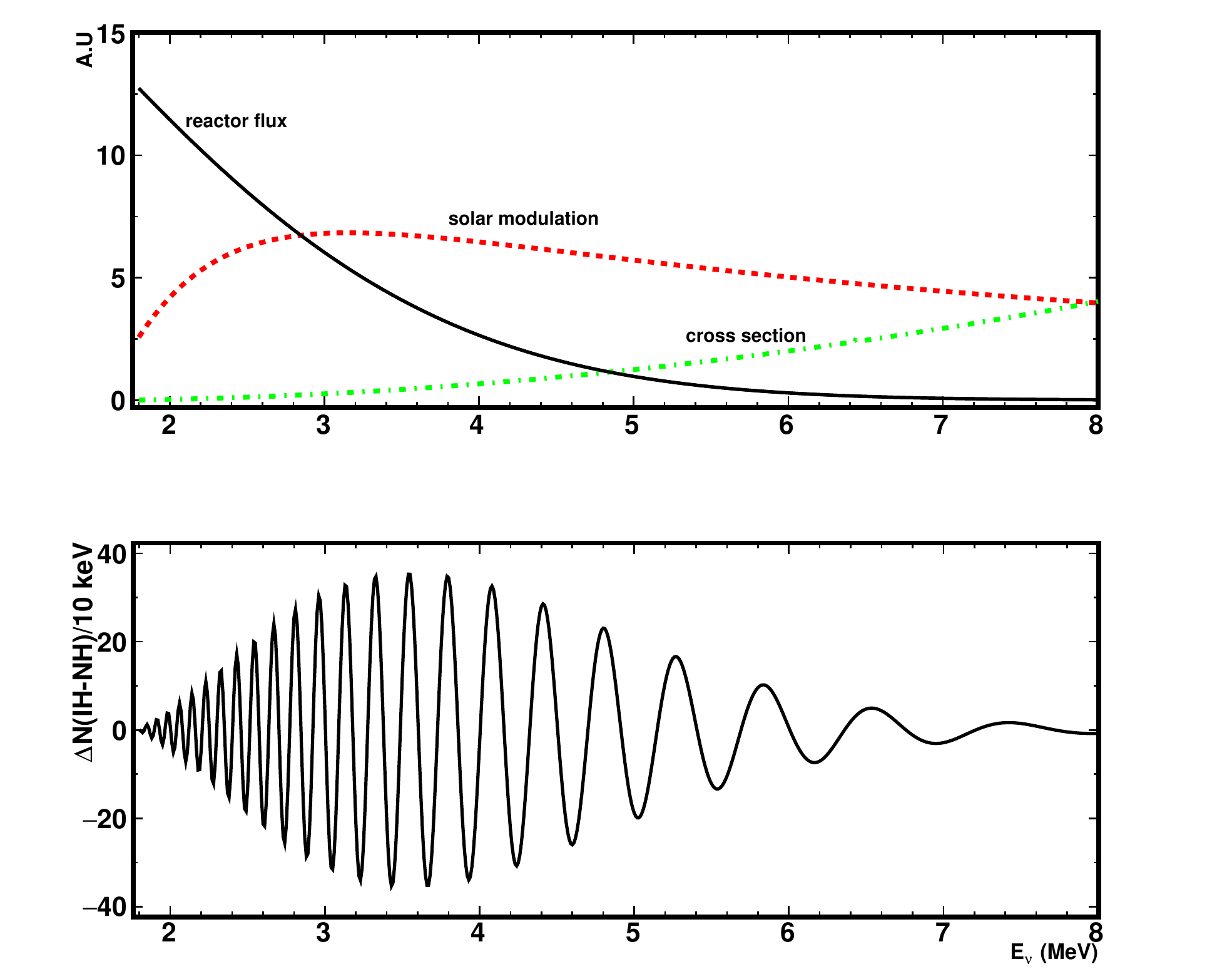}%
\caption{\label{fig:1}(color online) The breakdown of the detected anti-neutrino flux components as function of the \nubare
energy. The components corresponds to the factors in
equation~(\ref{eq:9}), for a baseline L=52.5 km and the neutrino oscillation parameters as in~Table~\ref{table:1}.
Top: reactor flux (continuous line), cross-section (dashed) and the part of the survival probability 
independent of $\Delta m^2_{atm}$ (dotted), i.e. the contribution of the first three terms in equations~(\ref{eq:6}, \ref{eq:7}).
Bottom: differential distribution of $\Delta N$ for six years of JUNO-like data taking, normalized as described in the text. 
The modulation is essentially due to $\Delta m^2_{atm}$.}
\end{center}
\end{figure}

Fig.~\ref{fig:1} reports the breakdown of each factor in equation~(\ref{eq:9})
for a 20 kton JUNO-like experiment and six years of data taking. The overall normalization is given by
the supposed signal event rate of 60 events/day, foreseen by JUNO with its pre-defined acceptance
analysis procedure~\cite{juno}. A total of 108 357 signal events have been used in our simulation, corresponding
to the ten reactor cores, each weighted by its baseline.
Results will be also given by adding the remote reactors, as well as 
for two years of exposure, 8 cores instead of 10.

\section{The new estimator}\label{sec:estim}

The F-estimator is based on the following definition:

\begin{eqnarray}\label{eq:3.1}
F_{MO}=\int_{1.8}^{8.0} \left|\Delta N(E_{\nu})\right| d E_{\nu},
\end{eqnarray}

\noindent the integral of the absolute value of the expression in~(\ref{eq:9}), in the energy range where the modulation due to 
$\Delta m^2_{atm}$
occurs. The underlying idea is to emphasize the energy intervals where one of the two mass hierarchies is expected to
produce more/less events than the opposite one. The intervals are constrained mainly by the combination of  two 
variables, the atmospheric mass $\Delta m^2_{atm}$ and the baseline $L$, all the other parameters acting either as scaling
factors or with a smooth dependence in $F_{MO}$.

$F$ is a  statistic computed on a data set discretized in energy bins of 10 keV  for an easy computation.
For each mass ordering hypothesis we inspect whether the 
``observed" data are above or below the expectation. In particular, for each bin, contribution to $F$ is added only when  
it follows the expectation. Assuming NH to be the true hierarchy, $I^+$ and $I^-$ energy intervals are defined:

\begin{eqnarray}
I^+(E_{\nu}) & \quad \mbox{for $N_{NH}(E_{\nu})>N_{IH}(E_{\nu})$}, \nonumber \\
I^-(E_{\nu}) & \quad \mbox{for $N_{NH}(E_{\nu})<N_{IH}(E_{\nu})$}, \nonumber 
\end{eqnarray}

\noindent where the $E_{\nu}$ dependence is explicitly given.
$F$ is computed via the integrals:

\begin{eqnarray}\label{eq:10}
F_{IH}&=\int_{1.8}^{8.0} \left(N_{obs}-N_{IH}\right) d E_{\nu} \quad \mbox{in $I^+$ when $N_{obs}>N_{IH}$} \\
&+\int_{1.8}^{8.0}\left(N_{IH}-N_{obs}\right)  d E_{\nu} \quad \mbox{in $I^-$ when $N_{obs}<N_{IH}$} \nonumber 
\end{eqnarray}
\begin{eqnarray}\label{eq:11}
F_{NH}&=\int_{1.8}^{8.0} \left(N_{obs}-N_{NH}\right) d E_{\nu} \quad \mbox{in $I^-$ when $N_{obs}>N_{NH}$} \\
&+\int_{1.8}^{8.0}\left(N_{NH}-N_{obs}\right)  d E_{\nu} \quad \mbox{in $I^+$ when $N_{obs}<N_{NH}$} \nonumber 
\end{eqnarray}

\noindent 
Equivalent expressions can be defined when IH is assumed true. 
Then, a 2-D estimator is obtained when, for each MO case, the couple ($F_{NH}$, $F_{IH}$) is computed.

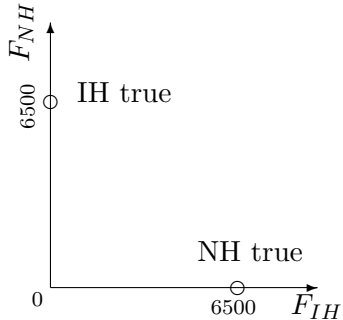
\begin{figure}[htbp]
\begin{center}
\begin{picture}(120,120)
\put(10,10){\vector(0,1){100}}
{\scriptsize \put(3,3){0} \put(70,0){6500}}
\put(10,10){\vector(1,0){100}}\put(100,0){$F_{IH}$}
\put(10,80){\circle{5}}\put(80,10){\circle{5}}
\put(20,80){IH true}\put(65,20){NH true}
\begin{rotate}{90}
\put(95,-2){$F_{NH}$}{\scriptsize \put(70,-4){6500}}
\end{rotate}
\end{picture}
    \caption{Behaviour of the 2-D $F$ estimator (not to scale) for six years of data taking in a JUNO-like experiment. Oscillation
    parameters, reactor power, baseline and normalization to JUNO-like event selection have been
    chosen as described in the text.}
    \label{fig:2}
\end{center}
\end{figure}

If NH is true the ``observed" data set will follow the $N_{NH}(E_{\nu})$ distribution. Ideally, 
$F_{NH} = 0$ and $F_{IH} \sim 6500$ for some choice of the oscillation parameters, reactor flux and
six years of data taking in a JUNO-like experiment,
and neglecting statistical and systematic errors. 
Equivalently, if IH is true for an ``observed" data set one would obtain $F_{IH} = 0$ and $F_{NH} \sim 6500$.
Note that the two mass hierarchy hypotheses are coupled through the definition of the $I^{\pm}$ energy intervals.
In Fig.~\ref{fig:2} a cartoon of the expectations for the two mass hierarchies is depicted.

One of us (A.L.), after the first release of this paper, found an interesting connection of $F$ from first principles
and the technique of the generalized moments.
The estimator F can be, in principle, constructed analytically
via the method of (quasi-)optimal weights~\cite{genmom1}, which provides
a recipe to derive statistical criteria for testing various
hypotheses (in our case the two MH hypotheses).
The resulting estimator is by construction efficient (in the statistical sense), i.e. it efficiently distinguishes between
the two alternative hypotheses.
This provides a rigorous demonstration that $F$ is an optimal estimator for the problem at hand.
At the same time the derived estimator or statistics can be further
modified to be more robust with respect to some systematics or
optimized for computational purposes without significant loss in
sensitivity.
The method itself was successfully used in data analysis of Troitsk
neutrino mass experiment that resulted in the best neutrino mass limit~\cite{genmom2}
and also in constructing specific statistical tests for searches of
anomalies in tritium $\beta$-spectrum~\cite{genmom3}.\footnote{A future paper will be devoted to
the mathematical aspects of the F estimator and its connection to the $\chi^2$ test statistic, too.}

\subsection{$F$-estimator and its coupling to $\Delta m^2_{atm}$}\label{subsec:3a}

\begin{figure}[htbp]
\begin{center}
\includegraphics[width=8cm,height=7cm]{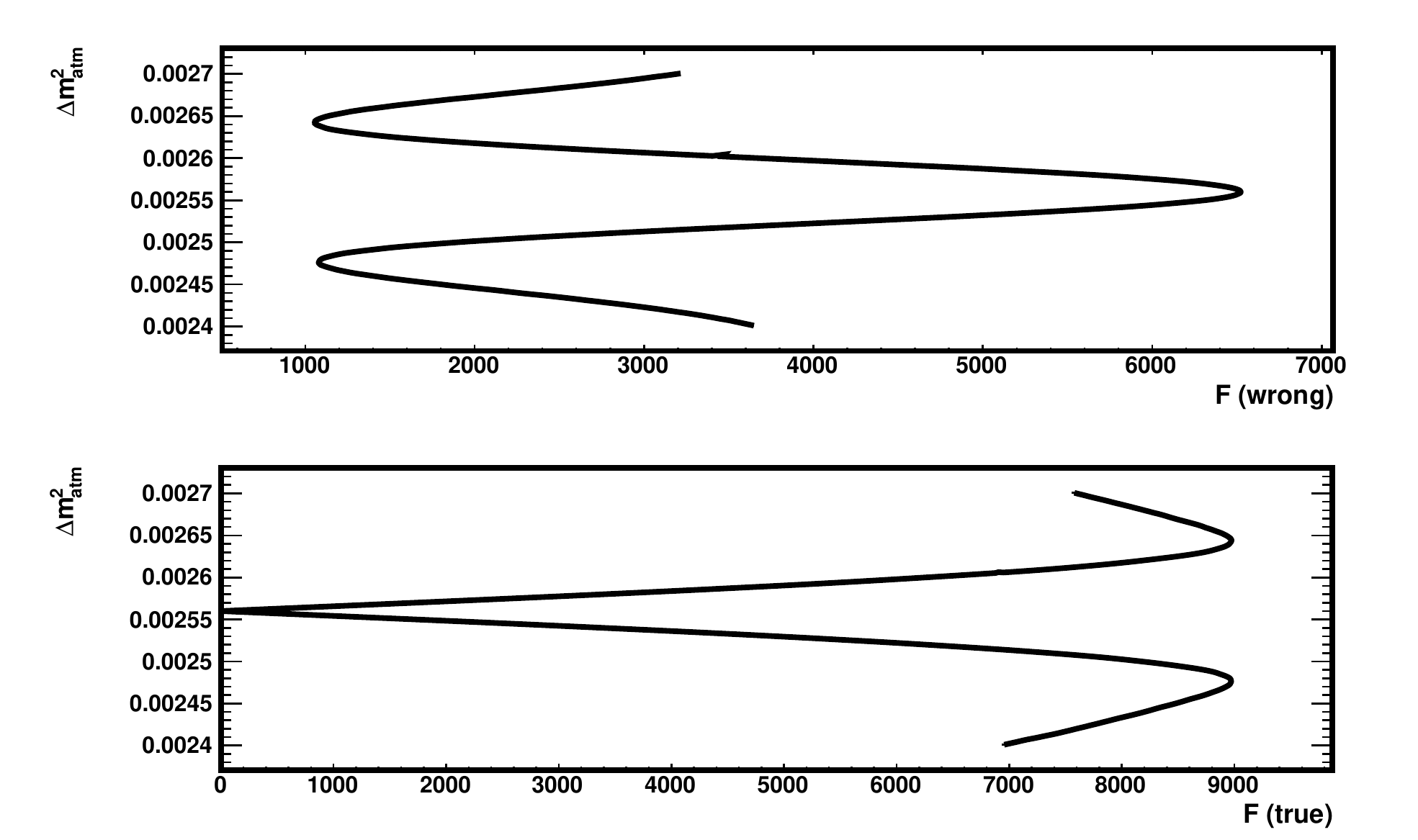}
\caption{\label{fig:3}(color online) Variation of $F$ for the wrong(true) hypothesis on the top(bottom) plot
as function of the chosen $\Delta m^2_{atm}$, for a specific true $\Delta m^2_{atm}$ and MO.
A baseline of L=52.5 km and a true $\Delta m^2_{atm}=0.00256$ eV$^2$ for the NH case have been selected, all the other 
parameters
being fixed  as described in the text to match six years of JUNO-like data taking. 
Note that going from $F_{max}$ to $F_{min}$ or vice-versa 
corresponds to changing $\Delta m^2_{atm}$ of a $\delta m^2_{sol}$ amount.
The non-symmetric behaviour of $F$(true) and $F$(wrong) is due to the initial choice of the model
that constraints the $I^{\pm}$ intervals.}
\end{center}
\end{figure}

The $F$ dependence on $\Delta m^2_{atm}$ enters the definition of the $I^{\pm}$ intervals.
In case the $\Delta m^2_{atm}$ used to construct the $I^{\pm}$ intervals is not the true one,
it is easy to convince ourselves that, given the way $F$ is defined, any difference between the chosen 
and the true $\Delta m^2_{atm}$ 
decreases (increases) the $F$ value of the wrong (true) hypothesis. Fig.~\ref{fig:3} shows the variation of $F$ as 
function of the difference between chosen and true $\Delta m^2_{atm}$.

Ideally, neglecting all other fluctuations, the $F$ estimator allows us to estimate the true $\Delta m^2_{atm}$ with
a remarkable precision. In the next section it will be shown that, even including statistical fluctuations and a finite
 energy resolution,
the dispersion of $F$ around its maximum and minimum, $F_{max}$ and $F_{min}$, 
corresponds to less than 1\% uncertainty on $\Delta m^2_{atm}$\footnote{That accuracy depends on the 
experiment configuration, in particular on its baseline, as analyzed in the appendix~\ref{app:dmatm}.}.

\begin{figure}[htbp]
\begin{center}
\vspace{-4cm}
\includegraphics[width=8cm,height=13cm]{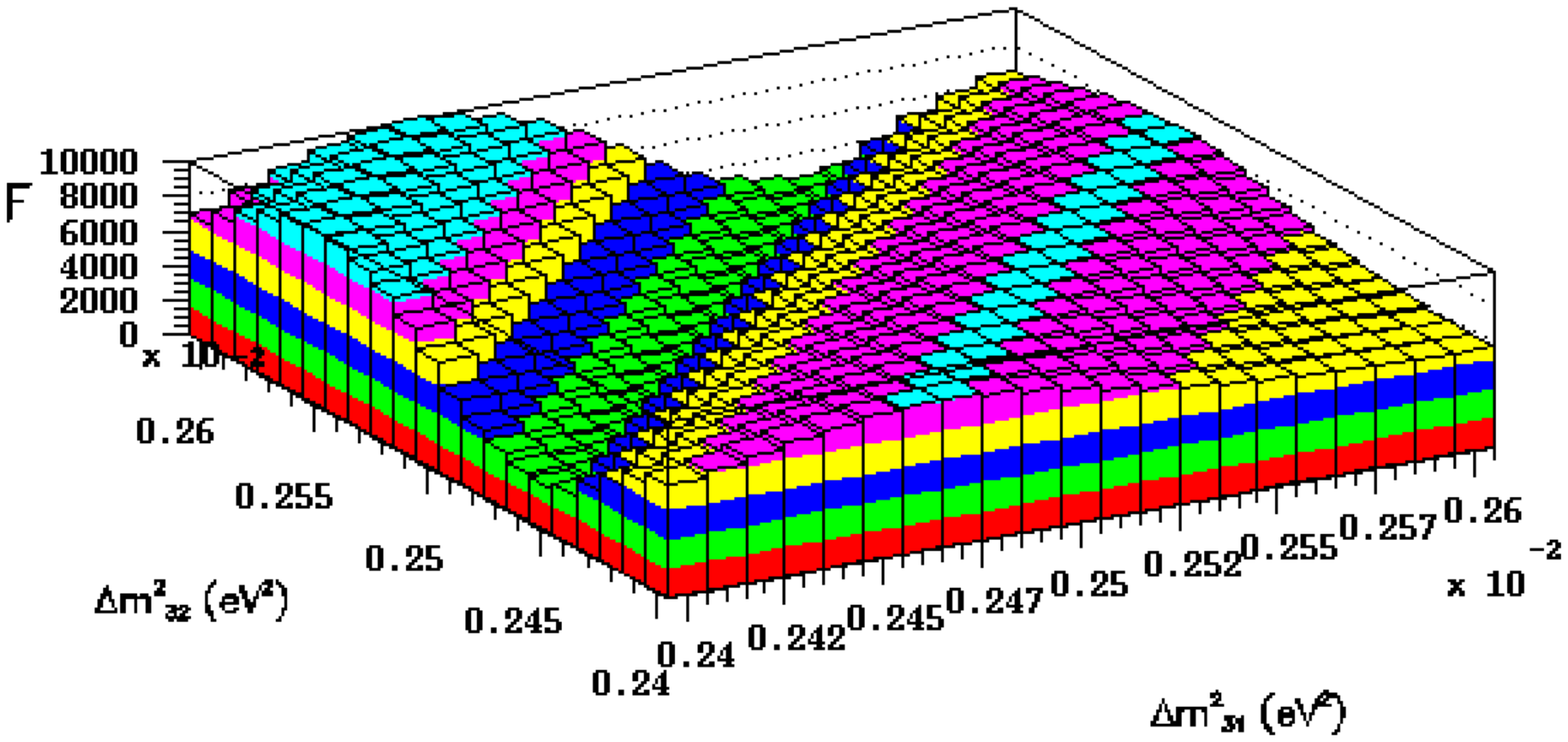}

\vspace{-4cm}
\caption{\label{fig:4}(color online) Variation of $F$(wrong) for $\Delta m^2_{31}$ (NH) vs $\Delta m^2_{23}$ (IH).
 Oscillation parameters and the other variables are chosen as described in the text, for a JUNO-like experiment
and a six years exposure.}
\end{center}
\end{figure}

Fig.~\ref{fig:3} illustrates the issue of which $\Delta m^2_{atm}$ has to be selected for either NH or IH.
In the standard $\Delta\chi^2$ procedure $\Delta m^2_{atm}$ is different for the NH or IH cases since the best fits
for NH and IH select the appropriate value for each of the two hypotheses. In principle, one could define $F$
starting from a $\Delta m^2_{31}$ for NH and a different $\Delta m^2_{23}$ for IH. 
Therefore, expressions~(\ref{eq:6}) and~(\ref{eq:7})
are computed with two different $\Delta m^2_{atm}$.
Fig.~\ref{fig:4} reports $F_{max}$(wrong) for a wide range
of $\Delta m^2_{31}$ vs $\Delta m^2_{23}$. The ciclic behaviour corresponds to the degeneracy 
$\Delta m^2_{31}(NH) = \Delta m^2_{23}(IH ) \pm \delta m^2_{sol}$. 

We suggest that $F$ has to be computed as a function of 
$\Delta m^2_{atm}=\Delta m^2_{31}(NH)=\Delta m^2_{23}(IH)$,
the optimal value being estimated by looking at its influence on $F$ itself (as from Fig.~\ref{fig:3}). 

$F_{max}$(wrong) is quite stable for different choices of the assumed $\Delta m^2_{atm}$ 
(i.e. along the bisector on Fig.~\ref{fig:4}), with a dispersion of about 0.02\%.
In the next section more details will be given on the procedure to be used to extract the correct value of 
$\Delta m^2_{atm}$. More technical aspects are also reported in the appendix~\ref{app:dmatm}, with
further motivations of the above assumption on $\Delta m^2_{atm}$.

\subsection{$F$-estimator and its coupling to baseline $L$}

\begin{figure}[tb]
\includegraphics[width=8cm,height=4cm]{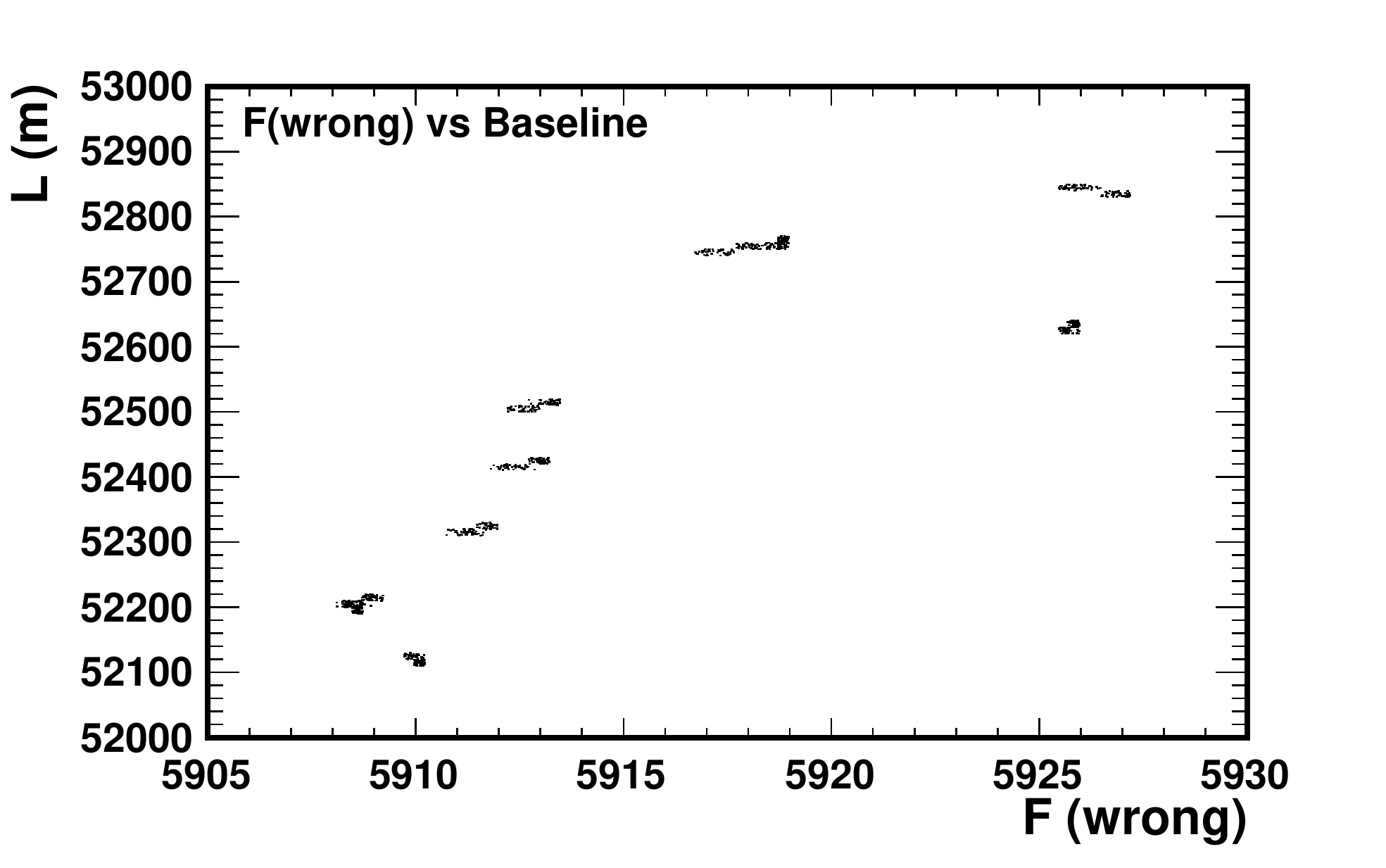}
\includegraphics[width=8cm,height=4cm]{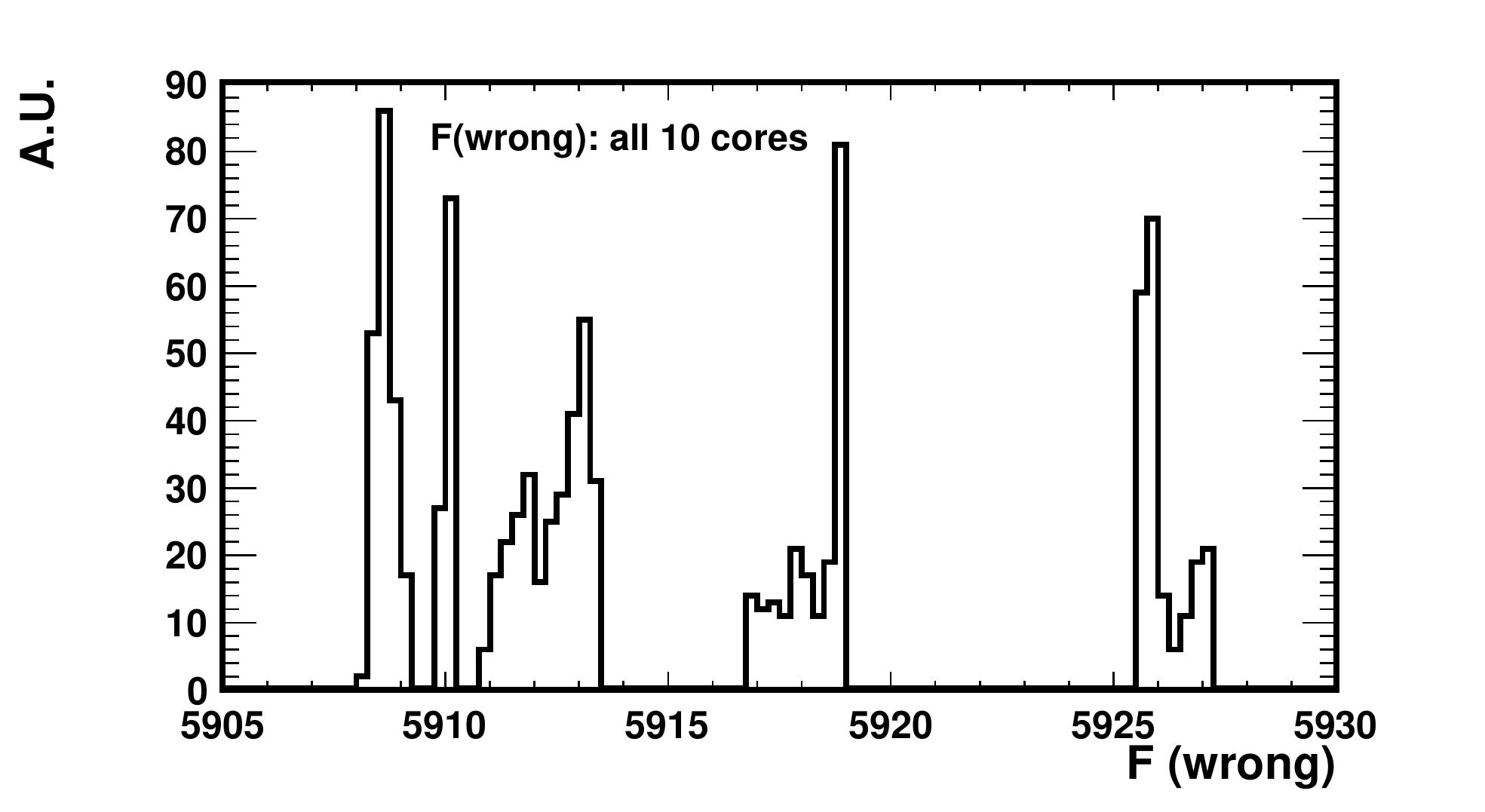}
\caption{\label{fig:5}(color online) Variation of $F$(wrong) due to different baselines' cores.
The 10  baselines correspond to the 10 reactor cores of the Yangjiang and Taishan sites.
A uniform $\pm$ 5 m dispersion for the baseline of each core have been included in the computation.
Oscillation parameters and the other variables are chosen as described in the text, for a JUNO-like experiment
six years long. A factor 0.91 has to be included in $F$(wrong) to scale to the total effective amount
for $L=52.5$ km.}
\end{figure}

$F$ is very sensitive to the baseline $L$. $F_{NH}$ shows a degeneracy with $F_{IH}$ when it is computed 
using an $L$ different from the assumed one ($\pm$1.5~km). In the JUNO experiment there are several reactors
with a baseline difference up to 0.64 km (Table~\ref{table:2}). Therefore, it is mandatory to properly 
handle the baseline composition
of the event sample. The most straightforward way is to weight the sub-samples with the thermal power of 
each single reactor core.
The procedure is acceptable as far as $F$ is stable under different choices of $L$,
as demonstrated in Fig.~\ref{fig:5}.

When the exact baseline of each core is taken into account and properly weighted to its power, $F$ drops from
6503 to 5918, due to the slight negative interference between the different reactor-baselines.
Taking into account the whole set of the ten reactor cores 
a global 0.3\% variation is obtained on $F$(wrong).
In~Fig.~\ref{fig:5} a uniform $\pm$~5 m dispersion has also been added to the baseline of each core.
The net result is totally negligible corresponding to about a factor 10 smaller effect (0.03\%).

One can conclude that $F$ is stable under different choices of the baseline $L$.
In the following analyses are performed weighting the single data sub-sample with the thermal power of each reactor core,
putting in eq.~(\ref{eq:9}) the corresponding $L$.

\subsection{Statistical fluctuations and energy resolution}

Toy Monte Carlo simulations have been implemented to include one
after the other the different sources of dispersion on $F$. A discretized version has been used with a 10 keV bin
width. The Monte Carlo samples have been constructed in two independent ways. 

In the first one the discretized cumulative function of
equation~(\ref{eq:2}) has been used to generate single events. The Poisson statistical fluctuation is automatically taken into account.
On the generated single event one can further add any of the foreseen systematic errors via a Gaussian distribution
centered at the expected mean and with the standard deviation of the estimated uncertainty. For example,
in JUNO a global 3\%/$\sqrt{E(MeV)}$ resolution on the energy reconstruction is foreseen. 

It is worth pointing out at this time that, as specified in~\cite{juno}, such global resolution encompasses already 
the two sources of 
detector response: on one hand the statistical fluctuations (in this particular case, the so-called `photo-statistics" in the 
LS),
on the other the systematic uncertainty associated to the correction of the linearity in energy of the detector response 
(for example from non-uniformity in the LS, photomultiplier manufacturing and the electronics)

In the toy Monte Carlo such overall uncertainty is added on a single
event basis. The migration effect over the energy bins is so correctly simulated.

The second implementation of the toy Monte Carlo is based on the standard procedure: the multi-parameter space
of the single event generation is set to reproduce the expected event distribution.
The two procedures, which are obviously equivalent, have been developed by different
authors of the paper to independently crosscheck the results.

\section{Results}\label{sec:result}

The following benchmark has been used: a JUNO-like configuration as described before, for six years
of data taking.

The distributions of $F$ have been first looked at by including only the statistical fluctuations 
for 1000 JUNO-like toy experiments, six years long, neglecting the $3\%/\sqrt{E(MeV)}$ energy resolution.

\begin{figure}[tb]
\begin{center}
\includegraphics[width=8cm,height=5cm]{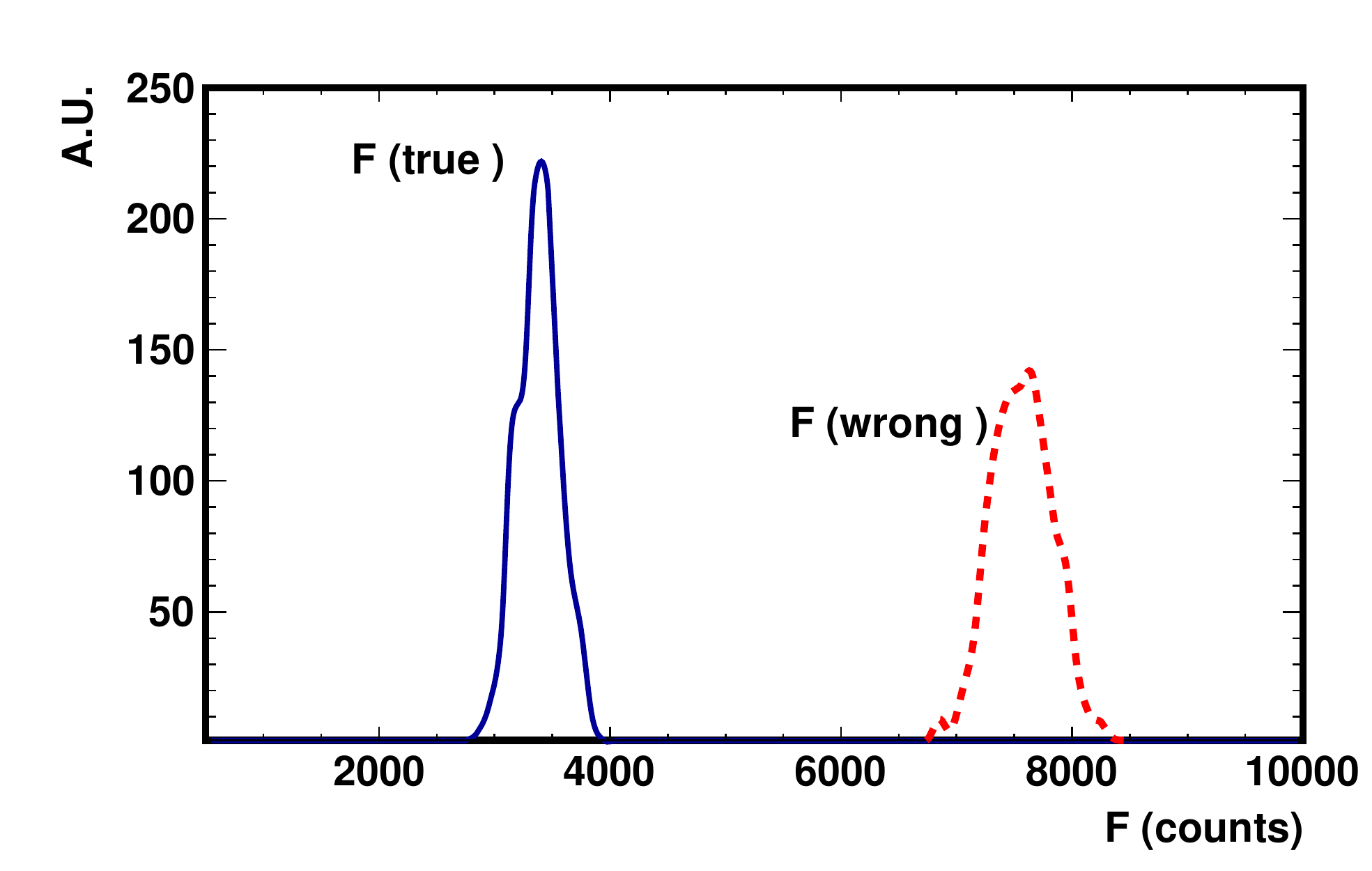}
\caption{\label{fig:6}(color online) $F_{NH}$ (left) and $F_{IH}$ (right) for 1000 JUNO-like toy experiments,
assuming NH and an infinite energy resolution.
The two distributions show the effect due to the statistical fluctuations in the collected number of events.
Oscillation parameters and the other variables are chosen as described in the text, for a JUNO-like experiment
six years long. The ten reactor cores of the Yangjiang and Taishan sites have been taking into account.}
\end{center}
\end{figure}

Given its definition (eq.~(\ref{eq:3.1})), $F$ is expected to be centered at larger values than the ideal one. Even for the true hierarchy choice, an
ideal  $F=0$
gets a certain positive amount
because under-fluctuations are not taken into account. 
Within the JUNO configuration defined above, $F$ is around 3300 counts for the true case and around 8000 counts 
for the wrong one (Fig.~\ref{fig:6}).

The two $F$ distributions are still very well separated. The addition of the energy resolution will move the two distributions
closer each other, as it will be shown later for the 2-D pattern of $F$.

\subsection
{Extracting the value of $\Delta m^2_{atm}$ by means of F}\label{bias}

\begin{figure}[tb]
\includegraphics[width=7.5cm,height=4.5cm]{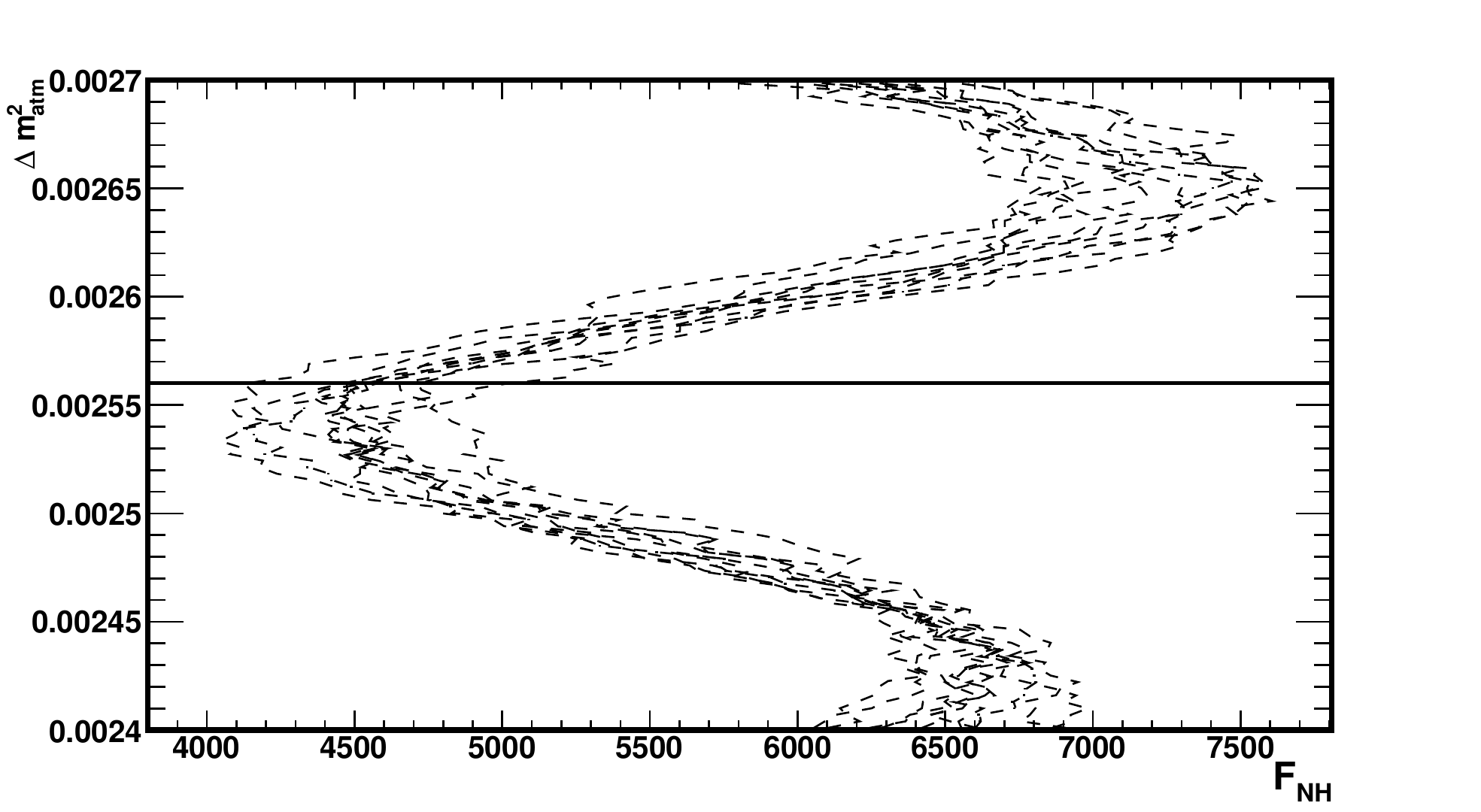}
\includegraphics[width=7.5cm,height=4.5cm]{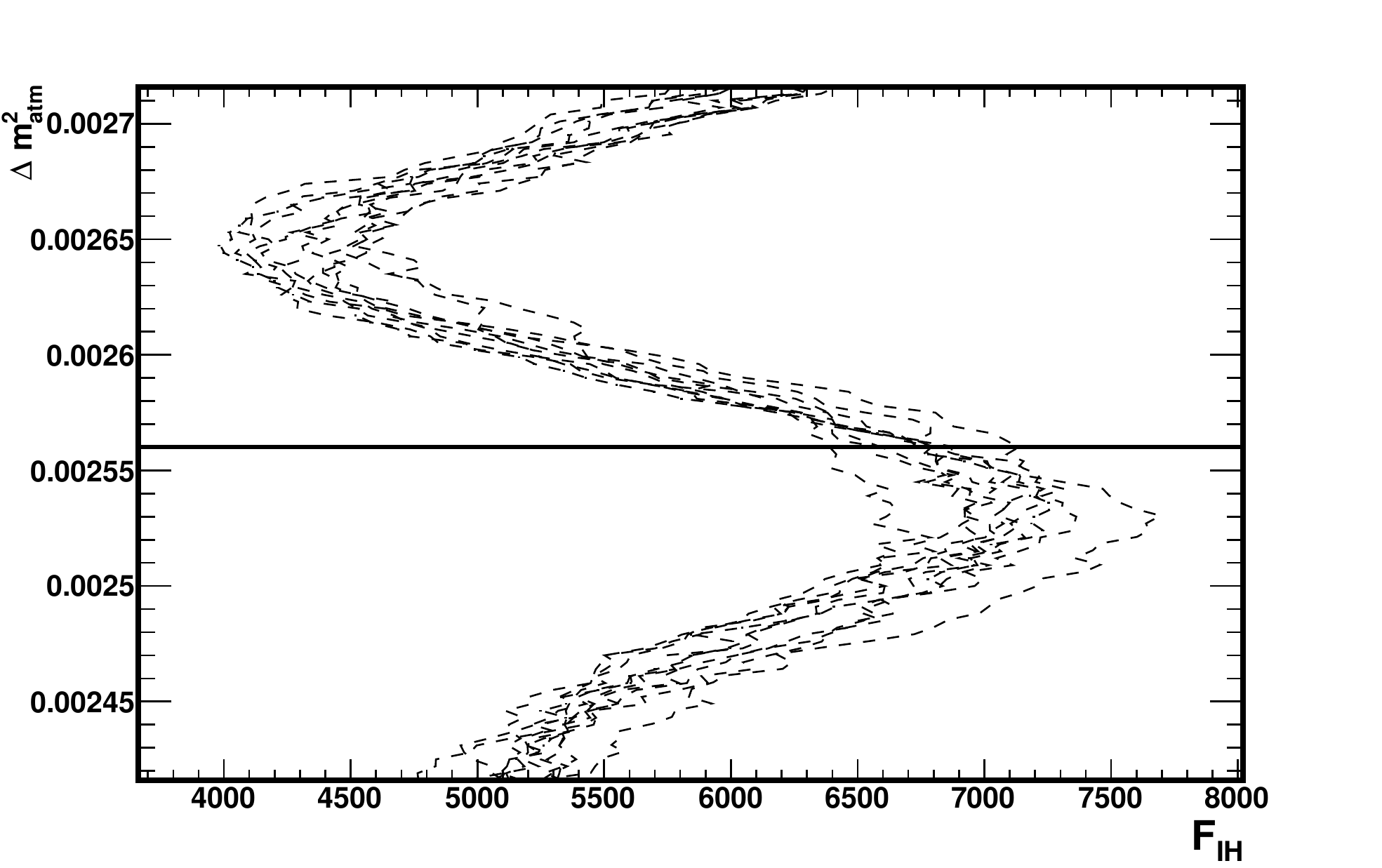}
\caption{\label{fig:7}(color online) $F_{MO}$ modulations due to the differences between the ``true'' $\Delta m^2_{atm}$ (NH) 
and the assumed one for ten JUNO-like toy experiments, including a 3\%/$\sqrt{E}$ energy resolution and the 
real distribution of the baselines ($\pm 5$ m) of the ten reactor cores in Yangjiang and Taishan sites.
On the top plot the dashed lines correspond to $F_{NH}$, whereas $F_{IH}$ modulations are shown in the bottom plot.
The horizontal lines correspond to the ``true''  $\Delta m^2_{atm}$. The observed bias is due to 
the the finite energy resolution, as explained in the text.}
\end{figure}

$F$ is very sensitive to the assumed value of $\Delta m^2_{atm}$, as already shown in section~\ref{subsec:3a}.
That dependence can be used to extract the best value for $\Delta m^2_{atm}$ when a scan is performed.
In Fig.~\ref{fig:7} the $F$ modulations due to the difference between the ``true'' $\Delta m^2_{atm}$  
and the assumed one are shown for few JUNO-like, six year long, toy experiments, including both the statistical fluctuation
and a $3\%/\sqrt{E(MeV)}$ energy resolution.

For each of the JUNO-like toys the minimum $F$ corresponds to the $\Delta m^2_{atm}$ closest to the 
truth, $\Delta m^2_{atm}$(best). A bias of 0.6\% is observed due to the finite energy resolution 
causing a distortion of the measured spectrum towards energies lower
than the true ones\footnote{A 3\%/$\sqrt{E(MeV)}$ corresponds for example to a dispersion of  33 keV 
for a 2 MeV neutrino energy and 80 keV for 8 MeV neutrinos.}. 
$F$ is sensitive to this energy spectrum deformation that, to our knowledge, has never been taken
into account so far. 
$F$ is the first estimator sensitive to such distortion, which can be cured at first order  
by shifting the reconstructed energy systematically by 25 keV upwards.
The bias could be studied with more detailed
analyses that go beyond the scope of the present paper. In any case, this bias is not relevant for the NH/IH study
since it does not produce asymmetric distortions of the $F$ patterns.

\begin{figure}[tb]
\begin{center}
\includegraphics[width=8cm,height=5cm]{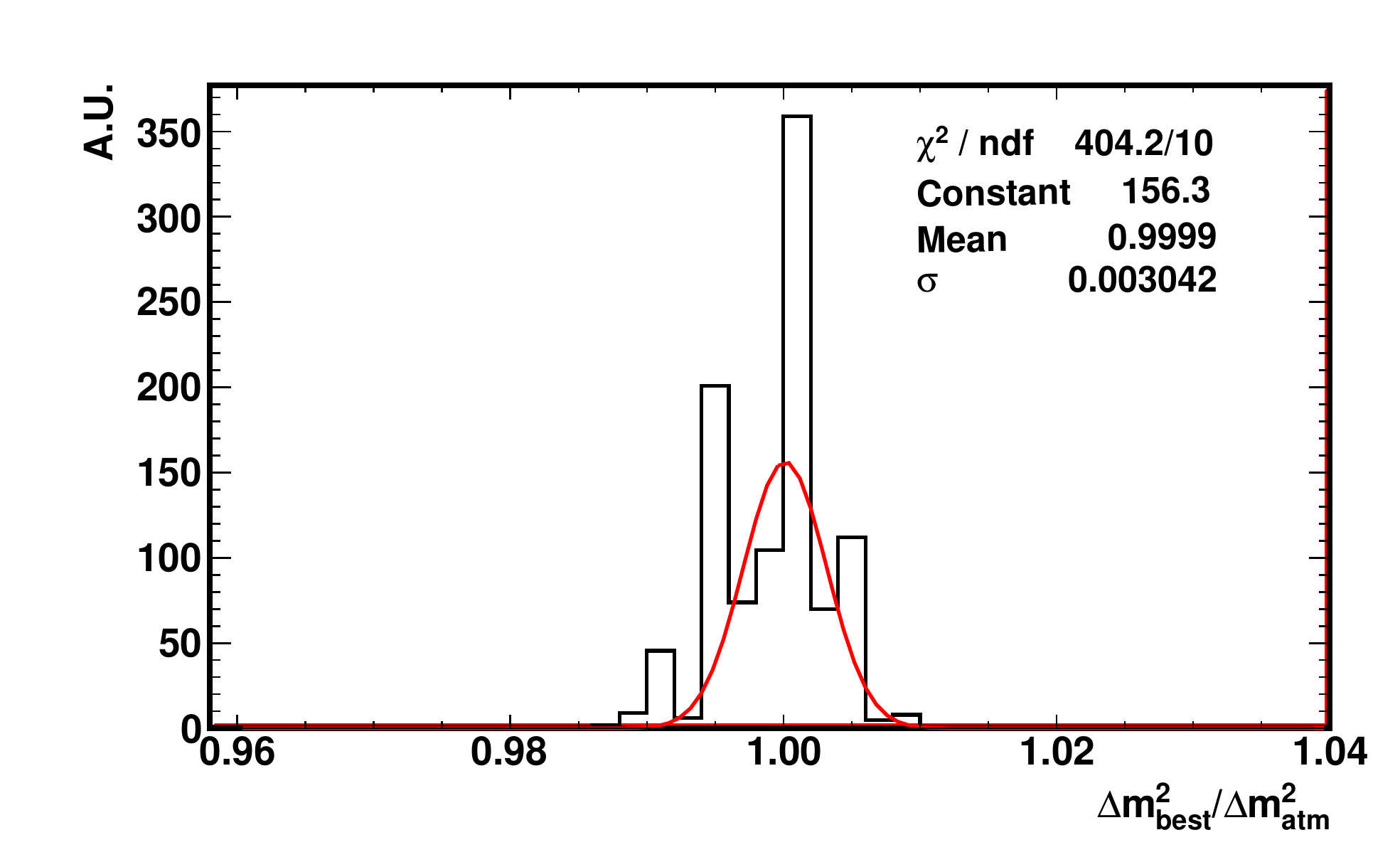}
\caption{\label{fig:8}(color online) The distribution of the ratio between $\Delta m^2_{atm}$(best) and the assumed value 
of the simulation, for one thousand JUNO-like toy experiments. $\Delta m^2_{atm}$(best) corresponds to $F_{min}$.
The 0.6\% bias due to the finite energy resolution
has been cured by adding a systematic 25 keV to the simulated energy of each neutrino event. 
A dispersion of 0.3\% is obtained. The poor gaussianity behaviour is due to the rough correction of the 0.6\% bias.
More refined analyses should be applied, but for the time being results are already far beyond the needs.}
\end{center}
\end{figure}

The distribution of $\Delta m^2_{atm}$(best) is reported in Fig.~\ref{fig:8} for one thousand JUNO-like toy experiments,
including a systematic addition of 25 keV to the simulated single-event neutrino-energy. 
The bias disappears and a 0.3\% precision is obtained for the estimated $\Delta m^2_{atm}$.
No significative change to this resolution has been observed for different choices of the neutrino-oscillation parameters.
For two years of data taking a 0.6\% precision is obtained on $\Delta m^2_{atm}$.

Fig.~\ref{fig:7} suggests that NH/IH can be distinguished through the $F$ patterns as a function of
 $\Delta m^2_{atm}$. 
 However, a ``quantitative''  determination will have to rely on a more sophisticated analysis, e.g. by
 extracting the phase of the oscillation pattern of Fig.~\ref{fig:7}, looking at the distance between largest and smallest counts in 
 $F$ or by a $\chi^2$ fit to the pattern itself. This goes beyond the extent of this article and rather can be the
 object of a later work~\cite{coming}. For now we concentrate on the MH information in the 2-D $F$ distributions 
 and proceed to the estimation of the sensitivities
%
on the NH/IH discrimination (see sub-section~\ref{subsec:4.2}).

The dispersion of $F$ around $\Delta m^2_{atm}$(best) is much smaller than
that due to the statistical fluctuations and the energy resolution. Thus,
decoupling of $F$ and $\Delta m^2_{atm}$ can be assumed.  This is a key point of the whole analysis.
Due to the strong correlation between $\Delta m^2_{atm}$ and the mass hierarchy determination, in case 
of a large uncertainty on $\Delta m^2_{atm}$, as it would be the case for baselines different 
from $\sim$ 50~km, the assumption is no more valid (see appendix~\ref{app:dmatm}).

\subsection{Backgrounds and systematics studies}

In JUNO-like experiments at reactors several sources of background and systematic errors are present.
We provide here only a sketch of the background issue, more details being available in~\cite{juno}.
The anti-neutrino interactions (charged-current mode) in liquid-scintillators are detected through the time coincidence
of the prompt positron annihilation with the delayed $\gamma$-ray released by the thermalized neutron with
a typical time $\tau=200\, \mu$s.
The major backgrounds for the reactor neutrino oscillation analysis are the accidentals, 
the $^8$He/$^9$Li cosmogenic, the fast neutron and ($\alpha$, n) interactions. 

All these background sources produce events uncorrelated in energy with the signal. They can be reduced with a set of 
selection cuts that, in JUNO, decreases the efficiency to about 70\%. 
The final yield corresponds to 60 signal events per day against a background of about 4 events, 
including geo-neutrinos.

Geo-neutrinos constitute an intrinsic background, at the level of 1.1 events per day after the selection cuts. However
their rate will be measured with very good precision by JUNO itself, reducing the current 30\% rate uncertainty to a few percent.
Anyhow, geo-neutrinos are detected up to about 3 MeV and they correspond to a negligible background for the present analysis.
In fact, an improved analysis on $F$ can be performed selecting an energy interval above 3 MeV,
so discarding the $1.8 - 3.0$ region where the very fast modulation of $\Delta N(IH-NH)$ is not really informative and
even reduces the sensitivity.\footnote{The parametrization introduced in~\cite{mee} reaches the opposite conclusion: the
information useful for the MO discrimination is limited to the low energy range, which is more suffering of the energy resolution
effect. That is a side benefit of the parametrization used instead for $F$.} 
Such issue has not been further studied in details since it goes beyond the scope of the paper.

\begin{figure}[tb]
\begin{center}
\includegraphics[width=8cm,height=5cm]{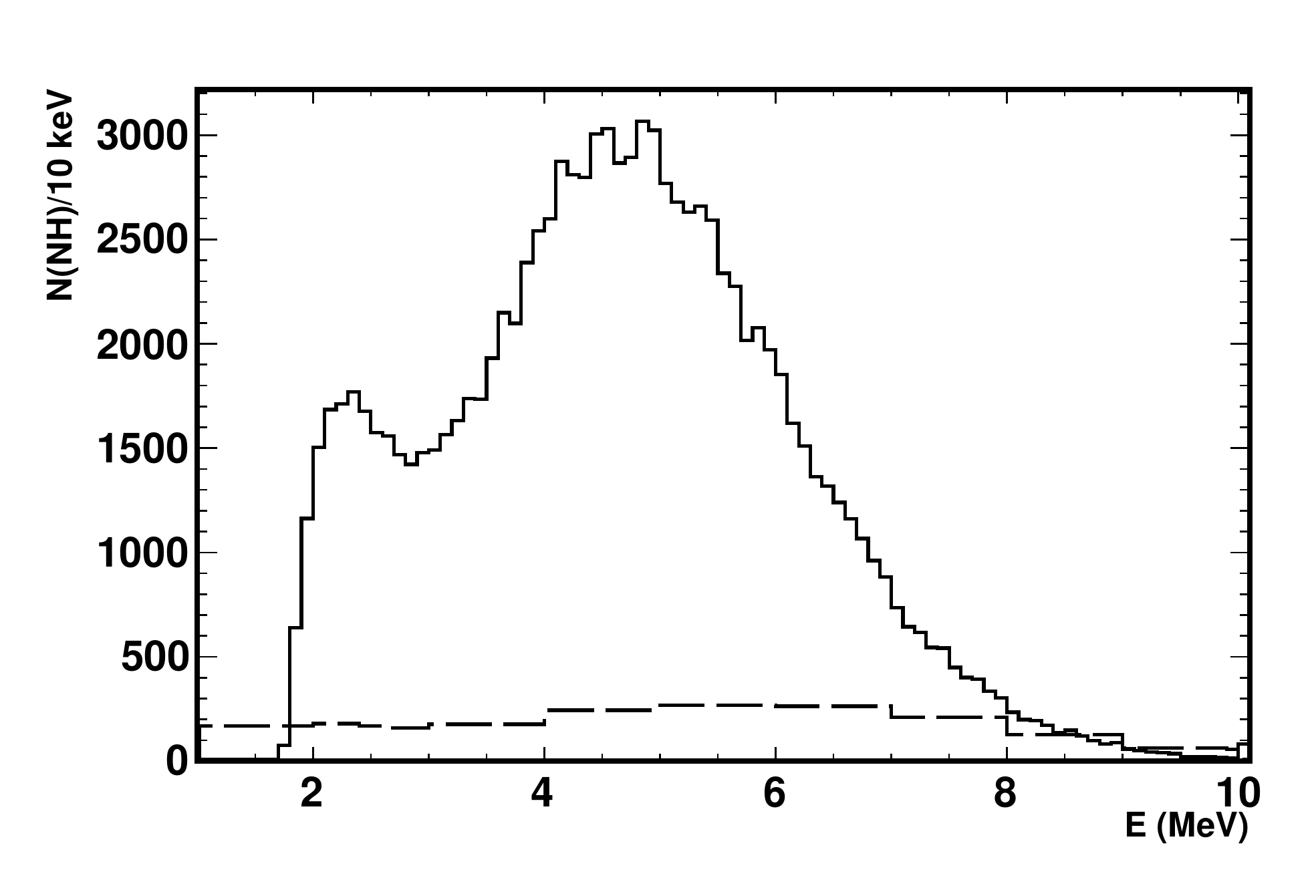}
\caption{\label{fig:9}(color online) The cosmogenic background distribution due to $^9$Li for a six year long JUNO-like 
experiment. This is the distribution, 
conservatively scaled to the total amount of expected incoherent background, used to extract the sensitivity on NH/IH
with this kind of background.}
\end{center}
\end{figure}

The background has been included in the $F$ analysis taking into account, conservatively, the $^9$Li shape with a total
rate that corresponds to the sum of the different sources of background (Fig.~\ref{fig:9}). As reported later, results on MH sensitivity 
are not affected by this kind of background due to its slow variation in energy that does not change the $F$ dispersions. 
The only net effect is a coherent shift of the counting peaks of NH and IH.

A different behaviour on $F$ is expected from systematic uncertainties. They are mainly due to the observed
$4\sim 6$ MeV ``bump'' (e.g.~\cite{daya} and references therein) and the reactor flux uncertainties. 
The 12\% contribution of the two remote reactors may also be
a relevant source of uncertainty.


The reactor flux uncertainties include several components, concerning both the overall normalization 
and the shape as a function of the anti-neutrino energy. 
As reference, the recent study performed by Daya-Bay~\cite{daya} has been used, where a summary table of the systematic uncertainties per reactor core is provided. 

Regarding the integrated flux normalization, we add in quadrature the relative uncertainties associated with the power, the energy produced per fission, the spent fuel, the non-equilibrium and the fission fraction at a given time. All but the last one can be assumed as correlated across all cores, because they derive from the same physical modeling uncertainty, intrinsic to a typical fission core. 
The latter is instead uncorrelated across cores, because the relative fractions of the four leading burning elements will likely be different for the cores considered at a given time (switching on/off and re-fueling at independent times). By taking the sum in quadrature of the maximum uncertainty coherently for all the considered cores the ``envelope'' total uncertainty is therefore used. Conservatively, we vary the overall flux by $\pm$ 3 $\%$ with respect to the expected nominal flux at JUNO. 



With regard to the energy dependence on the flux modeling uncertainty, from Ref.~\cite{daya} this is estimated to be less than 10$\%$ up to $E(\bar{\nu}_{e}$) = 10 MeV. However, a systematic uncertainty on the reactor flux is already considered
by introducing the ``bump'' between 4 and 6 MeV in the simulation.
That is, the deformation of the spectrum due to the bump is taken and studied as a source of systematic uncertainty, as an example of how a stretching of the energy spectrum would affect the F. The deformation is bin-dependent and lies 
between 5\% and 20\%. Since the bump contribution is neglected in the analytical calculation and in the $F$ definitions
it can be considered a genuine systematic effect of the energy shape.
Therefore, it is deemed safe to discard any further uncertainties on the reactor flux shape.

\begin{figure}[htbp]
\includegraphics[width=8cm,height=5cm]{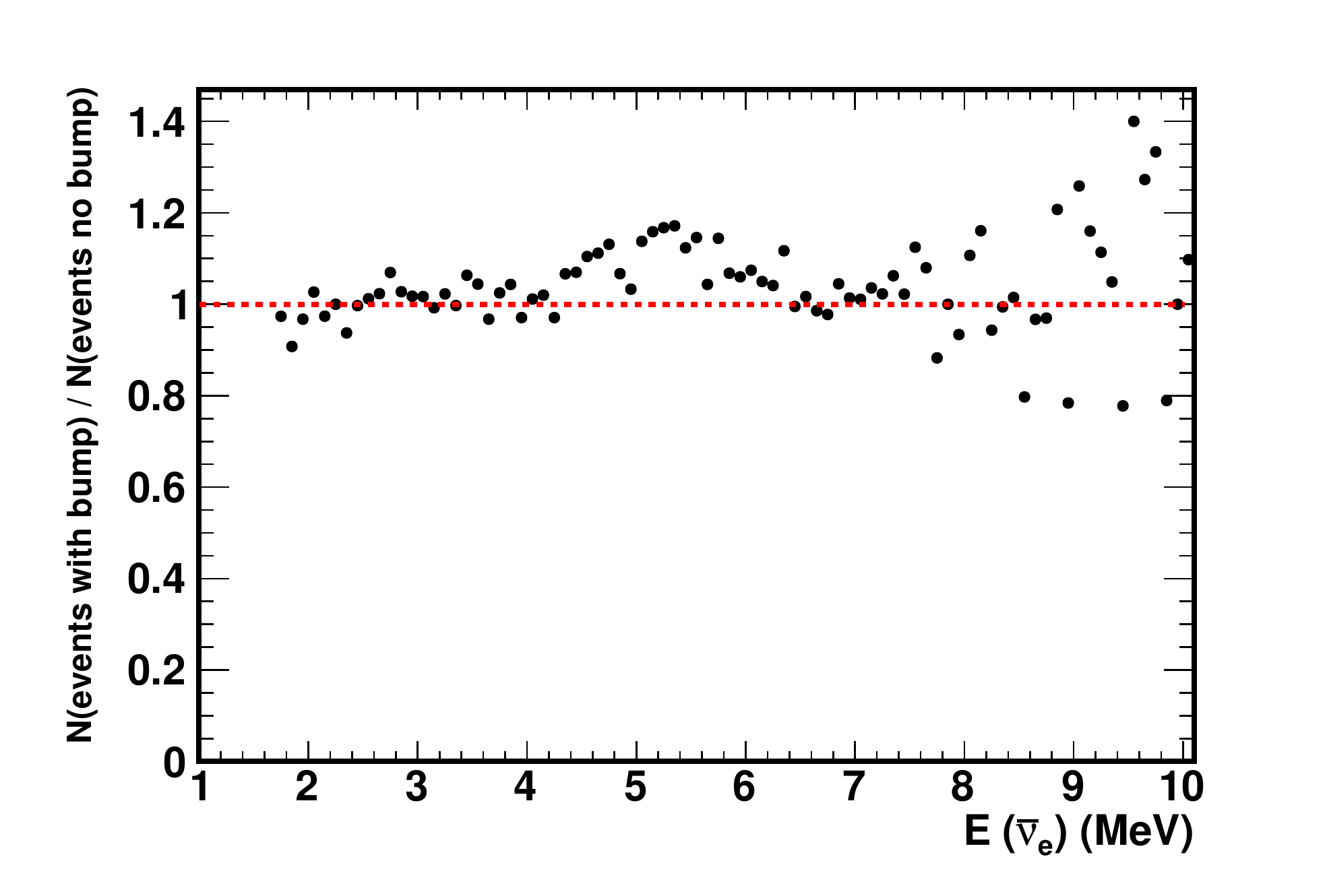}
\includegraphics[width=8cm,height=5cm]{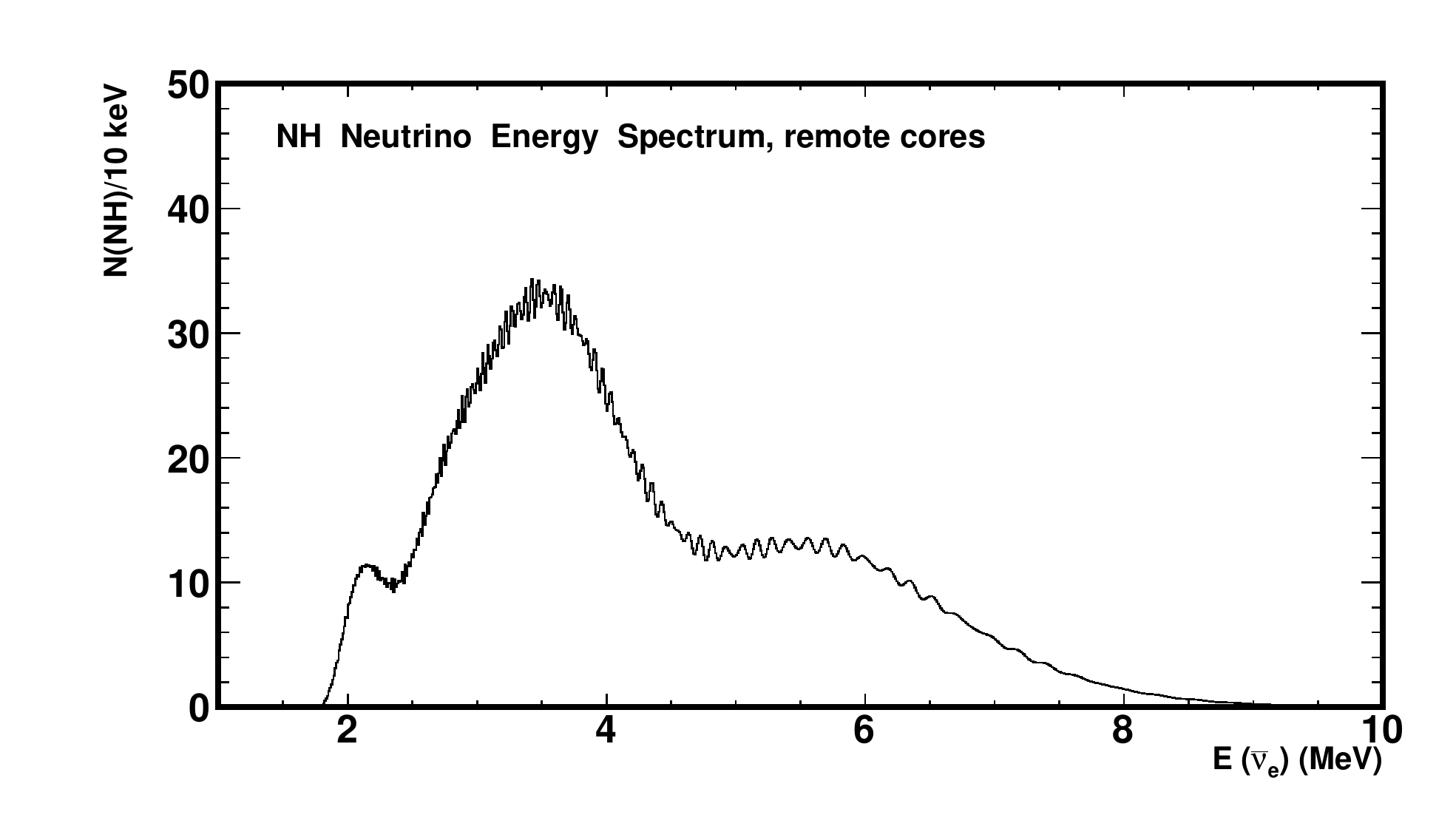}
\caption{\label{fig:10}(color online) The anti-neutrino event distribution when a modeled  bump around $4\sim 6$ MeV 
is added to the ten reactor production (top), is drawn. A single JUNO-like, six years exposure, toy Monte Carlo is shown in terms of the relative 
percentage with/without the addition.
In the bottom plot the distribution due to the two remote reactor production is shown (NH case).}
\end{figure}

The bump contribution and the reactor flux uncertainties are coherent in energy with
the expected signal (Fig.~\ref{fig:10}). Therefore they give a corresponding percentage of increase/decrease 
to the $F$ test-statistic sensitivity. In other words, they simply act as an increase/decrease of exposure.
Instead, the remote reactor contribution is incoherent due the large baselines. It decreases the sensitivity, as reported below.

Finally, we checked that other sources of incoherent systematic errors, related to the energy reconstruction,
will enter in quadrature with the energy resolution itself. The coherent systematic error due to a residual non-linear
energy scaling is a detector dependent effect. We verified that a parametric form as in~\cite{pet29}, eq. (13), would
produce a linear dependence on both $F_{NH}$ and $F_{IH}$. That linearity could be used to self-calibrate the energy
spectrum.

\subsection{Procedure for the NH/IH  sensitivity}\label{subsec:4.2}

To compute the sensitivity on the determination of the mass hierarchy a procedure has to be defined. We propose the approach
suggested in~\cite{lucas}. In the assumption that either NH or IH is true, data are expected to be compatible 
with one of the hypotheses. Compatibility should be defined a priori and a blind analysis be adopted.
The sensitivity is further given by the $p$-value of the other hypothesis, computed for the observed data.

The procedure is more complicated when one wants to evaluate the expectation, before data are available
as is the case for our study. Having in mind the 2-D nature of the $F$ estimator, the following procedure
has been adopted: assuming one hypothesis be true 
with a certain confidence level, an average $p$-value for the alternative hypothesis is evaluated weighted by the 
true hypothesis probability. The single $p$-value entering in the average is computed from the edge of the
true hypothesis confidence interval.
In formulas, being $f_{MO}(\vec{x})$ the MO density probability of $F_{MO}$ and $C.L.$ the chosen confidence interval for the
true hypothesis:
\begin{eqnarray}
&p-val(IH)=\int\limits_{\Omega_{NH}}\mathrm{d}\vec{x}\, f_{NH}(\vec{x}) \otimes \int\limits_{\Omega_{NH}(\vec{x})}
\mathrm{d}\vec{x}' \, f_{IH}(\vec{x}') , \nonumber \\
&\Omega_{NH}\perp  \int\limits_{\Omega_{NH}} \mathrm{d}\vec{x}\, f_{NH}(\vec{x}) = C.L., \nonumber \\
&\Omega_{NH}(\vec{x})\ni  f_{IH}(\vec{x}')\le\, f_{IH}(\vec{x})\quad \mathrm{for}\quad \vec{x}'\in \Omega_{NH}. \nonumber 
\end{eqnarray}

\noindent These expressions are valid for the IH case. The $p$-value for NH is equivalently computed.

This procedure is more conservative than the evaluation of the standard $p$-value on the median
of the true hypothesis (50\% C.L.). For larger C.L., e.g. at 95\% C.L. of the true hypothesis, 
our method is more conservative when the two hypotheses are distant from each others by at least  3-4 $\sigma$.
When the two hypothesis probabilities are closer than 3 $\sigma$ our method
gives slightly less conservative $p$-values than the standard ones. That is explained by the depletion 
due to the low probability of the true hypothesis in the far domain.

We believe the chosen procedure is the right one when trying to reproduce the behaviour of the real experiment, in the
assumption that the result of the experiment would be compatible with one of the two hypotheses.

The procedure to compute the sensitivity from the $F$-estimator is more delicate since $F$ is two-dimensional.
However, the same approach as before can be employed: the estimated $p$-value corresponds to
the weighted $p$-values over the domain of the true hypothesis. The domain in ($F_{IH}$, $F_{NH}$)
can be defined at a certain confidence level, e.g. 95\% C.L. 
Therefore, the computed $p$-value, $p$, has to be read as:
$p$ corresponds to the significance to reject the wrong hierarchy at the 95\% C.L. for the true one\footnote{The procedure
depicted e.g. in~\cite{pet33} defines a C.I from a C.L.. It is similar but more formal than the present one.
However, from a physical point of view it is more appropriate trying to reproduce the real experiment, weighting
the single possible outcome of the true hypothesis. In the more formal procedure a single $p$-value is evaluated in
the confidence region of the true hypothesis, assumed flat over the true hypothesis.}.

In the following, since the performed simulations could have some discrepancy with the real data
JUNO would obtain, we prefer to be very conservative and chose a 99.7\% C.L., which defines the domain 
in ($F_{IH}$, $F_{NH}$) where the true hypothesis is expected to be observed.


We  estimated the sensitivity on NH/IH for the JUNO-like benchmark, i.e. six years of data taking and a
3\%/$\sqrt{E}$ energy resolution. Near and remote reactors for twelve cores in total have been used, too,
with a $\pm 5$ m and $\pm 5$ km uniform dispersions, respectively, on the spatial generation of the single neutrino event. Backgrounds
from cosmogenic sources have been evaluated. Reactor flux uncertainty due to the $4\sim 6$ MeV bump 
has been checked in the simulation. 
The other uncertainties on the reactor fluxes have been
studied evaluating the corresponding $\pm \sigma$ bands.
Best fit values for the oscillation parameters~\cite{lisi-last} have been used. As discussed before, the 
$\Delta m^2_{atm}$ uncertainty can be internally recovered by studying the $F$ behaviour.

\begin{figure}[tb]
\begin{center}
\includegraphics[width=8cm,height=5cm]{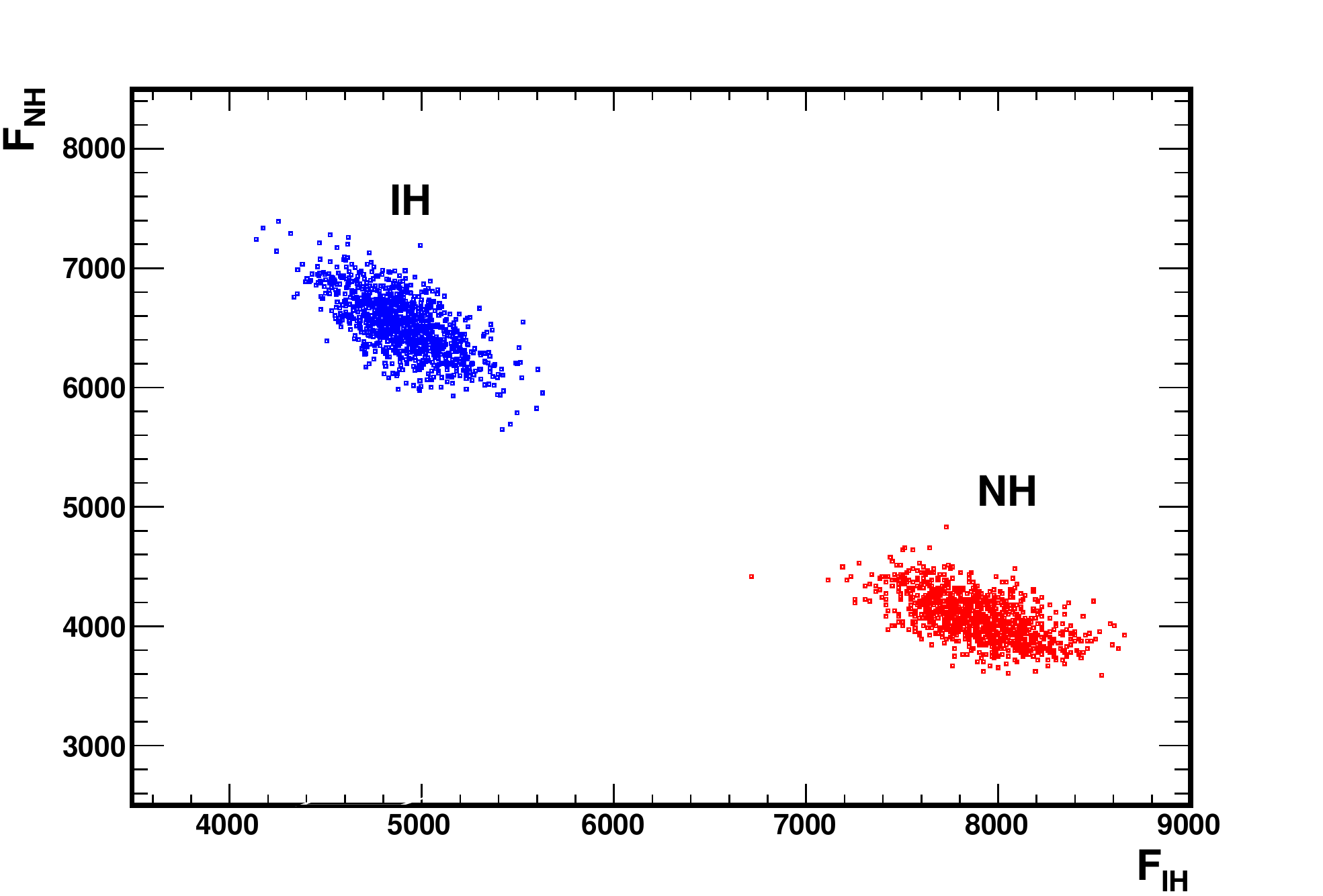}
\caption{\label{fig:11}(color online) $F_{NH}$ vs $F_{IH}$ distributions for 1000 (NH) $+$ 1000 (IH) toys JUNO-like
experiment, with six years of exposure and the ten near reactor cores, each with its own baseline. 
The ``island'' top-left (blue) corresponds to the IH simulation, while the one on the 
bottom-right (red) corresponds to the NH simulation. An infinite energy resolution is assumed. Note the asymmetric position
of NH and IH domains, due to the non symmetric behaviour of NH and IH in the oscillation model.}
\end{center}
\end{figure}

Fig.~\ref{fig:11} shows a typical distribution of $F_{NH}$ vs $F_{IH}$. The simulation corresponds
to six years of JUNO-like data taking, assuming a perfect energy reconstruction\footnote{A priori one should consider the 
neutron-recoil effect in the inverse-beta-decay, which prevents to attain the infinite resolution limit.
That is detailed e.g. in~\cite{pet30}. However, the statistical fluctuations  in our study are so largely overwhelming such effect
that it can be neglected, even for a 0\% detector energy resolution.} and the ten near detectors. 
The two ``islands'' are very well
separated and the identification of the right hierarchy is evident {\em per se}. Using the procedure described
before, we can any how quantify the sensitivity to more than 10 $\sigma$ ($p$-value less than $10^{-27}$).

To probe very small $p$-values 1000 toys are not sufficient. To cover the region  ($F_{IH}$, $F_{NH}$) (IH true)
where NH is expected to be observed at the pre-defined confidence level, several billions of events should
be simulated. Only in the context of real data, having identified all the uncertainty and background contributions,
such simulation could be afforded for the most precise configuration. 
That will be possibly made  within the JUNO experiment.
However, ($F_{IH}$, $F_{NH}$) islands are well fitted to a 2-D Gaussian (see appendix A). Consequently,
we proceed to extract the $p$-values from the Gaussian analytical expressions.

\begin{figure}[tb]
\begin{center}
\includegraphics[width=8cm,height=5cm]{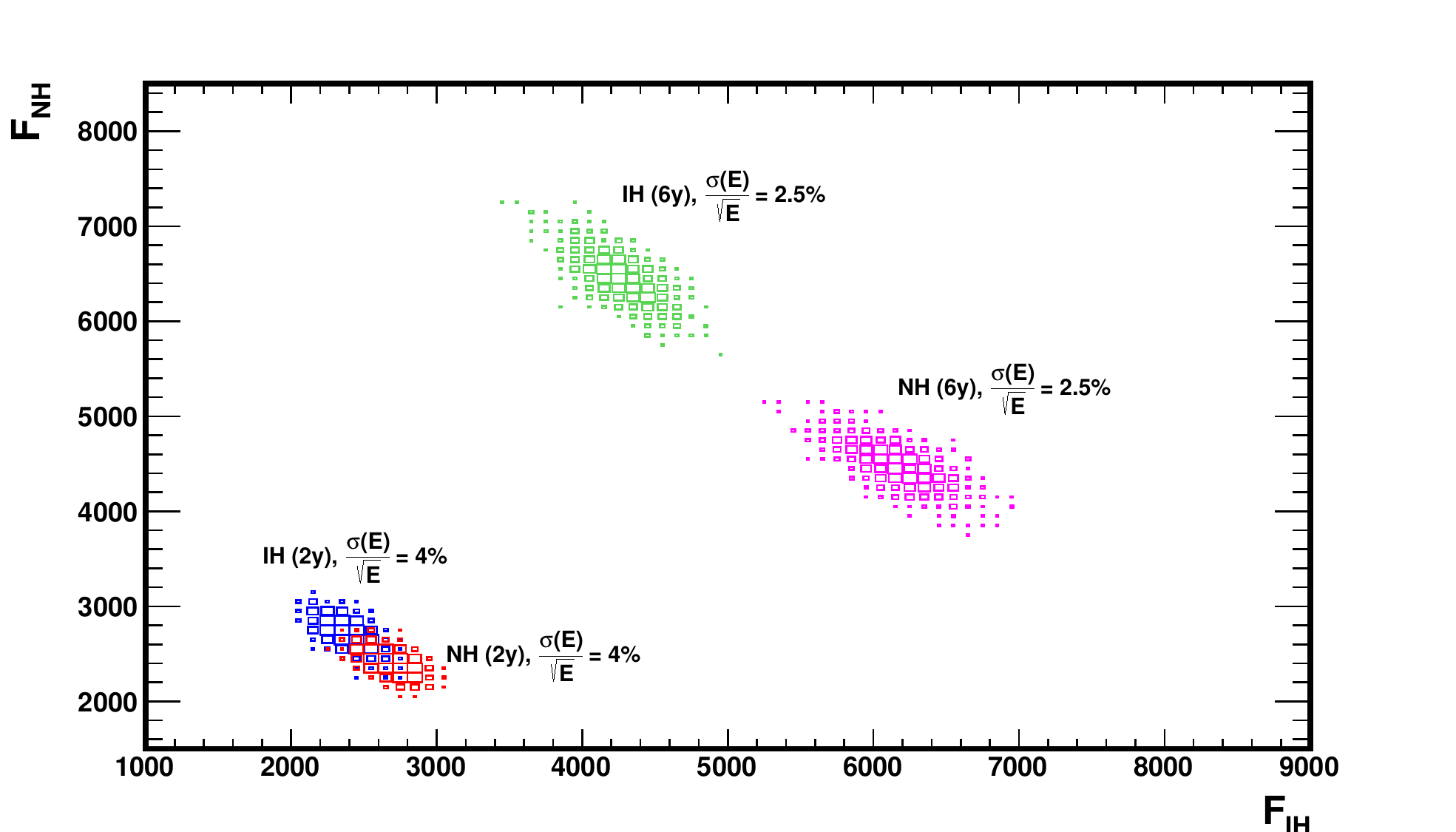}
\caption{\label{fig:12}(color online) $F_{NH}$ vs $F_{IH}$ distributions for 1000 (NH) $+$ 1000 (IH) toys JUNO-like
experiment, in two different configurations: the most favorable, six years of exposure with a $2.5\%/\sqrt{E}$ energy
resolution, against a short one for two years of exposure and $4\%/\sqrt{E}$.}
\end{center}
\end{figure}

When the energy resolution is included in the simulation, as well as any other source of the signal degradation,
or when the exposure is changed,
the two islands tend to get closer to each other, so reducing the sensitivity. Fig.~\ref{fig:12} describes such effect 
for two configurations: an  optimal one with $2.5\%/\sqrt{E}$ and six years of exposure and a limited one with
$4\%/\sqrt{E}$ and 2 years of exposure.

A campaign of simulations has been made for several JUNO-like configurations. In the appendix~\ref{app:A} the 
parameters and the errors of the 2-D Gaussian fits to the $F$ distributions are reported.
They describe the islands and the corresponding $p$-values.

\subsection{Expected sensitivities as a function of the energy resolution}

\begin{figure}[tb]
\begin{center}
\includegraphics[width=9cm,height=6cm]{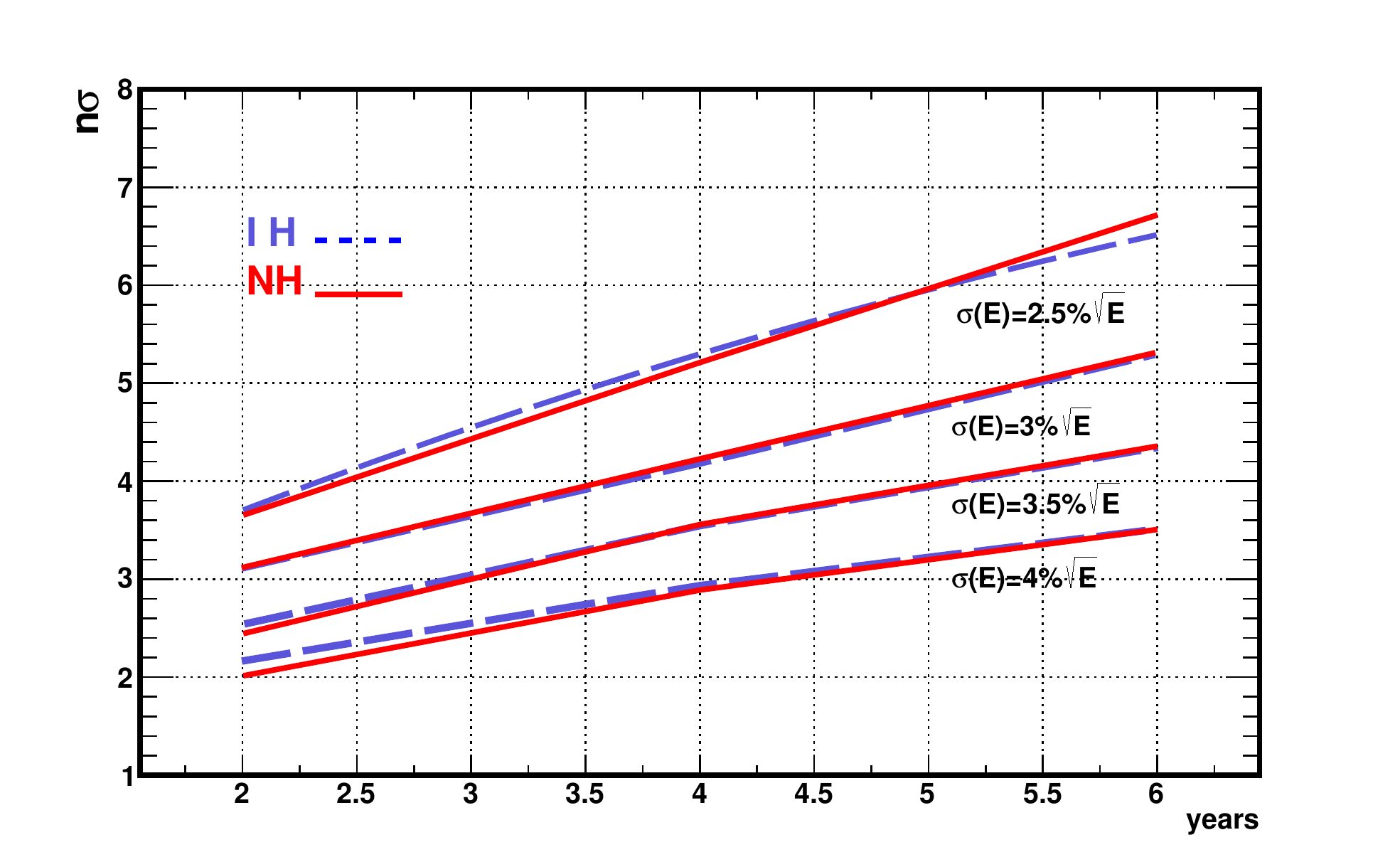}
\caption{\label{fig:13}(color online) Evolution of the NH/IH sensitivity as function of the data taking in JUNO-like
experiment. The different curves correspond to different energy resolution (continuous/dashed are for NH/IH hypothesis,
respectively).
The background has been conservatively assumed to be described by the $^9$Li component. Ten near reactor cores plus two remote cores have been used, each with a $\pm$ 5 m and $\pm$~5~km uniform dispersions for the relative baseline, respectively.}
\end{center}
\end{figure}

Table~\ref{table:3} of the appendix~\ref{app:A} reports the evolution of the NH/IH sensitivity as a function of the exposure
for a reference set of configurations. Background has been conservatively
assumed to be described by the $^9$Li component, properly scaled to reproduce the total expected amount,
and checked to have no influence on the results (Table~\ref{table:5}). 
Ten near reactor cores plus two remote cores have been used, each
with a $\pm$ 5 m and $\pm$~5~km uniform dispersion for the relative baseline, respectively, to extract 
the sensitivities in Table~\ref{table:4} and the corresponding curves of Fig.~\ref{fig:13}.

The result is very significant: at the foreseen six years of exposure JUNO will be able to set the mass hierarchy
at more than 5 $\sigma$ significance. The relevance of the energy resolution is confirmed.
The sensitivity curves are quite dependent on the resolution. A 4\%$\sqrt{E}$ energy resolution will be critical 
to get a definite answer on the mass hierarchy in the JUNO context.

Instead, already after two years of exposure (at a full reactor power)
JUNO would be able to provide first indications about the mass hierarchy,
at a level slightly above 3 $\sigma$, for an overall 3\%$\sqrt{E}$ energy resolution.

The bump around $4\sim 6$ MeV has been modeled
and included in the flux for one configuration. Contrary to intuitive expectations
it will slightly increase the significance by about 0.2 $\sigma$
(Fig.~\ref{fig:14} and Table~\ref{table:5} of the appendix~\ref{app:A}).

\begin{figure}[tb]
\begin{center}
\includegraphics[width=8cm,height=5cm]{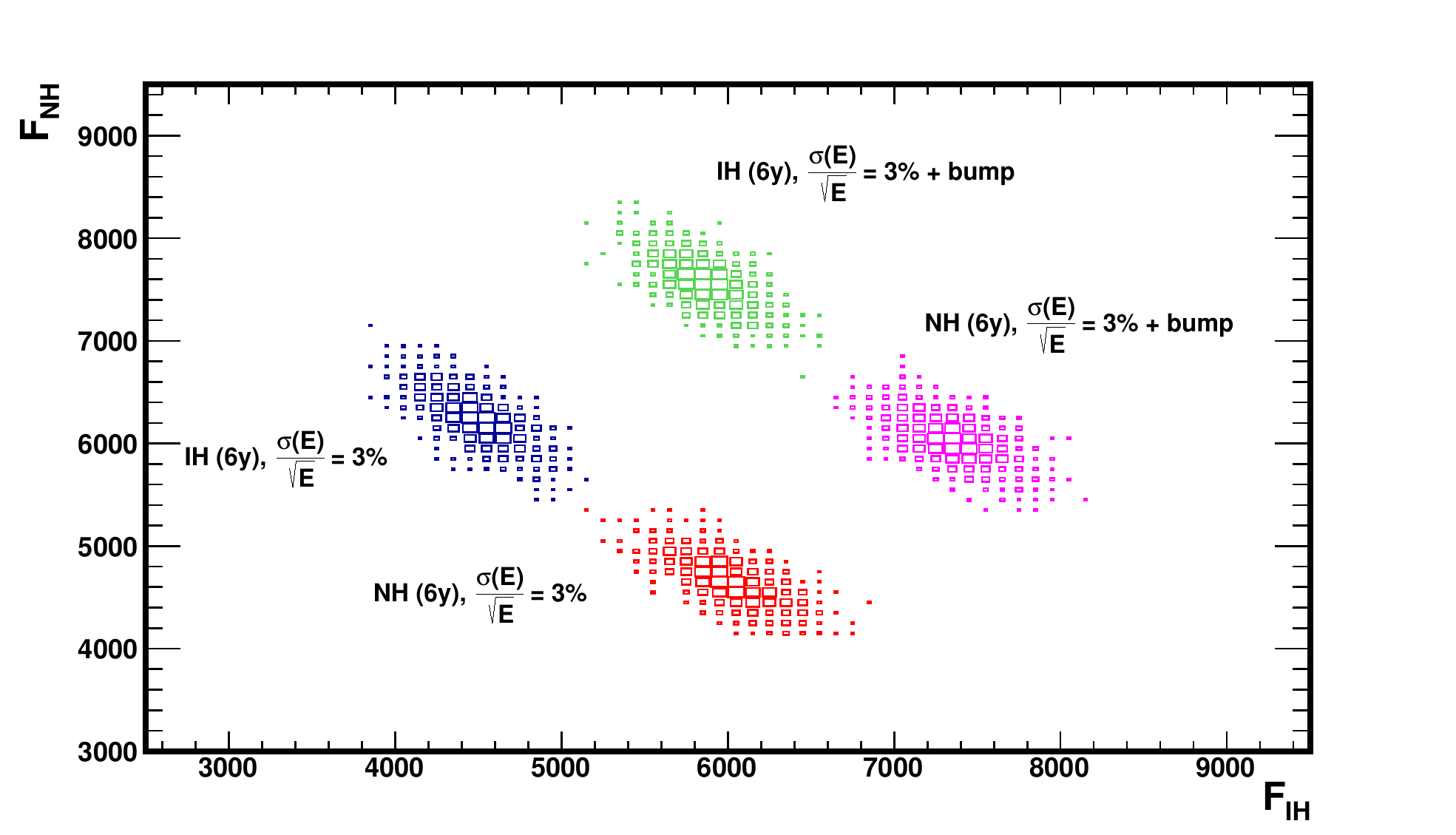}
\caption{\label{fig:14}(color online) $F_{NH}$ vs $F_{NH}$ distributions for 1000 (NH) $+$ 1000 (IH) toys JUNO-like
experiment, in two different configurations. The benchmark of six years exposure is compared to the same exposure
result when the $4\sim 6$ MeV bump is added.}
\end{center}
\end{figure}

\subsection{Possible results after two years from JUNO start up}

The $F$ estimator provides interesting perspectives already after two years of JUNO-like exposure.
However, when JUNO will start not all the reactors will be ready and less thermal power is foreseen~\cite{juno-less}.
We underwent the exercise of estimating the sensitivity in such realistic case. For a 3\%/$\sqrt{E}$ energy resolution
a 2.7 $\sigma$ significance is obtained (Table~\ref{table:7} of the appendix~\ref{app:A}).

\section{Conclusions}\label{sec:disc}

A new bi-dimensional estimator, $F$, has been introduced to determine the mass hierarchy with reactor anti-neutrinos
in a JUNO-like experiment. We demonstrated that $F$ owns several properties. First, it allows coupling
the two hypotheses, NH and IH, so granting the use of the simple approach suggested in~\cite{lucas}.
Under the assumption of the 3-neutrino oscillation paradigm only one of the two hypotheses is possible.
If the future experiments would be compatible with e.g. NH, then IH could be rejected via a simple
$p$-value computation. 

Furthermore, $F$ is so defined to factorize out all the incoherent background sources and their systematics.
The background components produce insensitive shifts of the two NH/IH regions in $F$. 
Systematic components, smoothly varying with the neutrino energy,
are responsible of a coherent depletion of $F$, which can be overcome with a corresponding (small)
increase in exposure. Only the coherent background due to the two remote reactor plants, at 215 and 265 km away,
gives a decrease of about 0.2 $\sigma$ on the sensitivity.

Two independent techniques have been used to simulate single Monte Carlo's events.
They incorporates all the relevant uncertainties,
excluding the detector response, which we leave to the experiment to implement; this is foreseen to be at the level of 1\%.
In any case, the latter source is believed not to be of great impact, due to its smooth correlation with energy. The only important
component is the residual non-linear energy scaling that may produce strong correlated variations of the 
estimator. This can be iteratively handled.

We introduced a conservative procedure to define the sensitivity in terms of $p$-values.
The results are very promising: the two mass hierarchies can be largely discriminated ($> 5\, \sigma$)
in JUNO after six years
of exposure, keeping the total energy resolution at 3\%. After two years of running and the foreseen 
initially-reduced available reactor power a little less than 3 $\sigma$ will still be possible. 

$F$ holds a $\Delta m^2_{atm}$ degeneracy at the level of $12\times 10^{-5}$ eV$^2$,
which is much larger than the current global fits uncertainty.
However, we argued that the new technique here outlined bears in itself a mean to resolve such ambiguity,
allowing to measure $\Delta m^2_{atm}$ with an unprecedented precision at
reactors, less than 1\%. That is due to the strong correlation of $F$ to $\Delta m^2_{atm}$, which will deserve more
refined analyses and promise even more interesting results. 
The study here presented contributes to clarify the impact of $\Delta m^2_{atm}$ on the mass hierarchy determination
at reactor experiments, and the corresponding limited sensitivity that can be obtained with 
a $\chi^2$ procedure.

Our study confirms the very positive perspectives for JUNO to determine the mass ordering in a
vacuum-oscillation dominated region.

\vspace{0.5cm}

\acknowledgments

We are particularly grateful to  J. Cao and D. Chiesa for valuable suggestions on the treatment of the reactor fluxes and uncertainties.
We also acknowledge E. Ciuffoli, D. Meloni and E. Vagnoni for some useful discussions.
S. Dusini provided relevant criticisms to the original manuscript.
We have to thank several JUNO colleagues and E. Lisi for critical discussions on the issue
about the $\Delta m^2_{atm}$ correlation.

\vskip 15pt
\appendix
\noindent{\large {\bf Appendices}}
\section{Sensitivity results}\label{app:A}

This appendix reports the details of the fits on the bi-dimensional $F$ distributions, and the relative
sensitivities for NH/IH discrimination, for several configurations of JUNO-like data taking, addition of
background and several systematic errors (Tables~\ref{table:3}-\ref{table:7}).

\begin{figure*}[tb]
\begin{center}
\includegraphics[width=13cm]{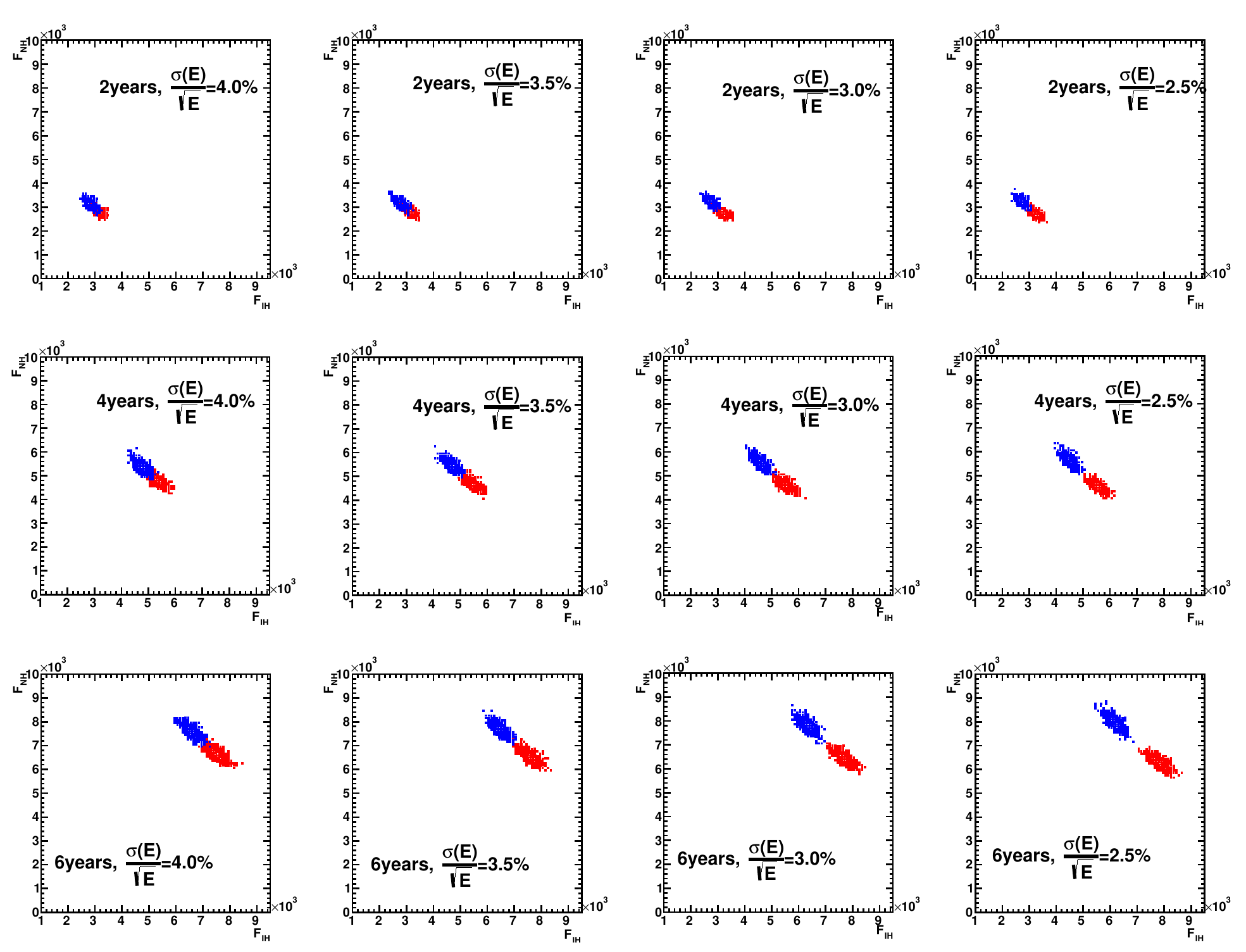}
\caption{\label{fig:15}(color online) $F_{NH}$ vs $F_{NH}$ distributions for 1000 (NH) $+$ 1000 (IH) toys JUNO-like
experiment, for different configurations in time exposure and energy resolution. The two remote reactor cores have been added.
The ``island'' top-left (blue) corresponds to the IH simulation, while the one on the bottom-right (red) corresponds to the NH
simulation. The corresponding fitted values to the 2-D Gaussians are those reported in table~\ref{table:4}.}
\end{center}\end{figure*}

Fig.~\ref{fig:15} shows the results for a full set of simulations for different years of data taking and energy resolution,
for the 10+2 reactor cores configuration.

The ($F_{IH}$, $F_{NH}$) islands are well fitted to a 2D-Gaussian distribution. In principle, given the definitions
(\ref{eq:10}) and (\ref{eq:11}), the islands are very well approximated by Gaussians if only the event statistical fluctuation
is considered. The Gaussianity is expected to be reduced when a finite energy resolution is added, and backgrounds and systematics as well.

To check the robustness of the bi-Gaussian fits a large Monte Carlo simulation has been performed, 50 000 $+$ 50 000 toys,
in the NH and IH hypothesis, respectively, for few configurations. Each toy corresponds to a JUNO-like experiment,
six year long in data taking. More than 10 billions of single anti-neutrino detected events were simulated,
for each selected configuration. The results of this large simulation turned out to be very stable in terms
of the fitted values of the Gaussian parameters. The $p$-values undergo a variation of 0.01-0.02~$\sigma$
in terms of sensitivity.
That confirms the reliability of the assumed Gaussian approximation for the used 1000 $+$ 1000 toy distributions.

\section{Details of the new method}\label{app:B}

In section~\ref{bias} we explored the relationship between $F$ and $\Delta m^2_{atm}$ and how to solve the ambiguity 
on the MH that arises from it. Then in section~\ref{subsec:4.2} and following we proceeded, by fixing the value of 
$\Delta m^2_{atm}$ in the context of the 2-D $F$ estimator, to evaluate its statistical power in discriminating the two MH hypotheses. 
In this appendix we elaborate further on these points and give some quantitative examples and additional suggestions 
on how to disentangle $F$ and the value of $\Delta m^2_{atm}$ at work in nature. 
So far people have been accustomed to relate the neutrino mass ordering
discrimination to the $\chi^2$ procedure. The $\chi^2$ performs a best fit over the multidimensional
space of the parameters' uncertainties. Therefore, one usually obtains two different sets of best fit values,
one for NH and one for IH. 

One should not confuse the experimental conditions with the physical context:  
there is obviously only one set
of true parameters, regardless of the MO established in nature. If it were possible
to single out such deconvolution
the result would be extremely efficient. The introduction of the $F$ estimator is an attempt in that direction.

The $F$ procedure operates distinctively being based on expressions with a factorized dependence on most of the parameters,
leaving $\Delta m^2_{atm}$ as the relevant parameter (eq.~\ref{eq:9})
 To be more precise, the factorization of the other parameters is only true at the leading order and approximately 
 for $\delta m^2_{sol}$. For example, if the mass terms dependences are included in the survival probabilities, the
 factorization of eq.~\ref{eq:9} does not hold. Nevertheless, the latter are minor effects that do not spoil the performances of $F$.
  
In the framework of the $F$ estimator introduced in this paper and relating to section~\ref{bias}, in the following 
appendix~\ref{app:B1}
we further study the interplay between $\Delta m^2_{atm}$ and $F$. Our aim is to clarify the potential of our technique, 
shed a light on this most critical issue and indicate how it can be resolved in the future with a detailed analysis. 
To do so we stem from the patterns of Fig.~\ref{fig:7} and probe the behaviour by means of a full simulation over all the current 
parameter uncertainties. 
We stress that our intent is solely to suggest how a JUNO-like experiment could proceed in resolving the issue using $F$ on its 
data. Our analysis has not been optimized: for example, no accurate treatment of the intrinsic systematic uncertainties and 
biases in energy mentioned in section~\ref{bias} is performed, which are better tackled in the context of the data analysis 
of a real experiment interested in using $F$.      

 
 Due to the very heavy computing power needed for this kind of simulation only one specific configuration of a
 JUNO-like experiment has been chosen. Being the most interesting outcome the sensitivity for six years
 of exposure with a 3\%$/\sqrt{E(MeV)}$ energy resolution and ten reactor cores
 at around 52.5 km distance, that is the chosen configuration. 
 
 In the second part of this appendix a brief study was performed on the optimization of the baseline for neutrino reactor experiments 
 to clarify the effect of the degeneracy with $\Delta m^2_{atm}$.
 
 \subsection{Extraction of MH sensitivity}\label{app:B1}
 
\begin{figure}[tb]
\begin{center}
\includegraphics[width=6.7cm,height=4cm]{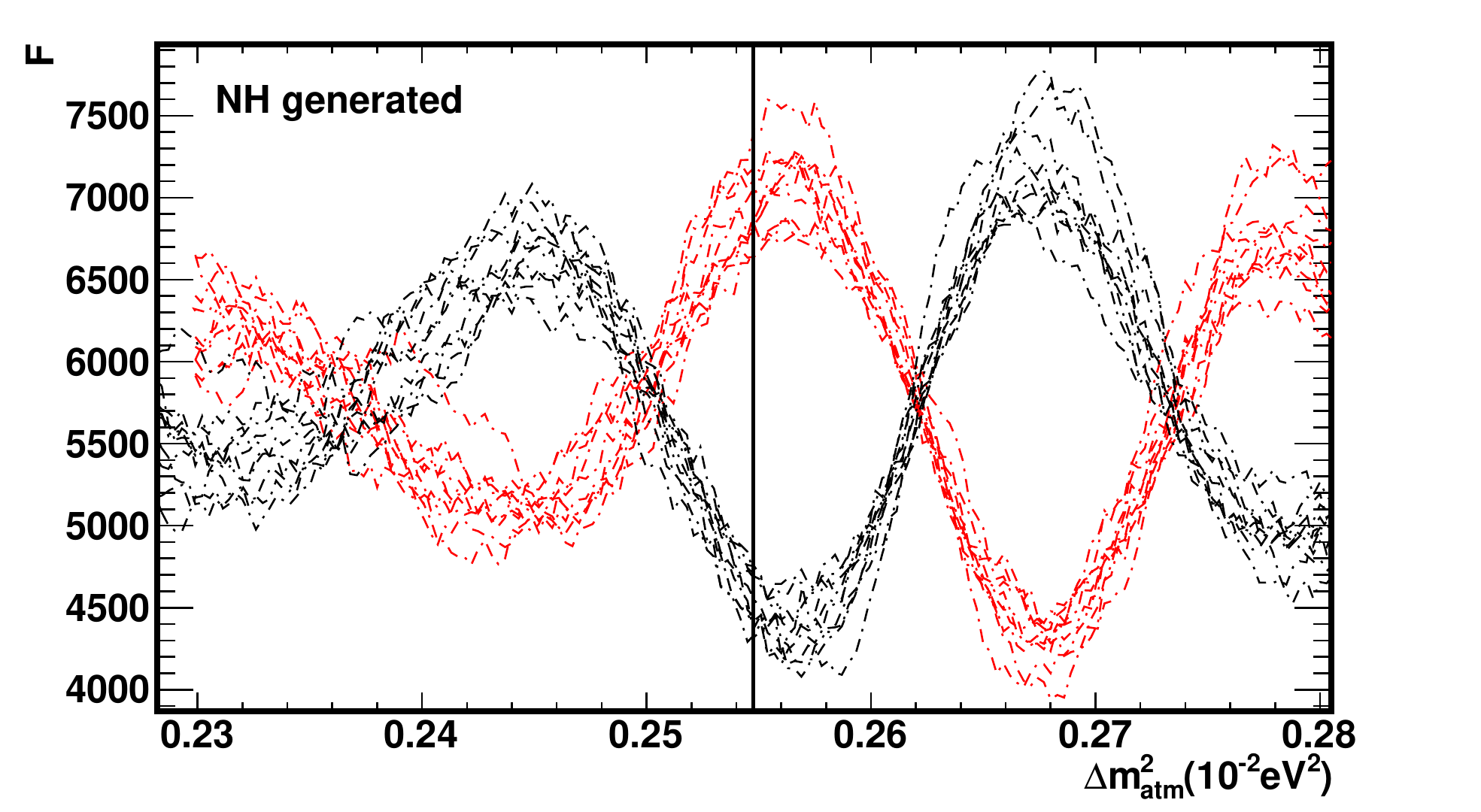}
\includegraphics[width=6.7cm,height=4cm]{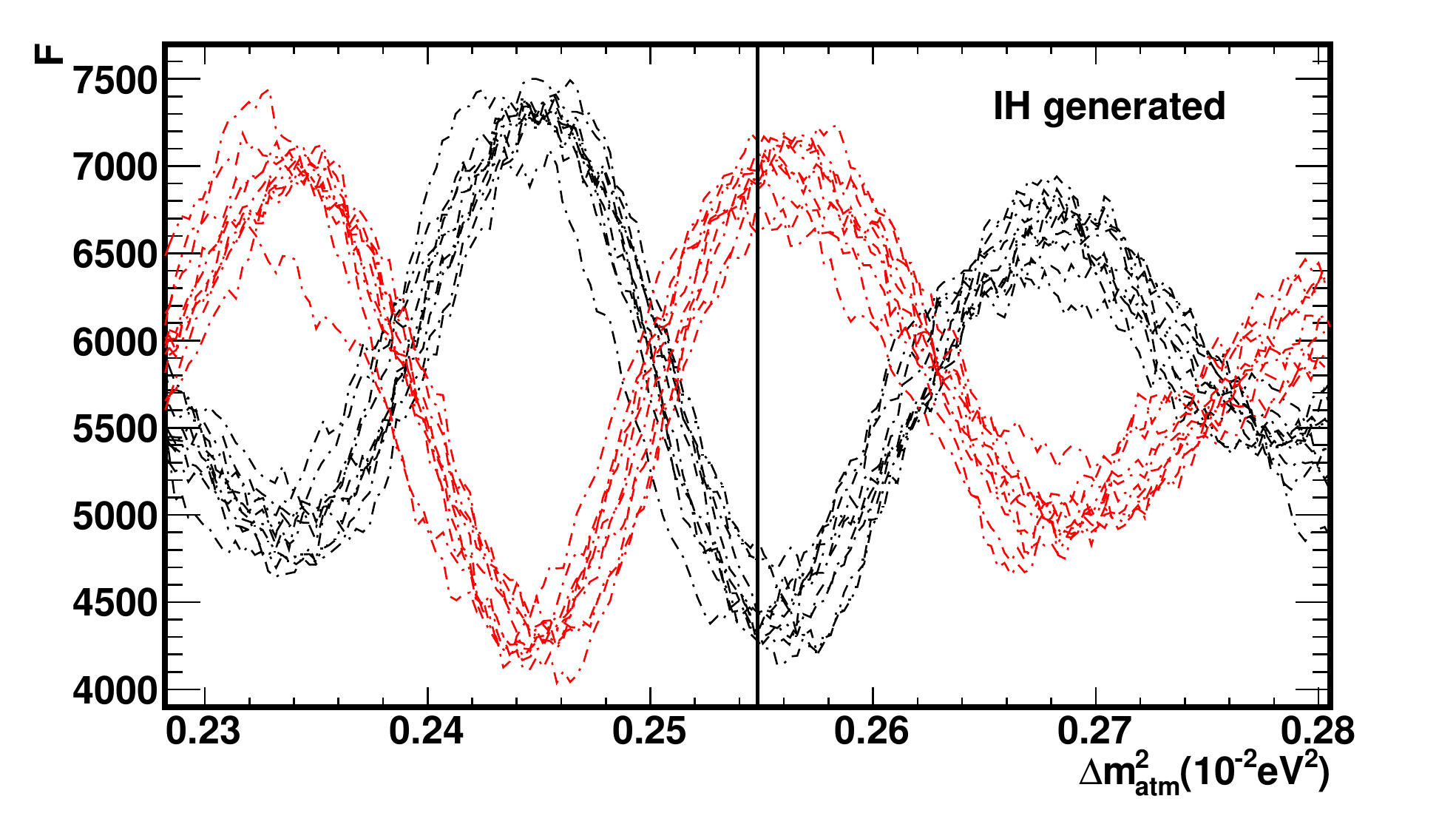}
\caption{\label{fig:17}(color online) $F$ vs $\Delta m^2_{atm}$ for ten generated JUNO-like toy experiments,
in the NH (top) and the IH (bottom) hypotheses. The black (red) curves correspond to the analysis when
the true (false) hypothesis is taken. The vertical lines indicate the selected $\Delta m^2_{atm}$.
The experimental conditions are the same as in Fig.~\ref{fig:7}.}
\end{center}
\end{figure}

The patterns of Fig.~\ref{fig:7}  are reported in Fig.~\ref{fig:17} for a larger  $\Delta m^2_{atm}$ interval.
The intrinsic bias has been adjusted by hand, not to confuse the reader. It has been checked that the adjustment
does not change with $\Delta m^2_{atm}$ and other parameters.

\begin{figure}[tb]
\begin{center}
\includegraphics[width=7.5cm]{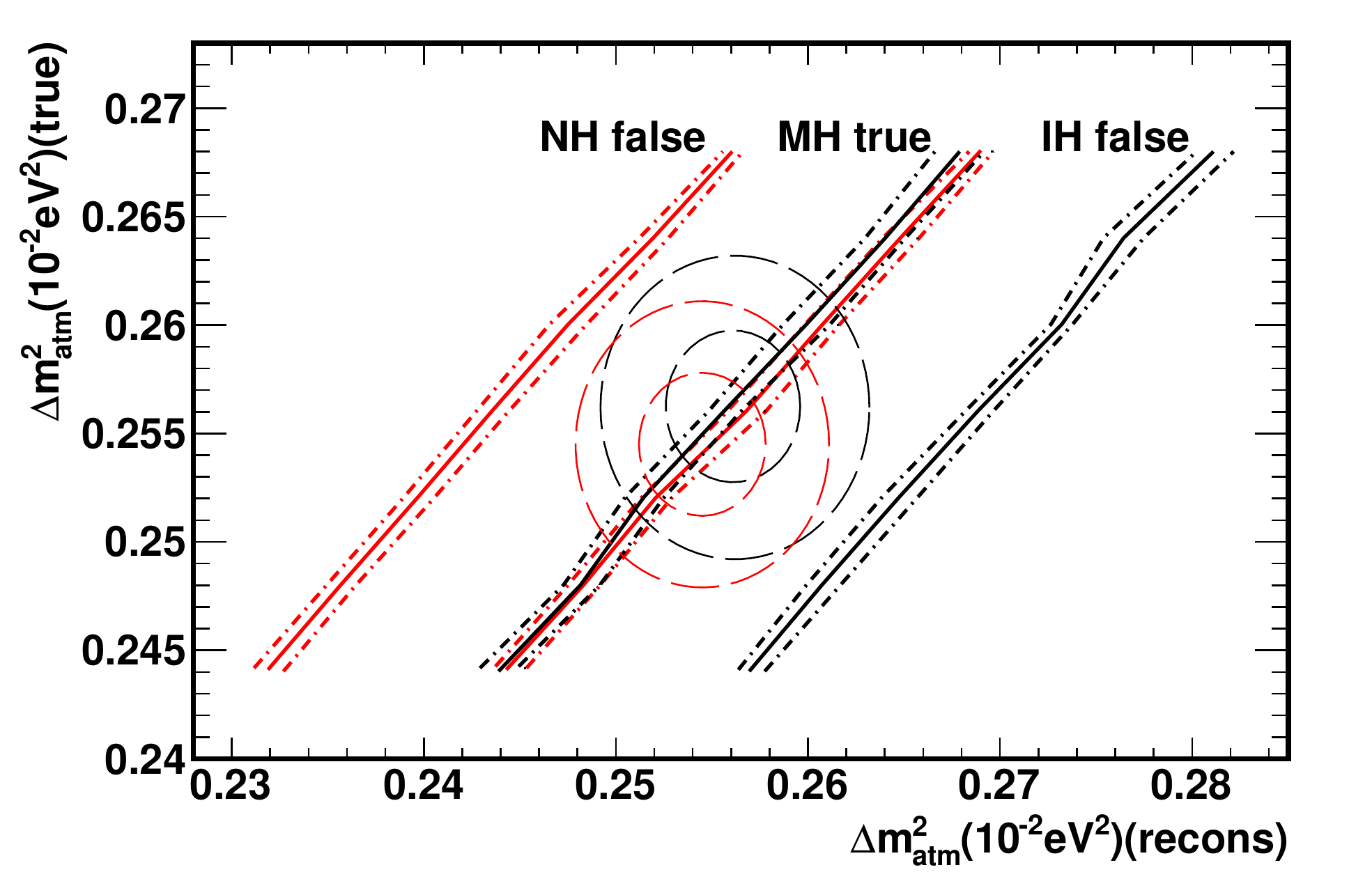}
\caption{\label{fig:18}(color online) $\Delta m^2_{atm}$(true) vs $\Delta m^2_{atm}$(recons) is drawn,
$\Delta m^2_{atm}$(recons) being obtained by the minimum $F$. The continuous lines correspond to the
central values, the dashed ones to the $\pm\,\sigma$ bands. Black (red) curves corresponds to the NH (IH) generation.
The central circles correspond to the 68\% and 95\% C.L.
contours of the current $\Delta m^2_{atm}$ uncertainties for NH and IH (from Table~\ref{table:1}).}
\end{center}
\end{figure}

For each neutrino-mass ordering case two full cycles are observed with maximum amplitude. 
Noticeably there is an absolute minimum for each hypothesis.
The two ordering can be discriminated when $\Delta m^2_{atm}$ varies less than $12\times 10^{-5}$ eV$^2$. 
That corresponds to the distance between 
a peak and the valley, or, equivalently, the distance between the two absolute minima of NH and IH.
The dispersions of the single minimum is less than $10^{-5}$ eV$^2$. 
The patterns are consistently reproduced across the whole range of $\Delta m^2_{atm}$:
defining $\Delta m^2_{atm}$(recons) as the $\Delta m^2_{atm}$ at the absolute minimum $F$, 
the central values and their $\pm\sigma$ bands are drawn in Fig.~\ref{fig:18} for a large $\Delta m^2_{atm}$(true)
range.

Fig.~\ref{fig:18} shows a strong linear correlation on $\Delta m^2_{atm}$. 
For a single reconstructed $\Delta m^2_{atm}$ two solutions can be identified.
For example, when $F$  identifies $\Delta m^2_{atm}{\rm (recons)}=0.00250$ there are two solutions:
$\Delta m^2_{atm}=0.00250\pm 0.00001$ eV$^2$ for NH and $\Delta m^2_{atm}=0.00262\pm 0.00001$ eV$^2$
for IH.
Due to this intrinsic degeneracy it is not possible to have a single solution if only one experiment is taken into account.
Information about $\Delta m^2_{atm}$ should be injected from external, like e.g. a global fit analysis, 
to definitively select between the two solutions. The final result depends on the knowledge about 
$\Delta m^2_{atm}$ from other neutrino oscillation experiments, although it is not so stringent because the two solutions
are $12\times 10^{-5}$ eV$^2$ far away.

We underline that, if one is just interested in the discrimination between the two degenerate solutions,
by taking into account the external $\Delta m^2_{atm}$ uncertainty a standard sensitivity around
4 $\sigma$ is roughly obtained.  
For example a 1\% uncertainty corresponds to a  $\sim 4.5\, \sigma$ significance, computed from the difference of
the two $F$ solutions, which are far away $12\times 10^{-5}$ eV$^2$. This applies to any kind of $\Delta m^2_{atm}$,
either true or reconstructed.
We would argue that the standard procedure is dominated by the strong $\Delta m^2_{atm}$ correlation
and it is then limited by its accuracy, externally provided to the experiment\footnote{This conclusion
should apply either to the parameterization used in this paper or to the $\Delta m^2_{ee}$ one.
Suppose that
$\Delta m^2_{atm}=0.00250\, {\rm eV}^2$ and NH be the true values. If the external measurement 
$\Delta m^2_{atm}=0.00250\pm 0.000025\, {\rm eV}^2$ is available, in the $F$ framework
one obtains $\chi^2_{NH}\, (min) =0$  with $\Delta m^2_{atm}(NH)=0.00250\, {\rm eV}^2$
and $\chi^2_{IH}\, (min) \sim16$ with $\Delta m^2_{atm}(IH)\sim 0.00262\, {\rm eV}^2$.
The $\Delta\chi^2$ would give the usual sensitivity.
That is almost equivalent to
compute the $\chi^2$ of the two ordering cases at the same $\Delta m^2_{atm}$(recons) including its
uncertainty. In other words, $\Delta\chi^2$ is equivalent to the $\chi^2_{MO-false} (min)$ of the wrong hypothesis
at $\Delta m^2_{atm}(true)$. We argue that is the asymptotic trend of the $\chi^2$ procedure at a JUNO-like experiment.}. 

Only a procedure that internally
estimates $\Delta m^2_{atm}$ would be able to achieve much higher sensitivities on the NH/IH discrimination.
Repeating the same computation, if $10^{-5}$ eV$^2$ is taken for the $\Delta m^2_{atm}$ uncertainty,
$\sim 8.5\, \sigma$ are reachable. However, this procedure is very rough since it does not include treatments
of systematics and backgrounds. That is the reason the 2-D approach has been used in the main text. It is surely
more robust and complete, even though more conservative.

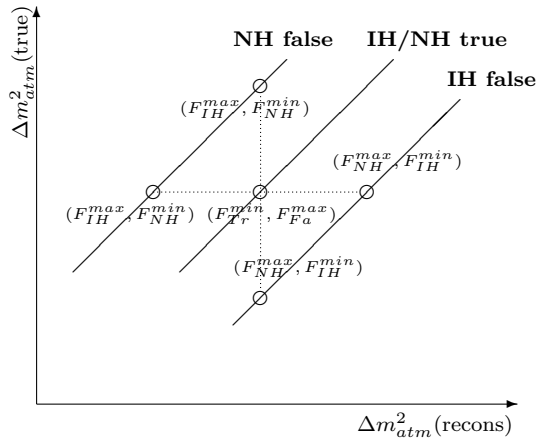
\begin{figure}[tb]
\begin{center}
\begin{picture}(200,170)
\put(10,10){\vector(0,1){150}}
\put(10,10){\vector(1,0){180}}
{\scriptsize \put(130,0){$\Delta m^2_{atm}$(recons)}}
\begin{rotate}{90} {\scriptsize \put(110,-3){$\Delta m^2_{atm}$(true)}} \end{rotate}
\put(20,60){\line(1,1){80}}\put(60,60){\line(1,1){80}}\put(80,40){\line(1,1){85}}
\put(50,90){\circle{5}}\put(90,90){\circle{5}}\put(90,130){\circle{5}}
\put(130,90){\circle{5}}\put(90,50){\circle{5}}
\multiput(50,90)(1.5,0){25}{\line(1,0){0.4}} \multiput(90,90)(0,1.5){25}{\line(1,0){0.4}}
\multiput(90,90)(1.5,0){25}{\line(1,0){0.4}} \multiput(90,90)(0,-1.5){25}{\line(1,0){0.4}}
{\tiny \put(70,80){($F_{Tr}^{min}, F_{Fa}^{max}$)}}
{\tiny \put(17,80){($F_{IH}^{max}, F_{NH}^{min}$)}}
{\tiny \put(60,120){($F_{IH}^{max}, F_{NH}^{min}$)}}
{\tiny \put(117,100){($F_{NH}^{max}, F_{IH}^{min}$)}}
{\tiny \put(80,60){($F_{NH}^{max}, F_{IH}^{min}$)}}
{\scriptsize \put(130,145){\bf IH/NH true}}
{\scriptsize \put(80,145){\bf NH false}}
{\scriptsize \put(160,130){\bf IH false}}
\end{picture}
    \caption{
    The couplings of $F$ to $\Delta m^2_{atm}$(true) and $\Delta m^2_{atm}$(recons) are pointed out.
    The quoted sensitivities
    corresponds to the probability to mis-identify ($F_{IH}^{min}, F_{NH}^{max}$) with 
    ($F_{IH}^{max}, F_{NH}^{min}$), either at the same $\Delta m^2_{atm}$(true) or the same $\Delta m^2_{atm}$(recons).
    An equivalent probability applies to ($F_{NH}^{min}, F_{IH}^{max}$).
}
    \label{fig:21}
\end{center}
\end{figure}

In Fig.~\ref{fig:18} the 68\% and 95\% C.L. contours of the current $\Delta m^2_{atm}$ 
uncertainty are drawn. It seems reasonable to assume that only one solution be selected by $F$, at least at 95\% C.L.
and for the current $\Delta m^2_{atm}$ uncertainty.
The sensitivity computed in this paper corresponds to the probability to discriminate between
NH/IH when only one solution is admitted. 
The quoted sensitivity owns a slightly different meaning from the standard one, which instead gives the probability
to distinguish NH/IH in the whole parameter space.
Since the $F$ technique identifies two solutions, one for NH and one for IH, each one at a different
$\Delta m^2_{atm}$, with less than a 0.5\% uncertainty, our sensitivity corresponds 
to the probability to mis-identify the two solutions at their own $\Delta m^2_{atm}$.
As can be inferred from Fig.~\ref{fig:18} computing that sensitivity is technically equivalent to 
use the same $\Delta m^2_{atm}$ value for both hypotheses. Fig.~\ref{fig:21} is a cartoon describing
that.

\begin{table}[tb]
\begin{center}
\begin{tabular}{lcc}
\hline
oscillation parameter  & best fit & 1 $\sigma$ range \\
\hline
$\sin^2\theta_{12}$ & 0.297 & $\pm 0.017$ \\
$\sin^2\theta_{13}$ & 0.0215 & $\pm 0.0007$ \\
$\delta m^2_{sol}$ & 7.37$\times 10^{-5}$ & 0.16$\times 10^{-5}$ \\
\hline
flux & & $\pm 3\%$ \\
cross-section & & $\pm$ 1\% \\
baselines && $\pm\,$ 5 m \\
\hline
$\Delta m^2_{atm}$ & 2.56$\times 10^{-3}$ &  \\
\hline
\end{tabular}
\caption{\label{table:8} 
The quantities used in the large simulation are listed. For each of them the chosen central values and their uncertainties are quoted.
They are allowed freely varying at the same time, each following a Gaussian distribution.
The baselines uncertainties follows a $\pm\, 5$ m uniform distribution.
The chosen value for $\Delta m^2_{atm}$ has been added to the list.
Central values for flux and cross-sections are taken from the computations described in the text.
The cross-section uncertainty is not realistic. It has been included just to show its possible correlation with the 
$F$ estimator.}
\end{center}
\end{table}

To confirm the $F$ properties with respect to the quantities other than $\Delta m^2_{atm}$,
a large Monte Carlo simulation has been done.
The parameters that are allowed to float freely in the simulation are reported in Table~\ref{table:8}.
500 different sets of parameters have been randomly selected. For each set of parameters
20 JUNO-like experiments, 6 years of exposure, have been simulated including a 3\%$/\sqrt{E(MeV)}$ energy 
resolution. A uniform uncertainty of $\pm$ 5 m for the baseline of each of the ten reactors at 52.5 km 
away has been considered. 
We have not included the two remote reactors as well as any background contribution.

The aim of the simulation is to demonstrate the independence of the evaluated sensitivity
from the parameters except $\Delta m^2_{atm}$, as argued from equation~(\ref{eq:9}). 
Specifically, the relative position of NH and IH in the ($F_{NH}$, $F_{IH}$) plane should not change.

\begin{figure}[tb]
\begin{center}
\includegraphics[width=7.5cm]{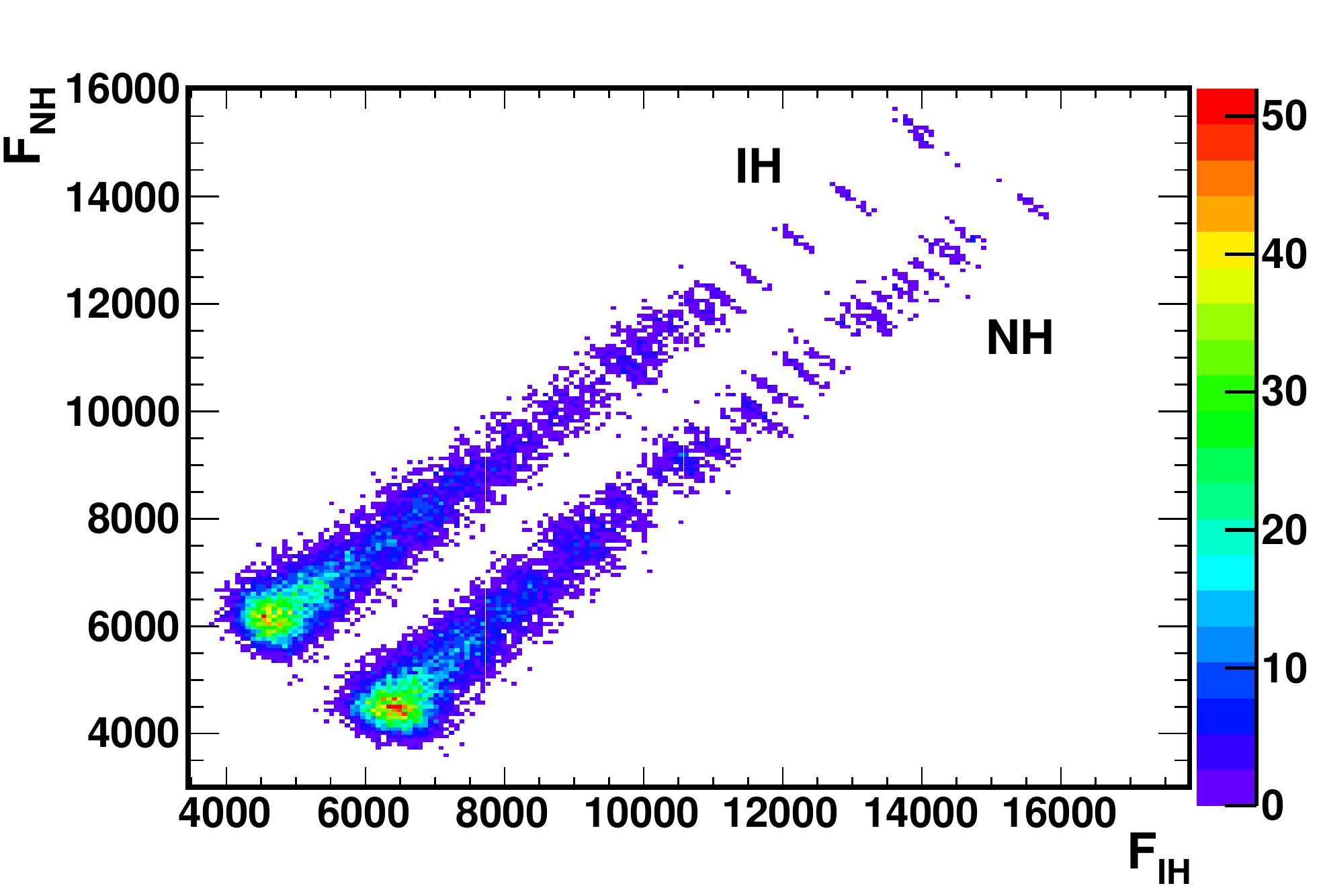}
\caption{\label{fig:19}(color online) $F_{NH}$ vs $F_{IH}$ distributions when all the parameters but 
$\Delta m^2_{atm}$ are letting free within their uncertainties, as in Table~\ref{table:8}.
The two populations correspond to the NH (bottom region) and the IH (top region) generations.
There is no observed change on the dispersions nor in the relative distance.}
\end{center}
\end{figure}

It is clear that 500 sets are not reproducing the full multi-parameter space. However, the generation can be considered
sufficient if no correlation is shown. The result is reported in Fig.~\ref{fig:19}: all the quantities except 
$\Delta m^2_{atm}$ are let fluctuate within their uncertainty. No correlation between $F_{NH}$ and $F_{IH}$ is evident, 
the net result being the linear increase of both $F$'s.
There is no observed change on the dispersions nor in the relative distance. That confirms the expectation.

\subsection{The $L/E$ and the not-$L/E$ dependent terms}\label{app:dmatm}
 
We consider here the developments presented in several papers
in the years 2012-2013. In particular, the possibility to use or not to use the odd-hierarchy terms for the determination
of the mass hierarchy in JUNO. For example, the analysis in~\cite{losecco} that make use 
of such terms found the optimized baseline around 20-30 km. 

 In the years 2012-2013 the conclusion that such terms were not useful to
 optimize the baseline for the next neutrino-reactor experiments and even to discriminate the hierarchy was a relevant 
 milestone~\cite{pet30}. At that time the uncertainty on $\Delta m^2_{atm}$ was rather large, about 4-5\%,
and the degeneracy, also observed for $F$ (see previous section), would not allow any firm conclusion. Only the critical
treatment (e.g.~\cite{pet30}) on the not-$L/E$ dependences allowed obtaining a discrimination of NH/IH.
Said in another way, only a $E$ dependence could break the degeneracy with $\Delta m^2_{atm}$~\cite{pet25}.
However, these conclusions did not  take into account either the increasing level of the $\Delta m^2_{atm}$ precision
obtained by the next measurements or the possibility to identify a more comprehensive estimator like $F$.

It is worth to note that equation~(\ref{eq:9}) does not define the $F$ estimator itself. It has been reported to enlighten the
 characteristics of $F$, and it is used to define the $I^+/I^-$ intervals. Specifically, $F$ does not come only from 
 the odd-hierarchy terms. It combines the NH/IH spectra with the neutrino energy through the observed data.
 A brief study of the baseline optimization for a JUNO-like reactor experiment
has been performed with $F$. We are grateful to E. Lisi who suggested this kind of study to prove the previous statements
about $F$, in particular its dependence on the neutrino energy variable.

\begin{figure*}[tb]
\begin{center}
\includegraphics[width=13cm]{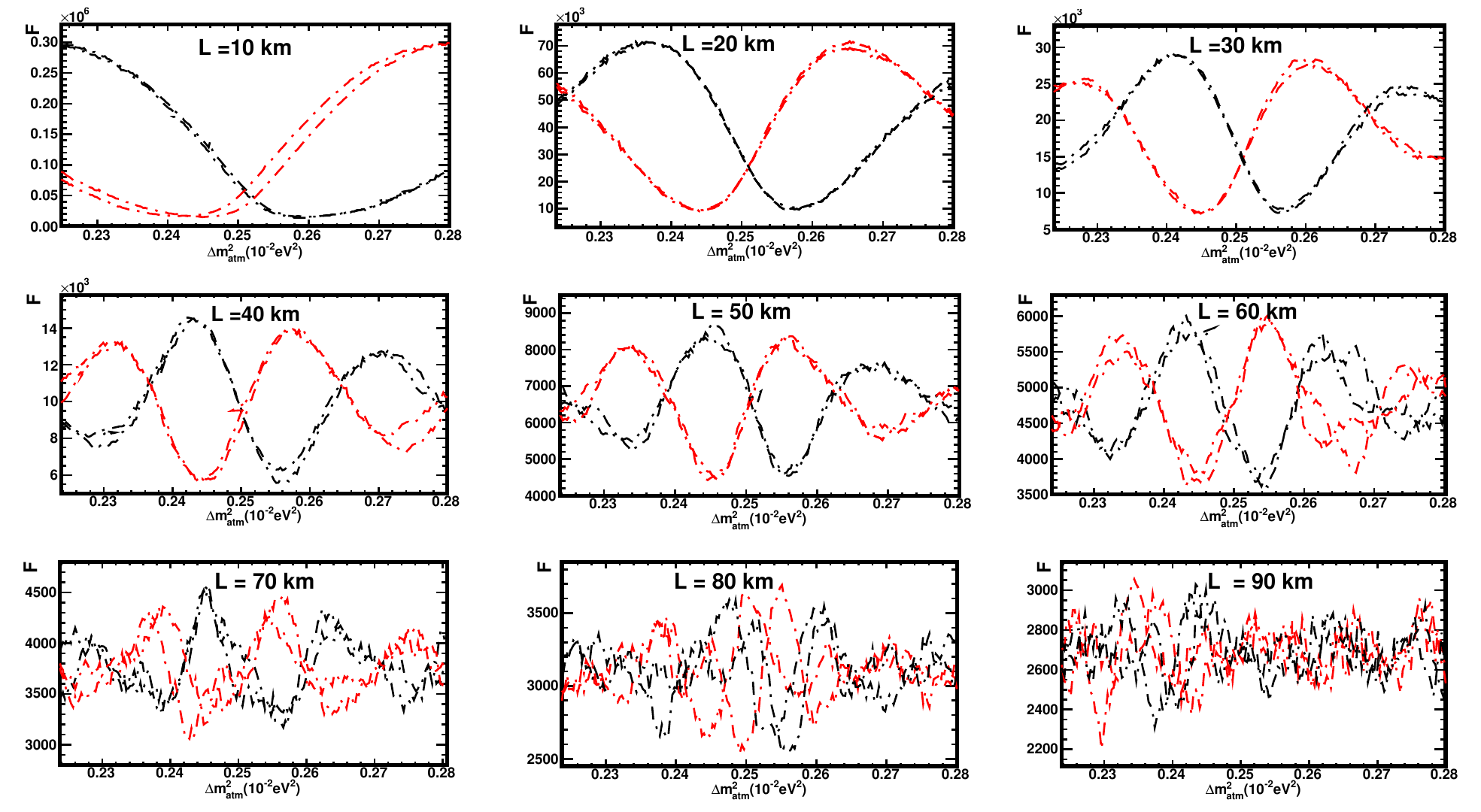}
\caption{\label{fig:20}(color online) $F$ vs $\Delta m^2_{atm}$ for a couple of generated JUNO-like toy experiments,
in the IH hypothesis. The black (red) curves correspond to the analysis when
the true (false) hypothesis is taken. The experimental configuration is the same as in Fig.~\ref{fig:7}.
Different baselines have been considered, as indicated in each plot.}
\end{center}
\end{figure*}

The performances of $F$ have been evaluated for baselines from 10 km to 90 km
for a JUNO-like experiment with 3\%$/\sqrt{E(MeV)}$ energy resolution and including the $1/L^2$ factor.
Results are shown in Fig.~\ref{fig:20}. At baselines smaller than 50 km the dispersion of
the minima are very large. For the 30 km baseline the precision is about 
$4\times 10^{-5}$ eV$^2$. Therefore, the bands of Fig.~\ref{fig:18} become four times wider, 
forbidding the use of $F$ due to the unbroken degeneracy with $\Delta m^2_{atm}$.
At baselines larger than 60 km the
NH/IH patterns loose accuracy and tend to overlap each other (the interference effect between \deltatre and  \deltadue
is vanishing). In the latter case the sensitivity decreases, up to disappear at 80-90 km.

Even if the analysis is only sketched out, we would any how
conclude that only the combined effect of a small uncertainty on $\Delta m^2_{atm}$ as extracted by the
$F$ procedure and the observed patterns for a specific experiment configuration, would allow us to obtain the quoted
sensitivities larger than 5 $\sigma$. Such combination is only possible for a reactor experiment with a
baseline around 50 km.

\section{Comparison with the $\chi^2$ method}\label{app:C}

There are at least two ways to parameterize the survival probability of reactor neutrinos. One is that used in this paper and
reported in eqs.~\ref{eq:two}-\ref{eq:7}. The second one follows the introduction of an effective atmospheric mass
$\Delta m^2_{ee}$ and a phase $\phi$~\cite{parke}. 

The latter one is quite useful to perform a correct $\chi^2$ analysis,
which lets the most critical parameter of the fit, i.e. the effective mass, vary freely.
The sign of the phase $\phi$ 
corresponds either to an advancement or a retardation of the phase oscillation, so providing
a non-degenerate solution of the MH sign. The achievable sensitivity is limited because most of the information
is kept into the $\Delta m^2_{atm}$ value, which is actually not used.

\begin{figure}[tb]
\begin{center}
\includegraphics[width=8.5cm]{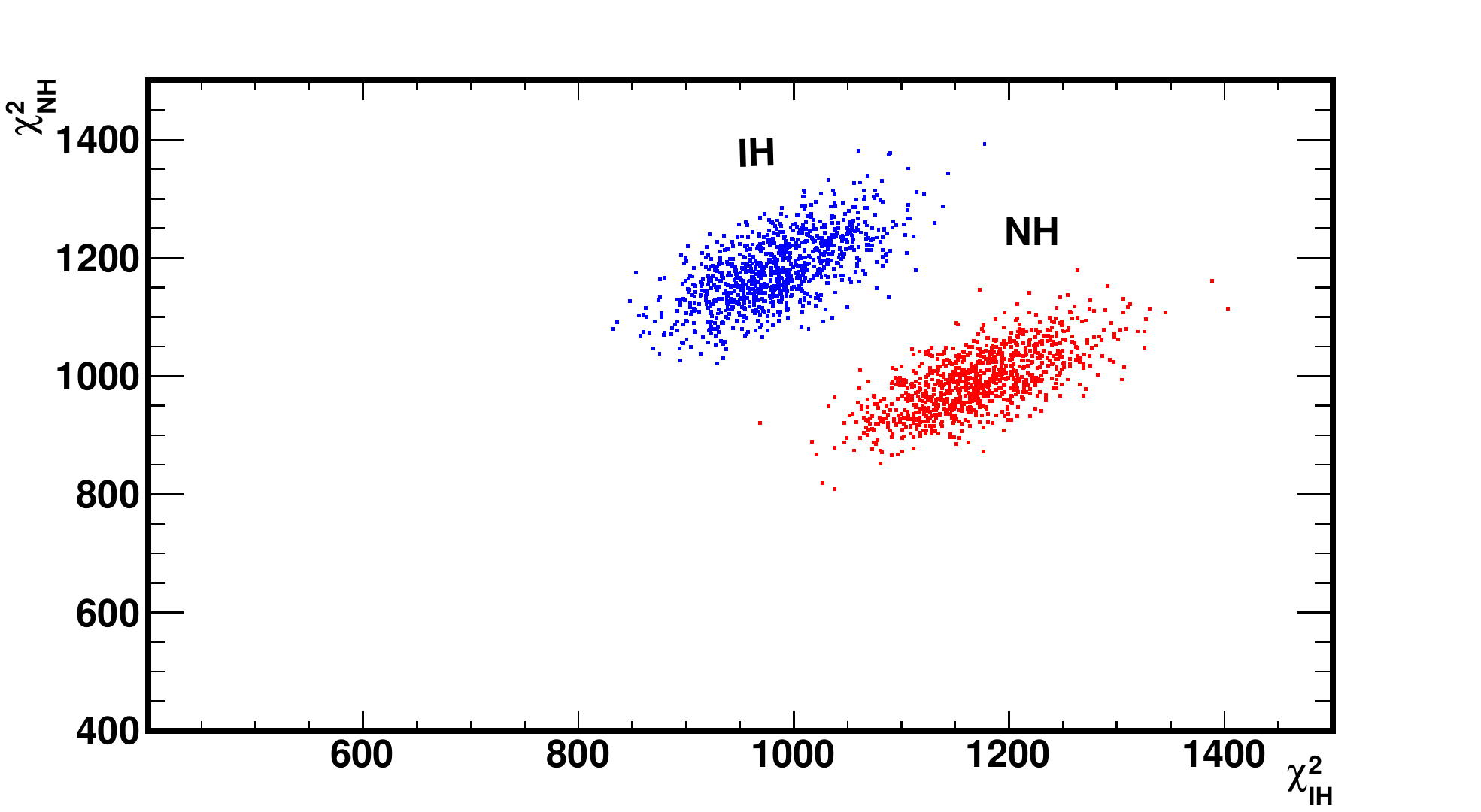}
\includegraphics[width=8.5cm]{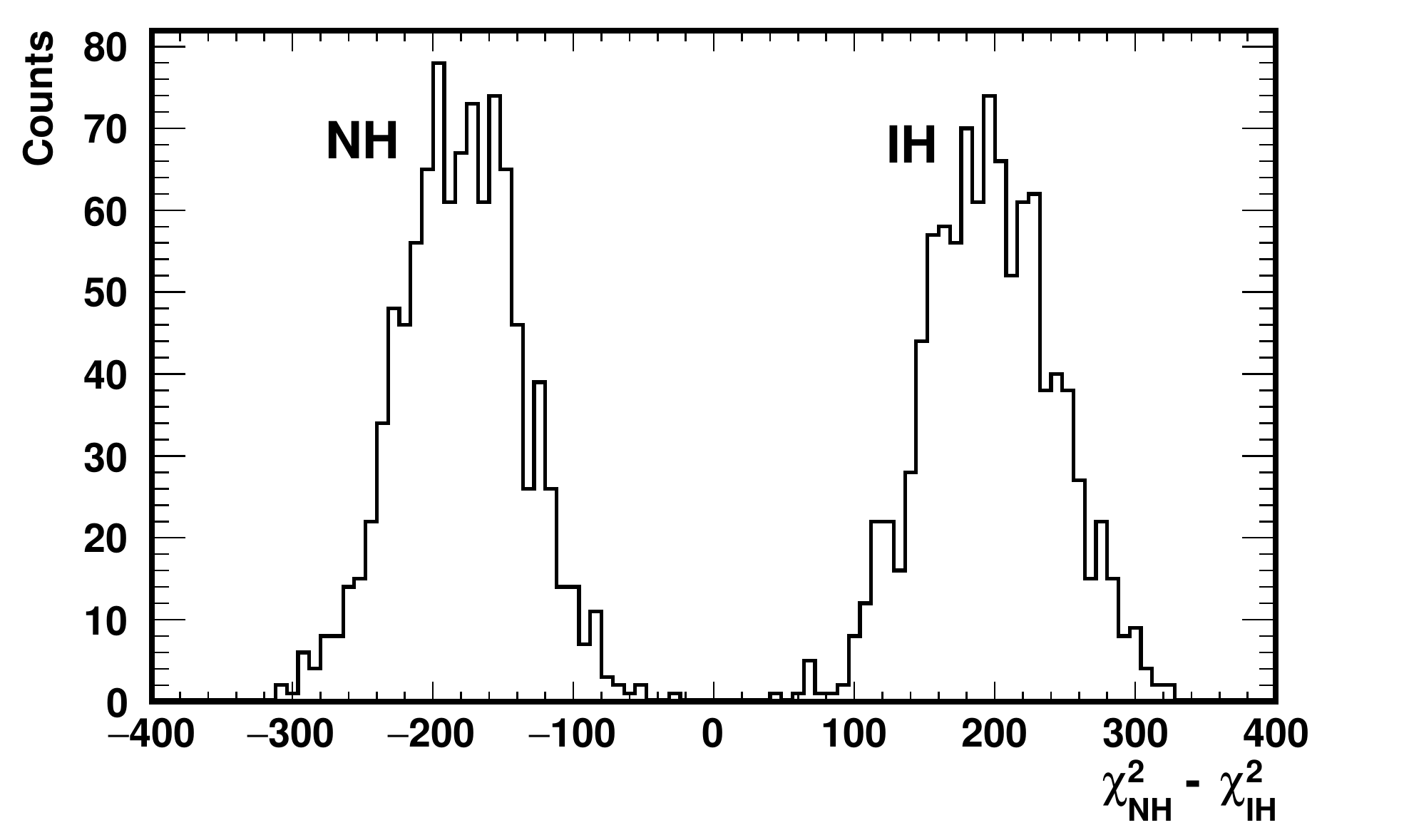}
\caption{\label{fig:211}(color online) (Top) $\chi^2_{min} (NH)$ vs $\chi^2_{min} (IH)$ 
distributions when all the parameters but $\Delta m^2_{atm}$ are letting free within their uncertainties, 
as in Table~\ref{table:8}.
The two populations correspond to the NH (bottom region) and the IH (top region) generations.
(Bottom) $\Delta\chi^2=\chi^2_{min} (NH)-\chi^2_{min} (IH)$ distributions for NH (right) and IH (left)
generation.}
\end{center}
\end{figure}

To partially demonstrate the previous conclusion a study on the $\chi^2$ has been developed, based on the
same toy samples of~\ref{app:B1}. For each JUNO-like toy the two $\chi^2_{min}$ have been evaluated for NH and
IH. The value of $\Delta m^2_{atm}$ for the result shown is kept fixed at $\Delta m^2_{atm}=0.00256\, {\rm eV}^2$.
Instead of computing the usual $\Delta\chi^2_{min}$  test statistic the two $\chi^2_{min}$ have been reported
in a scatter plot, similarly to $F$.  Two islands are found (top plot in Fig.~\ref{fig:211}), with a positive correlation.
The significance can be evaluated with the same technique applied to the $F$'s islands. 
For a fixed value of $\Delta m^2_{atm}$, the 2-D $\chi^2$ figure attains a similar significance level as the 2-D $F$,
i.e. around 5 $\sigma$ or more. In the bottom plot of Fig.~\ref{fig:211} the two
$\Delta\chi^2_{min}$ distributions are reported. The so called frequentist significance (i.e. the $p$-value
of the {\em wrong} hypothesis distribution at the mean of the {\em right} distribution) can be evaluated
and it corresponds to a similar level. 

In any case, it is well known (e.g.~\cite{pet22}) that the significance obtained from fitting with fixed $\Delta m^2_{atm}$ is overestimated.

\begin{figure}[bt]
\begin{center}
\includegraphics[width=7.5cm]{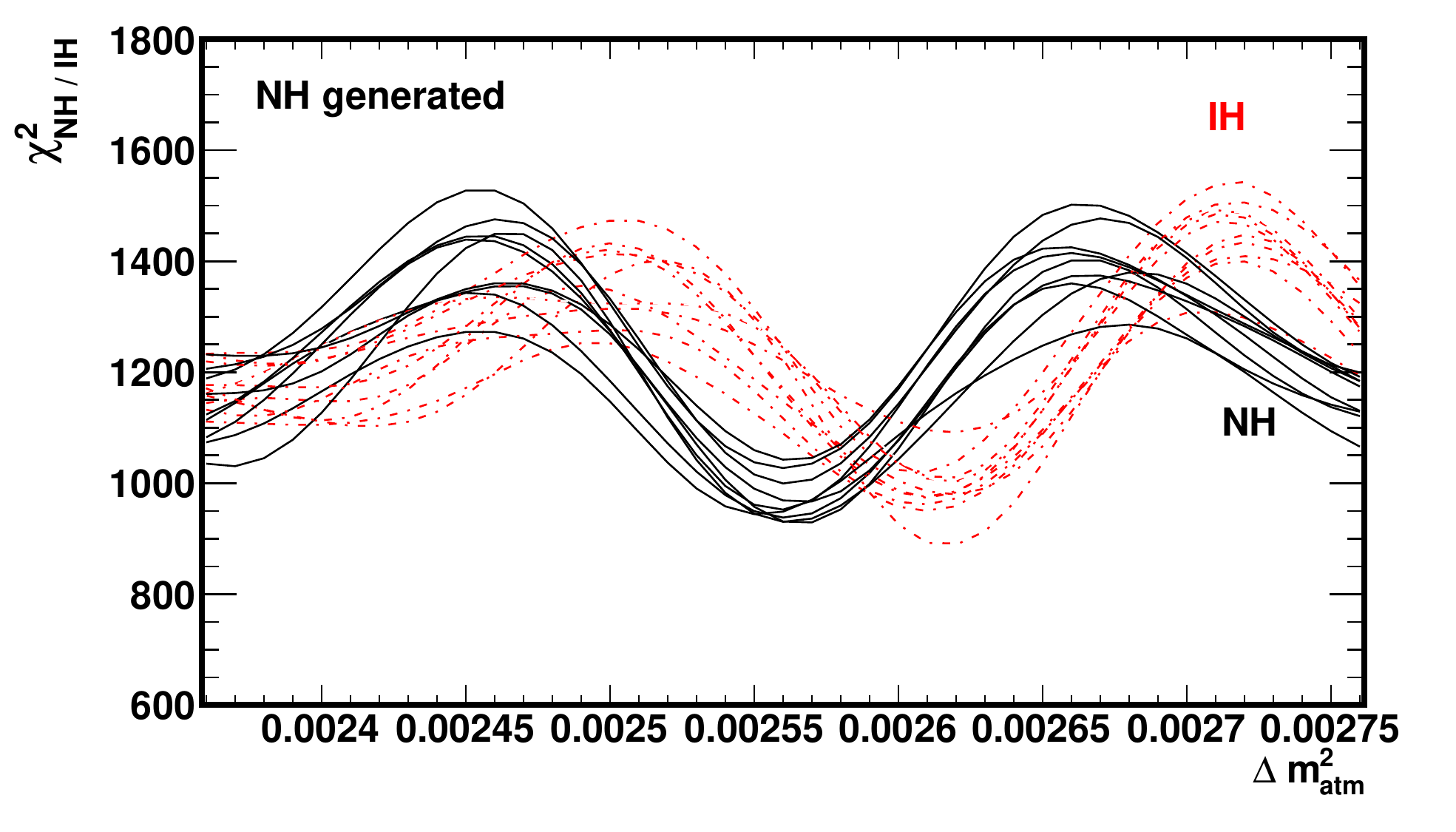}
\includegraphics[width=7.5cm]{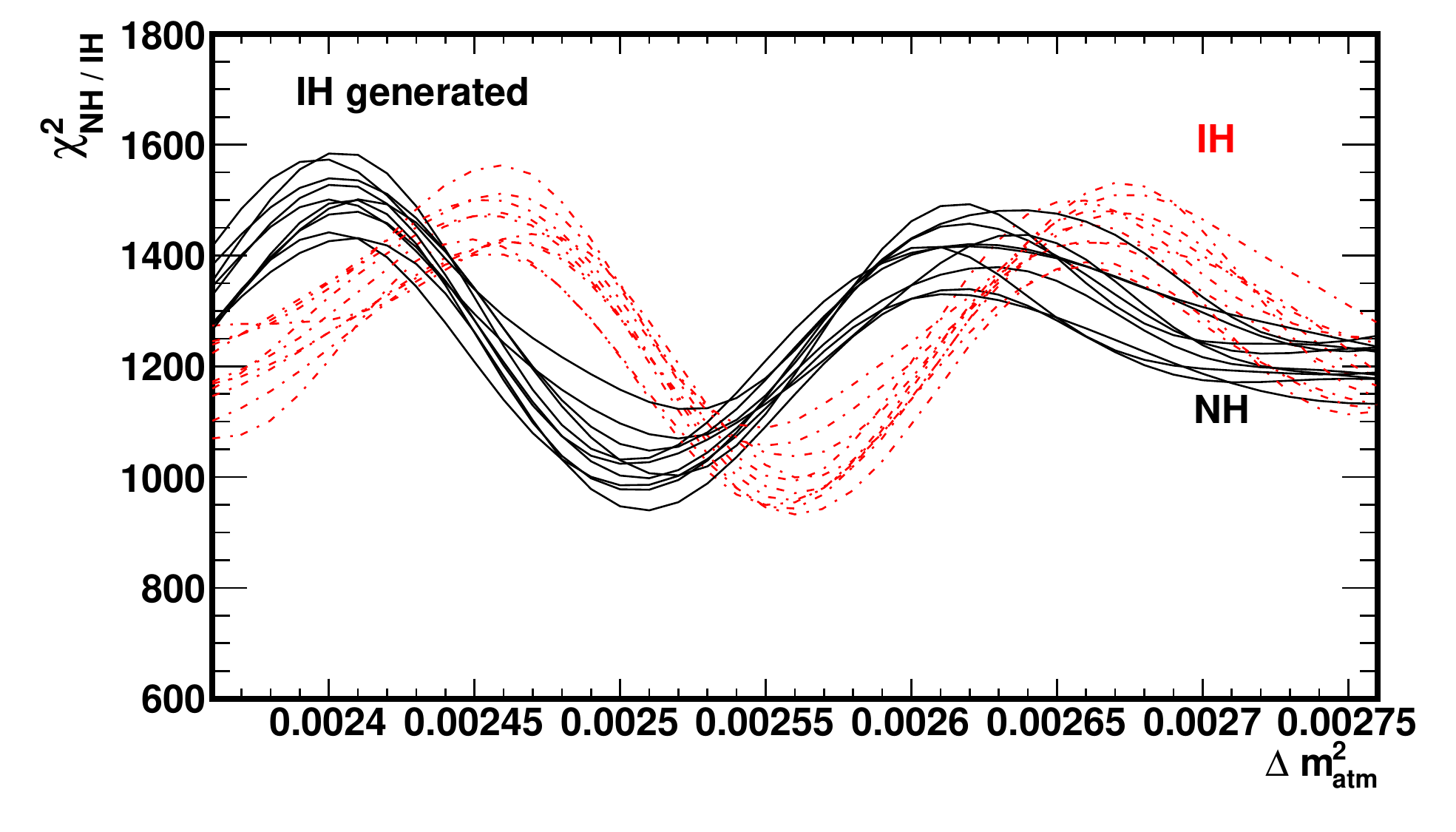}
\includegraphics[width=7.5cm]{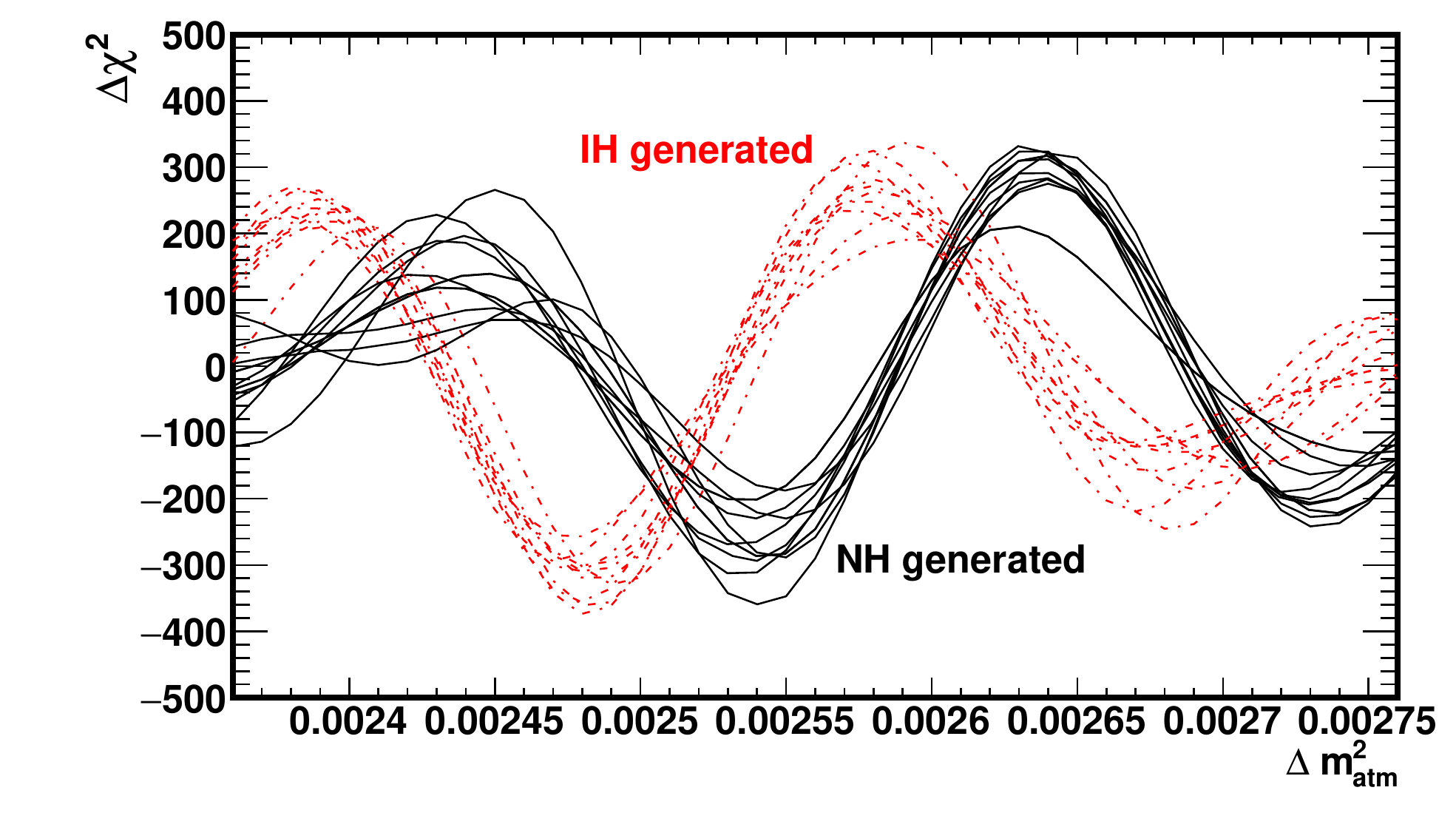}
\caption{\label{fig:22}(color online) $\chi^2_{min}$ vs $\Delta m^2_{atm}$ for ten generated NH (top-left) or
IH (top-right) JUNO-like toy experiments.
The black-continuous (red-dashed) curves correspond to the NH (IH) hypothesis in the analysis.
In the bottom picture the $\Delta\chi^2=\chi^2_{min} (NH)-\chi^2_{min} (IH)$ is drawn as function of 
$\Delta m^2_{atm}$ for NH generated toys 
(black, continuous line) and IH generated toys (red, dashed line). 
The experimental conditions are the same as in Fig.~\ref{fig:7}. }
\end{center}
\end{figure}

When the two minima of the $\chi^2$ are looked at as function
of $\Delta m^2_{atm}$ (top pictures in Fig.~\ref{fig:22}), different behaviours from those in Fig.~\ref{fig:17} for $F$ are obtained. 
Specifically, the two possible minima are rather close, around $6\times 10^{-5}$ eV$^2$, that is half of the 
distance with respect to the two $F$ minima.
 In general the two oscillatory shapes of the two models are much less distinguishable
than the corresponding ones of $F$. 
The dispersion of the single toys is so large that a simple numerical or analytical computation would not be accurate enough,
suggesting that 
a proper estimation of the sensitivity should proceed through a full Monte Carlo simulation.

A similar behaviour is obtained when the $\Delta\chi^2=\chi^2_{min}(NH)-\chi^2_{min}(IH)$ estimator is looked at 
(bottom picture in Fig.~\ref{fig:22}). The principal minimum (maximum) of $\Delta\chi^2$ corresponds to the NH (IH) 
solution for different values of $\Delta m^2_{atm}$. That is a bias induced by the finite energy resolution.
The opposite choices, principal maximum/minimum for NH/IH, gives the wrong solution. 
Considering the NH generation,
$\Delta\chi^2$ selects the NH solution at $\Delta m^2_{atm}=2.54\times 10^{-5}$ eV$^2$, and
the IH solution at $\Delta m^2_{atm}=2.63\times 10^{-5}$ eV$^2$. When the IH generation is considered, the 
$\Delta\chi^2$ selects the IH solution at $\Delta m^2_{atm}=2.56\times 10^{-5}$ eV$^2$, and
the NH solution at $\Delta m^2_{atm}=2.48\times 10^{-5}$ eV$^2$.
The relative distance in $\Delta\chi^2$ of the two minima/maxima roughly corresponds to the absolute
discrimination NH/IH as obtained from the standard $\Delta\chi^2$ analysis, without any degeneracy. 
As previously noted, the dispersion of the single toys is quite large even for $\Delta\chi^2$. 

In light of these results, we are temped to suggest that the $F$ estimator is more effective that the $\Delta\chi^2_{min}$ 
in the task of discriminating between the two MO hypotheses, at the same time identifying the correct one linked to the 
determination of the $\Delta m^2_{atm}$ value. In any case, it could be an interesting complementary option.


\begin{table*}[tb]
\begin{center}
{\footnotesize 
\begin{tabular}{c c | c c c c  }
\hline
\multicolumn{6}{l}{ten reactor cores, no background} \\
\hline
& & \multicolumn{4}{c}{2 years}  \\
\cline{3-6}
& & 4\% & 3.5\% & 3\% & 2.5\%   \\
\hline
\parbox[t]{2mm}{\multirow{5}{*}{\rotatebox[origin=c]{90}{NH true}}} 
& \multicolumn{1}{|c|}{$\mu_{NH}$} &2413.7 $\pm$ 3.8 & 2382.1 $\pm$ 3.7 & 2339.4 $\pm$ 3.8 & 2290.3 $\pm$ 3.8 \\
& \multicolumn{1}{|c|}{$\sigma_{NH}$} &117.4 $\pm$ 1.6 & 116.9 $\pm$ 2.4 & 120.2 $\pm$ 2.5 & 117.3 $\pm$ 1.6 \\
& \multicolumn{1}{|c|}{$\mu_{IH}$} &2680.9 $\pm$ 4.0& 2705.4 $\pm$ 4.1 &2742.3 $\pm$ 4.1 & 2784.7 $\pm$ 4.2 \\
& \multicolumn{1}{|c|}{$\sigma_{IH}$} &123.8 $\pm$ 1.7& 127.8 $\pm$ 2.6 & 128.1 $\pm$ 1.6 & 130.0 $\pm$ 1.7 \\
& \multicolumn{1}{|c|}{$r_{NH}$} &-0.557 $\pm$ 0.014 & -0.581 $\pm$ 0.0.18 & -0.585 $\pm$ 0.018 & -0.621 $\pm$ 0.012 \\
\hline 
\parbox[t]{2mm}{\multirow{5}{*}{\rotatebox[origin=c]{90}{IH true}}} 
& \multicolumn{1}{|c|}{$\mu_{NH}$} &2730.0 $\pm$ 4.3 & 2771.6 $\pm$ 4.3 & 2819.5 $\pm$ 4.2 & 2864.8 $\pm$ 4.2 \\
& \multicolumn{1}{|c|}{$\sigma_{NH}$} &135.6 $\pm$ 2.7 & 133.3 $\pm$ 2.7 & 130.5 $\pm$ 1.8 & 132.4 $\pm$ 2.4 \\
& \multicolumn{1}{|c|}{$\mu_{IH}$} &2377.4 $\pm$ 3.9 & 2327.5 $\pm$ 4.0 & 2268.3 $\pm$ 3.9 & 2226.2 $\pm$ 3.7 \\
& \multicolumn{1}{|c|}{$\sigma_{IH}$} &123.7 $\pm$ 2.5 & 125.6 $\pm$ 2.6 & 120.3 $\pm$ 1.6 & 117.4 $\pm$ 1.7 \\
& \multicolumn{1}{|c|}{$r_{IH}$} &-0.618 $\pm$ 0.017 & -0.591 $\pm$ 0.017 & -0.585 $\pm$ 0.013 & -0.594 $\pm$ 0.012 \\
\hline
& \multicolumn{1}{|c|}{$p-$value (IH)} & $2.73\times 10^{-2}$ & $7.66\times 10^{-3}$ & $1.07\times 10^{-3}$ & $1.07\times 10^{-4}$ \\
& \multicolumn{1}{|c|}{$n\, \sigma$ (IH)} &2.21 & 2.67 & 3.27 & 3.87 \\
& \multicolumn{1}{|c|}{$p-$value (NH)} & $3.64\times 10^{-2}$ & $1.00\times 10^{-2}$  & $1.12\times 10^{-3}$ & $1.11\times 10^{-4}$\\
& \multicolumn{1}{|c|}{$n\, \sigma$ (NH)} &2.09 & 2.58 & 3.26 & 3.86 \\
\hline
& & \multicolumn{4}{c}{4 years}  \\
\cline{3-6}
& & 4\% & 3.5\% & 3\% & 2.5\%   \\
\hline
\parbox[t]{2mm}{\multirow{5}{*}{\rotatebox[origin=c]{90}{NH true}}} 
& \multicolumn{1}{|c|}{$\mu_{NH}$} &3763.2 $\pm$ 5.6 & 3674.5 $\pm$ 5.7 & 3586.8 $\pm$ 5.8 & 3470.8 $\pm$ 5.4 \\
& \multicolumn{1}{|c|}{$\sigma_{NH}$} &176.8 $\pm$ 3.6 & 177.6 $\pm$ 3.8 & 180.7 $\pm$ 3.6 & 175.0 $\pm$ 2.9 \\
& \multicolumn{1}{|c|}{$\mu_{IH}$} & 4313.9 $\pm$ 6.0 & 4367.7 $\pm$ 6.1 & 4449.0 $\pm$ 6.4 & 4538.6 $\pm$ 6.0 \\
& \multicolumn{1}{|c|}{$\sigma_{IH}$} &186.5 $\pm$ 3.8 & 181.4 $\pm$ 3.9 & 201.5 $\pm$ 4.0 & 194.3 $\pm$ 2.6 \\
& \multicolumn{1}{|c|}{$r_{NH}$} &-0.650 $\pm$ 0.016 & -0.651 $\pm$ 0.010 & -0.668 $\pm$ 0.015 & -0.648 $\pm$ 0.010 \\
\hline 
\parbox[t]{2mm}{\multirow{5}{*}{\rotatebox[origin=c]{90}{IH true}}} 
& \multicolumn{1}{|c|}{$\mu_{NH}$} & 4428.6 $\pm$ 6.2& 4510.6 $\pm$ 6.1 & 4591.0 $\pm$ 6.2 & 4717.0 $\pm$ 6.3 \\
& \multicolumn{1}{|c|}{$\sigma_{NH}$} &195.0 $\pm$ 4.0 & 191.4 $\pm$ 3.7 & 190.1 $\pm$ 2.5 & 173.5 $\pm$ 3.2 \\
& \multicolumn{1}{|c|}{$\mu_{IH}$} & 3666.2 $\pm$ 5.8 & 3563.0 $\pm$ 5.7 & 3460.9 $\pm$ 5.9 & 3342.8 $\pm$ 5.3 \\
& \multicolumn{1}{|c|}{$\sigma_{IH}$} &181.6 $\pm$ 3.6 & 177.4 $\pm$ 0.6 & 178.8 $\pm$ 2.4 & 173.5 $\pm$ 3.2 \\
& \multicolumn{1}{|c|}{$r_{IH}$} &-0.629 $\pm$ 0.016 & -0.644 $\pm$ 0.013 & -0.663 $\pm$ 0.010 & -0.633 $\pm$ 0.011 \\
\hline
& \multicolumn{1}{|c|}{$p-$value (IH)} &$2.58\times 10^{-3}$ & $1.99\times 10^{-4}$ & $9.69\times 10^{-6}$ & $1.94\times 10^{-8}$ \\
& \multicolumn{1}{|c|}{$n\, \sigma$ (IH)} &3.01 & 3.73 & 4.42 & 5.21 \\
& \multicolumn{1}{|c|}{$p-$value (NH)} & $3.07\times 10^{-3}$ & $1.95\times 10^{-4}$ & $7.55\times 10^{-6}$ & $2.49\times 10^{-8}$ \\
& \multicolumn{1}{|c|}{$n\, \sigma$ (NH)} &2.96 & 3.73 & 4.48 & 5.16 \\
\hline
& & \multicolumn{4}{c}{6 years}  \\
\cline{3-6}
& & 4\% & 3.5\% & 3\% & 2.5\%   \\
\hline
\parbox[t]{2mm}{\multirow{5}{*}{\rotatebox[origin=c]{90}{NH true}}} 
& \multicolumn{1}{|c|}{$\mu_{NH}$} & 4956.5 $\pm$ 7.3 & 4815.3 $\pm$ 7.1 & 4667.2 $\pm$ 7.0 & 4486.2 $\pm$ 7.1\\
& \multicolumn{1}{|c|}{$\sigma_{NH}$} &230.6 $\pm$ 4.3 & 222.3 $\pm$ 4.4 & 227.9 $\pm$ 4.3 & 219.9 $\pm$ 3.1 \\
& \multicolumn{1}{|c|}{$\mu_{IH}$} & 5802.5 $\pm$ 7.9 & 5906.1 $\pm$ 7.7 & 6006.1 $\pm$ 7.4 & 6163.2 $\pm$ 8.3 \\
& \multicolumn{1}{|c|}{$\sigma_{IH}$} &248.8 $\pm$ 0.9 & 239.7 $\pm$ 4.8 & 252.0 $\pm$ 4.6 & 255.9 $\pm$ 3.1 \\
& \multicolumn{1}{|c|}{$r_{NH}$} &-0.674 $\pm$ 0.012 & -0.684 $\pm$ 0.010 & -0.682 $\pm$ 0.015 & -0.706 $\pm$ 0.009 \\
\hline 
\parbox[t]{2mm}{\multirow{5}{*}{\rotatebox[origin=c]{90}{IH true}}} 
& \multicolumn{1}{|c|}{$\mu_{NH}$} & 6001.9 $\pm$ 7.9 & 6110.2 $\pm$ 7.8 & 6242.9 $\pm$ 7.9 & 6440.7 $\pm$ 7.4  \\
& \multicolumn{1}{|c|}{$\sigma_{NH}$} &246.4 $\pm$ 4.8 & 241.9 $\pm$ 5.0 & 248.4 $\pm$ 4.9 &  233.5 $\pm$ 2.9 \\
& \multicolumn{1}{|c|}{$\mu_{IH}$} & 4791.1 $\pm$ 7.2 & 4644.1 $\pm$ 7.0 & 4491.3 $\pm$ 7.0 & 4280.0 $\pm$ 6.5 \\
& \multicolumn{1}{|c|}{$\sigma_{IH}$} & 226.8 $\pm$ 4.4 & 216.6 $\pm$ 4.5 & 219.2 $\pm$ 4.3 & 205.9 $\pm$ 2.7 \\
& \multicolumn{1}{|c|}{$r_{IH}$} &-0.692 $\pm$ 0.014& -0.673 $\pm$ 0.011 & -0.686 $\pm$ 0.14 & -0.664 $\pm$ 0.010 \\
\hline
& \multicolumn{1}{|c|}{$p-$value (IH)} &$3.48\times 10^{-4}$ & $3.51\times 10^{-6}$ & $3.98\times 10^{-8}$ & 
$7.96\times 10^{-11}$\\
& \multicolumn{1}{|c|}{$n\, \sigma$ (IH)} & 3.58 & 4.64 & 5.49 & 6.50 \\
& \multicolumn{1}{|c|}{$p-$value (NH)} & $3.30\times 10^{-4}$ & $3.21\times 10^{-6}$ & $3.12\times 10^{-8}$ & $1.08\times 10^{-11}$ \\
& \multicolumn{1}{|c|}{$n\, \sigma$ (NH)} & 3.59 & 4.66 & 5.53 &  6.79 \\
\hline
\end{tabular}
}
\end{center}
\caption{\label{table:3} The bi-gaussian fits of the $F$ distributions are reported, for the
JUNO-like configuration of 2, 4 and six years of data taking and different energy resolutions,
$\mu_{MH}$, $\sigma_{MH}$ and $r_{MH}$ being the means, the standard deviations and the correlation 
coefficients, respectively, of the fitted 2-D Gaussians.
The ten near reactor cores have been considered with a $\pm$ 5 m uniform dispersion 
on their relative baseline.
No background source has been included. The sensitivity 
has been computed from the $p$-values estimation as described in the text in terms of number of $\sigma$'s 
in the two-sided option. $n\, \sigma$ (IH) stays for the IH rejection significance, and equivalently for NH.}
\end{table*}

\begin{table*}[h]
\begin{center}
{\footnotesize 
\begin{tabular}{c c | c c c c  }
\hline
\multicolumn{6}{l}{10 reactor cores plus the 2 remote cores} \\
\hline
& & \multicolumn{4}{c}{2 years}  \\
\cline{3-6}
& & 4\% & 3.5\% & 3\% & 2.5\%   \\
\hline
\parbox[t]{2mm}{\multirow{5}{*}{\rotatebox[origin=c]{90}{NH true}}} 
& \multicolumn{1}{|c|}{$\mu_{NH}$} & 2834.2 $\pm$ 3.9 & 2807.4 $\pm$ 3.8 &2770.5 $\pm$ 3.9 & 2726.6 $\pm$ 3.7 \\
& \multicolumn{1}{|c|}{$\sigma_{NH}$} & 115.6 $\pm$ 1.5 & 117.2 $\pm$ 2.4 &121.1 $\pm$ 2.5 & 115.5 $\pm$ 1.5 \\
& \multicolumn{1}{|c|}{$\mu_{IH}$} & 3125.2 $\pm$ 3.9 & 3145.5 $\pm$ 4.1 & 3179.8 $\pm$ 4.0& 3220.9 $\pm$ 4.1 \\
& \multicolumn{1}{|c|}{$\sigma_{IH}$} & 123.5 $\pm$ 1.6 & 126.8 $\pm$ 2.7 & 125.7 $\pm$ 2.5 & 125.6 $\pm$ 1.6 \\
& \multicolumn{1}{|c|}{$r_{NH}$} & -0.585 $\pm$ 0.018 & -0.606 $\pm$ 0.013& -0.620 $\pm$ 0.014 & -0.644 $\pm$ 0.012\\
\hline 
\parbox[t]{2mm}{\multirow{5}{*}{\rotatebox[origin=c]{90}{IH true}}} 
& \multicolumn{1}{|c|}{$\mu_{NH}$} & 3138.9 $\pm$ 4.2 & 3177.9 $\pm$ 4.3 & 3225.0 $\pm$ 4.1 & 3264.5 $\pm$ 3.9   \\
& \multicolumn{1}{|c|}{$\sigma_{NH}$} & 131.7 $\pm$ 2.6 & 132.3 $\pm$ 1.8 & 126.0 $\pm$ 2.6 & 130.2 $\pm$ 2.8 \\
& \multicolumn{1}{|c|}{$\mu_{IH}$} & 2831.9 $\pm$ 4.0 & 2785.5 $\pm$ 4.1 & 2732.2 $\pm$ 3.8 & 2691.4 $\pm$ 3.8\\
& \multicolumn{1}{|c|}{$\sigma_{IH}$} & 124.6 $\pm$ 2.5 & 125.4 $\pm$ 1.5 & 118.9 $\pm$ 0.8 & 118.1 $\pm$ 2.3 \\
& \multicolumn{1}{|c|}{$r_{IH}$} & -0.632 $\pm$ 0.016 & -0.628 $\pm$ 0.010 & -0.610 $\pm$ 0.016 & -0.618 $\pm$ 0.018 \\
\hline
& \multicolumn{1}{|c|}{$p-$value (IH)} & $3.21\times 10^{-2}$ & $1.14\times 10^{-2}$ & $1.92\times 10^{-3}$ & $2.20\times 10^{-4}$\\
& \multicolumn{1}{|c|}{$n\, \sigma$ (IH)} & 2.14 & 2.53 & 3.10& 3.69 \\
& \multicolumn{1}{|c|}{$p-$value (NH)} & $4.23\times 10^{-2}$  & $1.46\times 10^{-2}$ & $1.85\times 10^{-3}$ & $2.59\times 10^{-4}$ \\
& \multicolumn{1}{|c|}{$n\, \sigma$ (NH)} & 2.03 & 2.44 & 3.11& 3.65 \\
\hline
& & \multicolumn{4}{c}{4 years}  \\
\cline{3-6}
& & 4\% & 3.5\% & 3\% & 2.5\%   \\
\hline
\parbox[t]{2mm}{\multirow{5}{*}{\rotatebox[origin=c]{90}{NH true}}} 
& \multicolumn{1}{|c|}{$\mu_{NH}$} & 4784.8 $\pm$ 3.1 & 4712.3 $\pm$ 5.5 & 4638.2 $\pm$ 5.9 & 4545.7 $\pm$ 5.2 \\
& \multicolumn{1}{|c|}{$\sigma_{NH}$} & 170.2 $\pm$ 3.0 & 170.6 $\pm$ 2.3 & 174.4 $\pm$ 3.6 & 162.5 $\pm$ 2.0 \\
& \multicolumn{1}{|c|}{$\mu_{IH}$} & 5357.8 $\pm$ 4.3 & 5411.3 $\pm$ 6.0 & 5483.2 $\pm$ 6.8 & 5560.4 $\pm$ 5.9 \\
& \multicolumn{1}{|c|}{$\sigma_{IH}$} & 178.8 $\pm$ 3.2 & 184.9 $\pm$ 2.1 & 190.9 $\pm$ 4.2 & 181.8 $\pm$ 2.3 \\
& \multicolumn{1}{|c|}{$r_{NH}$} & -0.656 $\pm$ 0.009 & -0.687 $\pm$ 0.010 & -0.696 $\pm$ 0.014 & -0.698 $\pm$0.009 \\
\hline 
\parbox[t]{2mm}{\multirow{5}{*}{\rotatebox[origin=c]{90}{IH true}}} 
& \multicolumn{1}{|c|}{$\mu_{NH}$} & 5399.6 $\pm$6.0 & 5475.5 $\pm$ 5.2 & 5557.6 $\pm$ 6.1 & 5653.7 $\pm$ 6.1 \\
& \multicolumn{1}{|c|}{$\sigma_{NH}$} & 184.6 $\pm$ 3.8 & 181.4 $\pm$ 3.1 & 187.9 $\pm$2.3 & 189.3 $\pm$ 2.3 \\
& \multicolumn{1}{|c|}{$\mu_{IH}$} & 4754.5 $\pm$ 5.4 & 4668.3 $\pm$ 4.9 & 4576.9 $\pm$5.4 & 4478.2 $\pm$ 5.4 \\
& \multicolumn{1}{|c|}{$\sigma_{IH}$} & 167.6 $\pm$ 3.5 & 170.4 $\pm$ 3.3 & 168.6 $\pm$ 2.0 & 167.0 $\pm$ 2.1 \\
& \multicolumn{1}{|c|}{$r_{IH}$} & -0.650 $\pm$ 0.010 & -0.670 $\pm$ 0.014 & -0.707 $\pm$ 0.010 & -0.687 $\pm$ 0.010 \\
\hline
& \multicolumn{1}{|c|}{$p-$value (IH)} & $3.50\times 10^{-3}$ & $4.14\times 10^{-4}$ & $3.06\times 10^{-5}$ &$1.23\times 10^{-7}$ \\
& \multicolumn{1}{|c|}{$n\, \sigma$ (IH)} & 2.92 & 3.53 & 4.17 & 5.29 \\
& \multicolumn{1}{|c|}{$p-$value (NH)} & $3.64\times 10^{-3}$ & $3.92\times 10^{-4}$ & $2.60\times 10^{-5}$ & $1.85\times 10^{-7}$\\
& \multicolumn{1}{|c|}{$n\, \sigma$ (NH)} & 2.91 & 3.55 & 4.21 & 5.21 \\
\hline
& & \multicolumn{4}{c}{6 years}  \\
\cline{3-6}
& & 4\% & 3.5\% & 3\% & 2.5\%   \\
\hline
\parbox[t]{2mm}{\multirow{5}{*}{\rotatebox[origin=c]{90}{NH true}}} 
& \multicolumn{1}{|c|}{$\mu_{NH}$} & 6650.4 $\pm$ 6.8 & 6538.2 $\pm$ 6.8 & 6427.2 $\pm$ 6.6 & 6292.0 $\pm$ 6.5 \\
& \multicolumn{1}{|c|}{$\sigma_{NH}$} & 213.1 $\pm$ 4.1 & 212.1 $\pm$ 2.2 & 203.8 $\pm$ 4.0 & 201.1 $\pm$ 2.5 \\
& \multicolumn{1}{|c|}{$\mu_{IH}$} & 7504.1 $\pm$ 7.2 & 7600.8 $\pm$7.1 & 7693.3 $\pm$ 7.3 & 7817.2 $\pm$ 7.5 \\
& \multicolumn{1}{|c|}{$\sigma_{IH}$} & 224.7 $\pm$ 4.3 & 220.4 $\pm$ 2.8 & 226.4 $\pm$ 4.4 & 232.1 $\pm$ 2.9 \\
& \multicolumn{1}{|c|}{$r_{NH}$} & -0.709 $\pm$ 0.013 & -0.726 $\pm$0.010 & -0.736 $\pm$ 0.009 & -0.719 $\pm$ 0.009  \\
\hline 
\parbox[t]{2mm}{\multirow{5}{*}{\rotatebox[origin=c]{90}{IH true}}} 
& \multicolumn{1}{|c|}{$\mu_{NH}$} & 7583.7 $\pm$ 7.1 &7685.1 $\pm$7.3 & 7801.1 $\pm$ 7.3 & 7964.8 $\pm$ 7.2 \\
& \multicolumn{1}{|c|}{$\sigma_{NH}$} & 222.0 $\pm$ 2.7 & 226.7 $\pm$ 2.7 & 227.3 $\pm$ 4.6 & 222.5 $\pm$ 4.3 \\
& \multicolumn{1}{|c|}{$\mu_{IH}$} & 6585.6 $\pm$ 6.7 & 6472.1 $\pm$ 6.5 & 6349.9 $\pm$ 6.4 & 6179.9 $\pm$ 6.2 \\
& \multicolumn{1}{|c|}{$\sigma_{IH}$} & 208.9 $\pm$ 2.5 & 202.8 $\pm$ 2.4 & 198.6 $\pm$ 4.0 & 193.5 $\pm$ 3.8 \\
& \multicolumn{1}{|c|}{$r_{IH}$} & -0.704 $\pm$ 0.009 & -0.731 $\pm$ 0.008 & -0.713 $\pm$ 0.010 & -0.708 $\pm$ 0.013 \\
\hline
& \multicolumn{1}{|c|}{$p-$value (IH)} & $4.67\times 10^{-4}$ & $1.46\times 10^{-5}$& $1.36\times 10^{-7}$ & $4.13\times 10^{-11}$ \\
& \multicolumn{1}{|c|}{$n\, \sigma$ (IH)} &  3.50 & 4.34 & 5.27 & 6.50\\
& \multicolumn{1}{|c|}{$p-$value (NH)} & $4.27\times 10^{-4}$ &$1.35\times 10^{-5}$ & $1.13\times 10^{-7}$ &$1.96\times 10^{-11}$ \\
& \multicolumn{1}{|c|}{$n\, \sigma$ (NH)} &3.52 & 4.35 & 5.30 & 6.71 \\
\hline
\end{tabular}
}
\caption{\label{table:4} The bi-gaussian fits of the $F$ distributions are reported, for the
JUNO-like configuration of 2, 4 and 6 years of data taking and different energy resolutions, 
$\mu_{MH}$, $\sigma_{MH}$ and $r_{MH}$ being the means, the standard deviations and the correlation 
coefficients, respectively, of the fitted 2-D Gaussians.
The 10 near reactor cores have been considered with a $\pm$ 5 m uniform dispersion 
on their relative baseline, as well as the two remote reactor plants with a $\pm$ 0.5 km uniform dispersion 
on their relative baseline. 
The sensitivity 
has been computed from the $p$-values estimation as described in the text in terms of number of $\sigma$'s 
in the two-sided option. $n\, \sigma$ (IH) stays for the IH rejection significance, and equivalently for NH.}
\end{center}
\end{table*}

\begin{table*}[tb]
\begin{center}
{\footnotesize 
\begin{tabular}{c c | c  c }
\hline
\multicolumn{4}{l}{10 reactor cores, 6 years } \\
\hline
& & \multicolumn{1}{c}{with uncorrelated background}  & \multicolumn{1}{c}{with correlated bump around $4\sim 6$ MeV} \\
\hline
\parbox[t]{2mm}{\multirow{5}{*}{\rotatebox[origin=c]{90}{NH true}}} 
& \multicolumn{1}{|c|}{$\mu_{NH}$} & 8908.2 $\pm$ 6.1 & 6039.3 $\pm$ 7.4 \\
& \multicolumn{1}{|c|}{$\sigma_{NH}$} &187.0 $\pm$ 3.4 & 226.9 $\pm$ 4.4\\
& \multicolumn{1}{|c|}{$\mu_{IH}$} & 10315.2 $\pm$ 6.5 & 7346.8 $\pm$ 7.7  \\
& \multicolumn{1}{|c|}{$\sigma_{IH}$} & 201.9 $\pm$ 3.6  & 238.9 $\pm$ 4.9 \\
& \multicolumn{1}{|c|}{$r_{NH}$} & -0.841 $\pm$ 0.005  & -0.573 $\pm$ 0.015\\
\hline 
\parbox[t]{2mm}{\multirow{5}{*}{\rotatebox[origin=c]{90}{IH true}}} 
& \multicolumn{1}{|c|}{$\mu_{NH}$} & 10230.0 $\pm$ 6.3 & 7573.7 $\pm$ 4.9   \\
& \multicolumn{1}{|c|}{$\sigma_{NH}$} & 196.5 $\pm$ 3.4  & 243.6 $\pm$ 4.7 \\
& \multicolumn{1}{|c|}{$\mu_{IH}$} & 8998.8 $\pm$ 5.9  & 5861.0 $\pm$ 5.5\\
& \multicolumn{1}{|c|}{$\sigma_{IH}$} & 181.5 $\pm$ 3.2 & 219.1 $\pm$ 3.1 \\
& \multicolumn{1}{|c|}{$r_{IH}$} & -0.835 $\pm$ 0.007 & -0.617 $\pm$ 0.010 \\
\hline
& \multicolumn{1}{|c|}{$p-$value (IH)} &  $2.88\times 10^{-8}$ & $1.57\times 10^{-8}$\\
& \multicolumn{1}{|c|}{$n\, \sigma$ (IH)} & 5.55 & 5.65 \\
& \multicolumn{1}{|c|}{$p-$value (NH)} &  $1.90\times 10^{-8}$ & $1.75\times 10^{-8}$ \\
& \multicolumn{1}{|c|}{$n\, \sigma$ (NH)} & 5.62 & 5.63 \\
\hline
\end{tabular}
}
\caption{\label{table:5} The bi-gaussian fits of the $F$ distributions are reported, for the
JUNO-like configuration of six years of data taking and a $3\%/\sqrt{E}$ energy resolutions,
$\mu_{MH}$, $\sigma_{MH}$ and $r_{MH}$ being the means, the standard deviations and the correlation 
coefficients, respectively, of the fitted 2-D Gaussians.
The ten near reactor cores have been considered with a $\pm$ 5 m uniform dispersion 
on their relative baseline.
In the first set of data the uncorrelated background has been included from the $^9$Li, scaled to the
total amount. In the second set the correlated bump around $4\sim 6$ MeV has been considered.
The sensitivity 
has been computed from the $p$-values estimation as described in the text in terms of number of $\sigma$'s 
in the two-sided option. $n\, \sigma$ (IH) stays for the IH rejection significance, and equivalently for NH.}
\end{center}
\end{table*}

\begin{table*}[tb]
\begin{center}
{\footnotesize 
\begin{tabular}{c c | c  c }
\hline
\multicolumn{4}{l}{10 reactor cores plus the 2 remote cores, 6 years } \\
\hline
& & \multicolumn{1}{c}{with -3\% on flux}  & \multicolumn{1}{c}{with +3\% on flux} \\
\hline
\parbox[t]{2mm}{\multirow{5}{*}{\rotatebox[origin=c]{90}{NH true}}} 
& \multicolumn{1}{|c|}{$\mu_{NH}$} & 5553.1 $\pm$ 5.9 & 7652.8 $\pm$ 6.2 \\
& \multicolumn{1}{|c|}{$\sigma_{NH}$} &212.5 $\pm$ 4.5 & 197.3 $\pm$ 3.6\\
& \multicolumn{1}{|c|}{$\mu_{IH}$} & 6823.3 $\pm$ 7.0 & 8915.8 $\pm$ 6.6  \\
& \multicolumn{1}{|c|}{$\sigma_{IH}$} & 240.2 $\pm$ 5.1  & 211.2 $\pm$ 3.9 \\
& \multicolumn{1}{|c|}{$r_{NH}$} & -0.607 $\pm$ 0.019  & -0.825 $\pm$ 0.006\\
\hline 
\parbox[t]{2mm}{\multirow{5}{*}{\rotatebox[origin=c]{90}{IH true}}} 
& \multicolumn{1}{|c|}{$\mu_{NH}$} & 6994.8 $\pm$ 4.9 & 8991.2 $\pm$ 7.1   \\
& \multicolumn{1}{|c|}{$\sigma_{NH}$} & 247.6 $\pm$ 5.2  & 214.8 $\pm$ 4.2 \\
& \multicolumn{1}{|c|}{$\mu_{IH}$} & 5426.0 $\pm$ 4.7  & 7603.9 $\pm$ 6.4\\
& \multicolumn{1}{|c|}{$\sigma_{IH}$} & 213.8 $\pm$ 4.1 & 196.3 $\pm$ 3.7 \\
& \multicolumn{1}{|c|}{$r_{IH}$} & -0.603 $\pm$ 0.012 & -0.815 $\pm$ 0.005 \\
\hline
& \multicolumn{1}{|c|}{$p-$value (IH)} &  $7.53\times 10^{-8}$ & $1.48\times 10^{-7}$\\
& \multicolumn{1}{|c|}{$n\, \sigma$ (IH)} & 5.38 & 5.25 \\
& \multicolumn{1}{|c|}{$p-$value (NH)} &  $9.68\times 10^{-8}$ & $1.59\times 10^{-7}$ \\
& \multicolumn{1}{|c|}{$n\, \sigma$ (NH)} & 5.34 & 5.24 \\
\hline
\end{tabular}
}
\end{center}
\caption{\label{table:6} The bi-gaussian fits of the $F$ distributions are reported, for the
JUNO-like configuration of six years of data taking and a $3\%/\sqrt{E}$ energy resolutions,
$\mu_{MH}$, $\sigma_{MH}$ and $r_{MH}$ being the means, the standard deviations and the correlation 
coefficients, respectively, of the fitted 2-D Gaussians.
The ten near reactor cores have been considered with a $\pm$ 5 m uniform dispersion 
on their relative baseline, as well as the two remote reactor plants with a $\pm$ 0.5 km uniform dispersion 
on their relative baseline. 
The two data sets correspond to a - 3\% and +3\% on the reactor flux.
The sensitivity 
has been computed from the $p$-values estimation as described in the text in terms of number of $\sigma$'s 
in the two-sided option. $n\, \sigma$ (IH) stays for the IH rejection significance, and equivalently for NH.
As expected the results are not so sensitive to $\pm$ 3 variations of the flux, although the $F$ islands are
shifted. The slight increase(decrease) of the sensitivity corresponding to -3\%(+3\%) is due to the 
two remote reactor cores contribution. The counter-intuitive effect is explained by the greater (smaller) influence of the
destructive oscillation pattern of the two remote cores than the increase (decrease) of the flux.
}
\end{table*}

\begin{table*}[tb]
\begin{center}
{\footnotesize 
\begin{tabular}{c c | c c c c  }
\hline
\multicolumn{6}{l}{8 near reactor cores plus the 2 remote cores} \\
\hline& & \multicolumn{4}{c}{2 years}  \\
\cline{3-6}
& & 4\% & 3.5\% & 3\% & 2.5\%   \\
\hline
\parbox[t]{2mm}{\multirow{5}{*}{\rotatebox[origin=c]{90}{NH true}}} 
& \multicolumn{1}{|c|}{$\mu_{NH}$} & 4192.6 $\pm$ 6.9 & 4149.4 $\pm$ 6.8 & 4083.4 $\pm$ 6.9 & 4016.1 $\pm$ 6.6\\
& \multicolumn{1}{|c|}{$\sigma_{NH}$} & 212.4 $\pm$ 5.0  & 214.6 $\pm$ 4.4 & 215.2 $\pm$ 4.5 & 209.2 $\pm$ 4.5 \\
& \multicolumn{1}{|c|}{$\mu_{IH}$} & 4606.3 $\pm$ 7.2 & 4639.1 $\pm$ 7.4 & 4699.3 $\pm$ 7.2 &  4747.1 $\pm$ 7.2\\
& \multicolumn{1}{|c|}{$\sigma_{IH}$} & 225.2 $\pm$ 4.7 & 232.1 $\pm$ 4.8 & 226.6 $\pm$ 4.7 & 228.1 $\pm$ 4.7 \\
& \multicolumn{1}{|c|}{$r_{NH}$} &-0.570 $\pm$ 0.012 & -0.555 $\pm$ 0.019& -0.545 $\pm$ 0.020& -0.566 $\pm$ 0.013\\
\hline 
\parbox[t]{2mm}{\multirow{5}{*}{\rotatebox[origin=c]{90}{IH true}}} 
& \multicolumn{1}{|c|}{$\mu_{NH}$} & 4637.6 $\pm$ 6.7 & 4711.0 $\pm$ 7.6 & 4775.1 $\pm$ 7.0 & 4841.5 $\pm$ 7.5 \\
& \multicolumn{1}{|c|}{$\sigma_{NH}$} & 239.8 $\pm$ 3.7 & 236.9 $\pm$ 4.8 & 219.9 $\pm$ 4.5 & 232.2 $\pm$ 4.9 \\
& \multicolumn{1}{|c|}{$\mu_{IH}$} & 4167.6 $\pm$ 5.2 & 4095.3 $\pm$ 7.2 & 4018.7 $\pm$ 6.8 & 3951.7 $\pm$ 6.8 \\
& \multicolumn{1}{|c|}{$\sigma_{IH}$} & 224.3 $\pm$ 3.7 & 225.1 $\pm$ 4.6 & 212.6 $\pm$ 4.4 & 211.5 $\pm$ 4.4  \\
& \multicolumn{1}{|c|}{$r_{IH}$} & -0.588 $\pm$0.019 & -0.597 $\pm$ 0.017 & -0.552$\pm$ 0.019 & -0.601 $\pm$ 0.013\\
\hline
& \multicolumn{1}{|c|}{$p-$value (IH)} &$7.60\times 10^{-2}$ & $3.14\times 10^{-2}$ & $6.71\times 10^{-3}$ & $1.57\times 10^{-3}$\\
& \multicolumn{1}{|c|}{$n\, \sigma$ (IH)} & 1.77 & 2.15 & 2.71& 3.16\\
& \multicolumn{1}{|c|}{$p-$value (NH)} &$9.44\times 10^{-2}$ & $3.59\times 10^{-2}$ & $6.08\times 10^{-3}$ & $1.72\times 10^{-3}$\\
& \multicolumn{1}{|c|}{$n\, \sigma$ (NH)} & 1.67 & 2.10 & 2.74 & 3.13 \\
\hline\hline
\multicolumn{6}{l}{8 reactor cores without the two remote cores} \\
\hline& & \multicolumn{4}{c}{2 years}  \\
\cline{3-6}
& & 4\% & 3.5\% & 3\% & 2.5\%   \\
\hline
\parbox[t]{2mm}{\multirow{5}{*}{\rotatebox[origin=c]{90}{NH true}}} 
& \multicolumn{1}{|c|}{$\mu_{NH}$} & 4026.9 $\pm$ 6.5 & 3979.9 $\pm$ 4.1 & 3912.9 $\pm$ 6.6 & 3845.6 $\pm$ 5.5\\
& \multicolumn{1}{|c|}{$\sigma_{NH}$} & 203.9 $\pm$ 4.2  & 204.7 $\pm$ 3.2 & 203.6 $\pm$ 3.2 & 198.4 $\pm$ 2.9 \\
& \multicolumn{1}{|c|}{$\mu_{IH}$} & 4438.1 $\pm$ 6.9 & 4474.3 $\pm$ 3.6 & 4537.4 $\pm$ 7.1 &  4590.8 $\pm$ 4.7\\
& \multicolumn{1}{|c|}{$\sigma_{IH}$} & 215.8 $\pm$ 4.4 & 221.9 $\pm$ 3.2 & 219.3 $\pm$ 2.7 & 220.9 $\pm$ 4.8 \\
& \multicolumn{1}{|c|}{$r_{NH}$} &-0.575 $\pm$ 0.018 & -0.564 $\pm$ 0.015& -0.553 $\pm$ 0.014& -0.575 $\pm$ 0.013\\
\hline 
\parbox[t]{2mm}{\multirow{5}{*}{\rotatebox[origin=c]{90}{IH true}}} 
& \multicolumn{1}{|c|}{$\mu_{NH}$} & 4474.2 $\pm$ 7.3 & 4546.0 $\pm$ 7.1 & 4614.9 $\pm$ 6.8 & 4680.5 $\pm$ 7.2 \\
& \multicolumn{1}{|c|}{$\sigma_{NH}$} & 225.7 $\pm$ 4.8 & 223.7 $\pm$ 4.6 & 212.9 $\pm$ 4.4 & 224.5 $\pm$ 3.9 \\
& \multicolumn{1}{|c|}{$\mu_{IH}$} & 3998.8 $\pm$ 6.8 & 3928.5 $\pm$ 6.9 & 3849.7 $\pm$ 6.4 & 3782.6 $\pm$ 5.6 \\
& \multicolumn{1}{|c|}{$\sigma_{IH}$} & 210.1 $\pm$ 4.5 & 215.4 $\pm$ 4.4 & 200.9 $\pm$ 4.1 & 202.1 $\pm$ 4.1  \\
& \multicolumn{1}{|c|}{$r_{IH}$} & -0.597 $\pm$0.012 & -0.572 $\pm$ 0.018 & -0.571$\pm$ 0.018 & -0.580 $\pm$ 0.018\\
\hline
& \multicolumn{1}{|c|}{$p-$value (IH)} &$6.48\times 10^{-2}$ & $2.33\times 10^{-2}$ & $4.33\times 10^{-3}$ & $8.34\times 10^{-4}$\\
& \multicolumn{1}{|c|}{$n\, \sigma$ (IH)} & 1.85 & 2.27 & 2.85& 3.34\\
& \multicolumn{1}{|c|}{$p-$value (NH)} &$7.41\times 10^{-2}$ & $2.65\times 10^{-2}$ & $3.93\times 10^{-3}$ & $9.26\times 10^{-4}$\\
& \multicolumn{1}{|c|}{$n\, \sigma$ (NH)} & 1.79 & 2.22 & 2.88 & 3.31 \\
\hline
\end{tabular}
}
\end{center}
\caption{\label{table:7} 
The bi-gaussian fits of the $F$ distributions are reported, for the
JUNO-like configuration of two years of data taking and different energy resolutions,
$\mu_{MH}$, $\sigma_{MH}$ and $r_{MH}$ being the means, the standard deviations and the correlation 
coefficients, respectively, of the fitted 2-D Gaussians.
The 8 near reactor cores foreseen to be available at the starting of JUNO 
have been considered with a $\pm$ 5 m uniform dispersion 
on their relative baseline, as well as the 2 remote reactor plants with a $\pm$ 0.5 km uniform dispersion 
on their relative baseline. No background has been considered. 
The sensitivity 
has been computed from the $p$-values estimation as described in the text in terms of number of $\sigma$'s 
in the two-sided option. $n\, \sigma$ (IH) stays for the IH rejection significance, and equivalently for NH.}
\end{table*}




\end{document}